
\def\today{\number\day\ \ifcase\month\or January \or February \or March \or
April \or May \or June \or July \or August \or September \or October \or
November \or December\fi\space\number\year}

\input harvmac.tex
\input epsf.tex
\overfullrule=0mm

\catcode`\^^Q=11 \uccode`\^^Q=`\^^Q \lccode`\^^Q=`\^^Y
\catcode`\^^C=11 \uccode`\^^C=`\^^C \lccode`\^^C=`\^^K
\catcode`\^^P=11 \uccode`\^^P=`\^^P \lccode`\^^P=`\^^X
\catcode`\^^W=11 \uccode`\^^W=`\^^W \lccode`\^^W=`\^^_
\catcode`\^^Y=11 \uccode`\^^Y=`\^^Q \lccode`\^^Y=`\^^Y
\catcode`\^^?=11 \lccode`\^^?=`\^^? \uccode`\^^?=`\^^_
\catcode`\^^>=11 \lccode`\^^>=`\^^> \uccode`\^^>=`\^
\catcode`\^^K=11 \uccode`\^^K=`\^^C \lccode`\^^K=`\^^K
\catcode`\^^X=11 \uccode`\^^X=`\^^P \lccode`\^^X=`\^^X
\catcode`\^^_=11 \uccode`\^^_=`\^^W \lccode`\^^_=`\^^_
\lccode`\_=`\^^?
\lccode`\^=`\^^>
\font\cyr=wncyr10

  \def\J{^^Q} 
\def\I{\accent'44I}

\newcount\figno
\figno=0
\def\fig#1#2#3{
\par\begingroup\parindent=0pt\leftskip=1cm\rightskip=1cm\parindent=0pt
\baselineskip=11pt
\global\advance\figno by 1
\midinsert
\epsfxsize=#3
\centerline{\epsfbox{#2}}
\vskip 12pt
{\bf Fig. \the\figno:}{ #1}\par
\endinsert\endgroup\par
}
\def\figlabel#1{\xdef#1{\the\figno}}
\def\encadremath#1{\vbox{\hrule\hbox{\vrule\kern8pt\vbox{\kern8pt
\hbox{$\displaystyle #1$}\kern8pt}
\kern8pt\vrule}\hrule}}

\def\blank#1{}



\def\za{\alpha} \def\zb{\beta} \def\zg{\gamma} \def\zd{\delta}
\def\ze{\varepsilon}   
\def\zk{\kappa} \def\zl{\lambda}

\def\bl{\bar{\lambda}}


\def\CA{{\cal A}}        
    \def\CE{{\cal E}}    
    \def\CH{{\cal H}}    \def\CI{{\cal I}}
        \def\CL{{\cal L}}
        
        \def\CR{{\cal R}}
        
\def\CV{{\cal V}}        
    
%
%


\def\IR{\relax{\rm I\kern-.18em R}}
\font\cmss=cmss10 \font\cmsss=cmss10 at 7pt
\def\IZ{\relax\ifmmode\mathchoice
{\hbox{\cmss Z\kern-.4em Z}}{\hbox{\cmss Z\kern-.4em Z}}
{\lower.9pt\hbox{\cmsss Z\kern-.4em Z}}
{\lower1.2pt\hbox{\cmsss Z\kern-.4em Z}}\else{\cmss Z\kern-.4em Z}\fi}
\def\inbar{\,\vrule height1.5ex width.4pt depth0pt}
\def\IB{\relax{\rm I\kern-.18em B}}
\def\ID{\relax{\rm I\kern-.18em D}}
\def\IE{\relax{\rm I\kern-.18em E}}
\def\IF{\relax{\rm I\kern-.18em F}}
\def\IG{\relax\hbox{$\inbar\kern-.3em{\rm G}$}}
\def\IH{\relax{\rm I\kern-.18em H}}
\def\II{\relax{\rm I\kern-.18em I}}
\def\IK{\relax{\rm I\kern-.18em K}}
\def\IL{\relax{\rm I\kern-.18em L}}
\def\IM{\relax{\rm I\kern-.18em M}}
\def\IN{\relax{\rm I\kern-.18em N}}
\def\IO{\relax\hbox{$\inbar\kern-.3em{\rm O}$}}
\def\IP{\relax{\rm I\kern-.18em P}}
\def\IQ{\relax\hbox{$\inbar\kern-.3em{\rm Q}$}}
\def\IGa{\relax\hbox{${\rm I}\kern-.18em\Gamma$}}
\def\IPi{\relax\hbox{${\rm I}\kern-.18em\Pi$}}
\def\ITh{\relax\hbox{$\inbar\kern-.3em\Theta$}}
\def\IOm{\relax\hbox{$\inbar\kern-3.00pt\Omega$}}

\def\Z{\IZ}

\def\R{\IR}

\def\un{{\bf 1}}
\def\bra{\langle}\def\ket{\rangle}
\def\Fo{{{}^{(1)}\!F}}
\def\tFo{{}^{(1)}\!\tilde{F}}

\def\Fot{{{}^{(2)}\!F}}
\def\tFot{{{}^{(2)}\!\tilde{F}}}
\def\Exp{{\rm Exp}}
\def\tExp{\widetilde{\rm Exp}}
\def\slh{\widehat{sl}}
\def\omit#1{{}}

\def\tF{\tilde{F}}

\def\IC{\relax\hbox{$\inbar\kern-.3em{\rm C}$}}

\input amssym.def
\input amssym.tex
\def\IZ{\Bbb Z}\def\IR{\Bbb R}\def\IC{\Bbb C}\def\IN{\Bbb N}
\def\II{\Bbb I}\def\IP{\Bbb P}
\def\gg{\goth g} \def\gh{\bar{\goth h}}
\def\gA{\goth A} \def\gAe{{\goth A}^{\rm ext}}


\def\Ge{\epsilon}
\newdimen\xraise \newcount\nraise
\def\xpoint{\hbox{\vrule height .45pt width .45pt}}
\def\udiag#1{\vcenter{\hbox{\hskip.05pt\nraise=0\xraise=0pt
\loop\ifnum\nraise<#1\hskip-.05pt\raise\xraise\xpoint
\advance\nraise by 1\advance\xraise by .4pt\repeat}}}
\def\ddiag#1{\vcenter{\hbox{\hskip.05pt\nraise=0\xraise=0pt
\loop\ifnum\nraise<#1\hskip-.05pt\raise\xraise\xpoint
\advance\nraise by 1\advance\xraise by -.4pt\repeat}}}
\def\bdiamond#1#2#3#4{\raise1pt\hbox{$\scriptstyle#2$}
\,\vcenter{\vbox{\baselineskip12pt
\lineskip1pt\lineskiplimit0pt\hbox{\hskip10pt$\scriptstyle#3$}
\hbox{$\udiag{30}\ddiag{30}$}\vskip-1pt\hbox{$\ddiag{30}\udiag{30}$}
\hbox{\hskip10pt$\scriptstyle#1$}}}\,\raise1pt\hbox{$\scriptstyle#4$}}
\def\badiamond#1#2#3#4{\raise1pt\hbox{$\scriptstyle#2$}
\,\vcenter{\vbox{\baselineskip12pt
\lineskip1pt\lineskiplimit0pt\hbox{\hskip10pt$\scriptstyle#3$}
\hbox{$\swarrow\!\searrow$}\vskip4pt
\hbox{$\searrow\swarrow$}
\hbox{\hskip10pt$\scriptstyle#1$}}}\,\raise1pt\hbox{$\scriptstyle#4$}}
\def\bj{\bar{j}}
\def\bt{\bar{t}}
\def\bp{\bar{p}}
\def\bk{\bar{k}}

\def\I{I}\def\J{J}\def\K{K}\def\Jp{J'}

\def\cn{{\check n}}

\def\tm{\widetilde m}
\def\tn{\tilde n}
\def\td{\tilde d}

\def\tV{{\tilde V}} 
\def\tCV{{\widetilde {\cal V}}}

\def\tA{{\widetilde A}}
\def\tG{{\widetilde G}}
\def\tM{{\widetilde M}}
\def\tN{{\widetilde N}}
\def\tP{{\widetilde P}}
\def\bi{\bar i}
\def\bl{\bar l}
\def\bj{\bar j}

\def\bw{\bar w}
\def\bz{\bar z}

\def\hN{{\hat N}}
\def\hA{\hat{\CA}}
\def\hCA{\hat{\CA}}
\def\hV{\hat{V}}
\def\hB{\hat{B}}

\def\H{ \hbox{\raise -0.5mm\hbox{\epsfxsize=4mm\epsfbox{H.eps}}}}
\def\dH{ \hbox{\raise -0.5mm\hbox{\epsfxsize=4mm\epsfbox{dH.eps}}}}

\def\cmp#1#2#3{{\it Comm. Math. Phys.} {\bf  #1} (#2) #3}
\def\ijmpa#1#2#3{{\it Int. J. Mod. Phys.} {\bf A  #1} (#2) #3}
\def\ijmpb#1#2#3{{\it Int. J. Mod. Phys.} {\bf B  #1} (#2) #3}
\def\jpa#1#2#3{{\it J. Phys.} {\bf A  #1} (#2) #3}
\def\jsp#1#2#3{{\it J. Stat. Phys.} {\bf  #1} (#2) #3}
\def\lmp#1#2#3{{\it Lett. Math. Phys.} {\bf  #1} (#2) #3}
\def\mpla#1#2#3{{\it Mod. Phys. Lett.} {\bf A  #1} (#2) #3}

\def\npb#1#2#3{{\it Nucl. Phys.} {\bf B #1} (#2) #3}

\def\plb#1#2#3{{\it Phys. Lett.} {\bf B #1} (#2) #3}
\def\tmp#1#2#3{{\it Teor. Mat. Fiz.} {\bf  #1} (#2) #3}

\def\mathph#1{{\tt math-ph/#1}}
\def\hepth#1{{\tt hep-th/#1}}
\def\qalg#1{{\tt q-alg/#1}}
\def\mathOA#1{{\tt math.OA/#1}}
\def\mathQA#1{{\tt math.QA/#1}}


\lref\AO{A. Ocneanu, {\it Paths on Coxeter diagrams: From
Platonic solids and  
singularities to minimal models and subfactors},  
in {\sl Lectures on Operator Theory, }
 {Fields Institute, Waterloo, Ontario, April 26--30, 1995, }{(Notes taken by
S. Goto)}
{Fields Institute Monographies, AMS 1999,}{ Rajarama Bhat et al, eds.}}

\lref\AOb{A. Ocneanu, 
{\sl Quantum symmetries for $SU(3)$ CFT Models, }
{Lectures at Bariloche Summer School, }
{Argentina, Jan 2000}, to appear 
in AMS Contemporary Mathematics, R. Coquereaux, A. Garcia and
R. Trinchero, eds.}

\lref\AOold{A. Ocneanu, 
{\sl Quantized group string algebras and Galois theory for algebras}, 
in {\it Operator Algebras and applications}, vol 2, (Warwick 1987), London
Math. Soc. Lect. Note Series vol 136, Cambridge U. Pr. p 119-172.}

\lref\BE{J. B\"ockenhauer and D.E.  Evans,
\cmp{205}{1999}{183-228}, \hepth{9812110}.}

\lref\BEb{J. B\"ockenhauer and D.E.  Evans,
\cmp{200}{1999}{57-103}, \hepth{9805023}.}

\lref\BEc{J. B\"ockenhauer and D.E.  Evans, 
{\it Modular invariants from subfactors: Type I coupling matrices
 and intermediate subfactors}, \mathOA{9911239}.}

\lref\Xu{F. Xu,  \cmp{192}{1998}{349-403}. }

\lref\BEK{
J. B\"ockenhauer, D. E. Evans and Y. Kawahigashi,
\cmp{208}{1999}{429-487}, \mathOA{9904109}; 
\cmp{210}{2000}{733-784}, \mathOA{9907149}. }

\lref\BSz{ G. B\"ohm and K. Szlach\'anyi, \lmp{200}{1996}{437-456}, 
\qalg{9509008}; 
{\sl Weak $c^*$-Hopf algebras: The coassociative symmetry of 
non-integral dimensions}, in: {\it Quantum Groups and Quantum Spaces},
Banach Center Publ. v. 40, (1997) 9-19
\semi G. B\"ohm,  {\sl Weak $C^*$-Hopf Algebras and their
Application to Spin Models}, PhD Thesis, Budapest 1997.}

\lref\NTV{D.  Nikshych, V.  Turaev and L.  Vainerman, 
 {\sl Invariants of knots and 3-manifolds from quantum groupoids}, 
 \mathQA{0006078}.}

\lref\McL{S. Mac Lane, Categories for the  Working Mathematician, 2th ed., 
GTM {\bf 5}, Springer, 1998. }
 

\lref\MR{G. Moore  and N.Yu. Reshetikhin, \npb{328}{1989}{557-574}.}

\lref\GS{C. G\'omez and H. Sierra, \plb{240}{1990}{149-157}; 
\npb{352}{1991}{791-828}; {\sl A brief history of hidden quantum
symmetries in Conformal Field Theories}, \hepth{9211068}.  }

\lref\MSch{G. Mack and V. Schomerus, \npb{370}{1992}{185-230}. }

\lref\MSchb{G. Mack and V. Schomerus, \cmp{134}{1990}{139-196}.}

\lref\BFMT{D. Buchholz, I. Frenkel, G. Mack and I.T. Todorov,
as reported in : I.T. Todorov, 
{\it Lect. Notes in Physics} {\bf 370} (1990)  231-277,
(Springer).}

\lref\FGP{P. Furlan, A.Ch. Ganchev and V.B. Petkova, 
\ijmpa{6}{1991}{4859-4884}.} 

\lref\KR{A.N. Kirillov and N.Yu. Reshetikhin,  
{\sl Representations of the algebra $U_q(sl(2))$, $q$-orthogonal
polynomials and invariants of links }, 
{\it  Adv. Series in Math. Phys.} {\bf 7} (1989), 285-339.}

\lref\Reh{ C.-H. Rehren, \cmp{116}{1988}{675-688}.}


\lref\CIZ{A. Cappelli, C. Itzykson and J.-B. Zuber,
\npb{280}{1987}{445-465}; 
\cmp{113}{1987}{1-26} \semi 
 A. Kato, \mpla{2}{1987}{585-600}.} 

\lref\DIFZ{P. Di Francesco and J.-B. Zuber,
\npb{338}{1990}{602-646}.}


\lref\Pas{V. Pasquier, \npb{285}{1987}{162-172}.}

\lref\Pasq{V. Pasquier, \jpa{20}{1987}{5707-5717}.}

\lref\Vpa{V. Pasquier, {\sl Mod\`eles Exacts Invariants Conformes},
 Th\`ese d'Etat, Orsay 1988. }

\lref\VP{V. Pasquier, \cmp{118}{1988}{355-364}.  }

\lref\DIFZb{P. Di Francesco and J.-B.  Zuber, 
{\sl $SU(N)$ Lattice Integrable Models and Modular Invariance,
 Recent Developments in Conformal Field Theories}, Trieste
Conference, (1989), S. Randjbar-Daemi, E. Sezgin and J.-B. Zuber eds.,
World Scientific (1990); 
P. Di Francesco, \ijmpa{7}{1992}{407-500}.} 

\lref\BPO{R.E. Behrend, P.A. Pearce and D.L. O'Brien, 
\jsp{84}{1996}{1-  } .}

\lref\BP{R.E. Behrend and P.A.  Pearce, \jpa{29}{1996}{7827-7836}.}

\lref\KRS{P.P. Kulish, N.Yu. Reshetikhin and E.K. Sklyanin,
\lmp{5}{1981}{393-403}.}

\lref\Kul{P.P. Kulish,  \hepth{9507070}. }

\lref\Ch{I.V. Cherednik, \tmp{61}{1984}{35-44}.}
%
%

\lref\Soch{N. Sochen, \npb{360}{1991}{613-640}.}

\lref\Roch{P. Roche, \cmp{127}{1990}{395-424}.}

\lref\JMO{M. Jimbo, T.  Miwa  and M. Okado, \lmp{14}{1987}{123-131}.}

\lref\JKMO{M. Jimbo, A.  Kuniba, T.  Miwa and M. Okado, 
\cmp{119}{988}{543-565}.}

\lref\Wen{H. Wenzl, {\it Inv. Math.} {\bf 92} (1988) {349-383}.}

\lref\PZh{P.A. Pearce and Y.K. Zhou, \ijmpb{B7}{1993}{3649-3705},
\hepth{9304009}.}

\lref\BPa{R.E. Behrend and P.A.  Pearce,  
{\sl Integrable and conformal boundary conditions for sl(2) A-D-E
lattice models and unitary minimal conformal field theories}, 
\jsp{}{to appear}{} \hepth{0006094}. }


\lref\MS{G. Moore and N. Seiberg, \cmp{123}{1989}{177-254}.}

\lref\DV{R. Dijkgraaf and E. Verlinde, {\it Nucl. Phys.} (Proc. Suppl.)
{\bf 5B}  (1988) 87.}

\lref\Reh{ C.-H. Rehren, \cmp{116}{1989}{675-688}.}



\lref\PZa{V.B. Petkova and J.-B. Zuber, 
\npb{438}{1995}{347-372}, \hepth{9410209}.}

\lref\PZb{V.B. Petkova and J.-B.  Zuber, 
\npb{463}{1996}{161-193}, \hepth{9510175}; 
{\sl Conformal Field Theory and Graphs},
 \hepth{9701103}.}


\lref\VPlon{V.B. Petkova, {\sl Chiral approach to the (non)-diagonal
2D conformal models},  Proceedings of the
13th International Congress in Mathematical Physics (ICMP 2000),
July 2000, London, to appear.}

\lref\PZtmr{ V.B. Petkova and J.-B. Zuber,
%
{}PRHEP-tmr2000/038  (Proceedings of the TMR network conference
{\it Nonperturbative Quantum Effects 2000}),
 \hepth{0009219}.}

\lref\PZtw{V.B. Petkova and J.-B. Zuber,
\plb{504}{2001}{157-164}, \hepth{0011021}.}

\lref\PZbud{V.B. Petkova and J.-B. Zuber, Budapest, August 2000. }


\lref\BPZ{R.E. Behrend, P.A. Pearce  and J.-B.  Zuber,
\jpa{31}{1998}{L763-L770}, \hepth{9807142}.}

\lref\BPPZ{R.E. Behrend, P.A. Pearce, V.B. Petkova and J.-B.  Zuber, 
\plb{444}{1998}{163-166}, hep-th/9809097; 
 \npb{579}{2000}{707-773}, \hepth{9908036}. 
}

\lref\Car{J.L. Cardy, \npb{270}{1986}{186-204}.}

\lref\Ishi{N. Ishibashi, \mpla{4}{1987}{251-264}.}

\lref\Lew{D.C. Lewellen, \npb{372}{1992}{654-682}.}

\lref\CL{J.L. Cardy and D.C. Lewellen, \plb{259}{1991}{274-278}. }

\lref\R{I. Runkel, \npb{549}{1999}{563-578},  \hepth{9811178}.}

\lref\Rb{I. Runkel, \npb{579}{2000}{561-589}, \hepth{9908046}. }

\lref\RSch{A. Recknagel and V. Schomerus, \npb{531}{1998}{185-225}, 
\hepth{9712186}, 
\npb{545}{1999}{233-282}, \hepth{9811237}.}

\lref\FS{J. Fuchs and C. Schweigert, \npb{530}{1998}{99-136},
\hepth{9712257}.}

\lref\PSS{G. Pradisi, A.  Sagnotti and  Ya.S. Stanev, 
\plb{381}{1996}{97-104}, \hepth{9603097}.}


\lref\BI{E. Bannai and T. Ito, {\it Algebraic combinatorics I:  
Association schemes}, New York: Benjamin/Cummings, 1984.}

\lref\Fen{P. Fendley,  \jpa{22}{1989}{4633-4642}.}

\lref\Coq{R. Coquereaux, {\it Notes on the quantum tetrahedron},
 \mathph{0011006}.}

\lref\Hon{A. Honecker, \npb{400}{1993}{574-596}, \hepth{9211130}.}
   
\lref\DJKMO{E. Date, M. Jimbo, A. Kuniba, T. Miwa and M. Okado,
{\it Adv. Stud. Pure Math.} {\bf 16} (1988) 17-122.} 

\lref\WDA{M. Wadati, T. Deguchi and Y. Akutsu, {\it Phys. Rep.} {\bf 180} 
(1989) 247-332. }

\lref\PZhf{P.A. Pearce and Y.K. Zhou, \ijmpb{B8}{1994}{3531-3577}.}

\lref\Dub{B. Dubrovin,
{\it Nucl. Phys.} {\bf B 379} (1992) 627-689:
{ hep-th/9303152};
Springer Lect. Notes in Math. {\bf 1620} (1996) 120-348:
{hep-th/9407018}}
\lref\many{W. Lerche and N.P. Warner, 
in {\it Strings \& Symmetries, 1991}, N. Berkovits, H. Itoyama et al. eds,
World Scientific 1992 \semi
P. Di Francesco, F. Lesage and J.-B. Zuber,
        \npb{408}{1993}{600-634},  \hepth{9306018}.
}
\lref\HT{L.K.  Hadjiivanov, I.T. Todorov, {\it Monodromy representations
of the braid group}, \hepth{0012099}.}


\lref\Hon{A. Honecker, \npb{400}{1993}{574-596}, \hepth{9211130}.}

\lref\FSoo{J. Fuchs and C. Schweigert, 
\plb{490}{2000}{163-172}, \hepth{0006181}.}

\lref\Ca{J.L. Cardy, \npb{275}{1986}{200-218}.}

\lref\Z{ J.-B.  Zuber, \plb{176}{1986}{127-129}.}

\lref\VBPb{V.B. Petkova, \ijmpa{3}{1988}{2945-2958};
\plb{225}{1989}{357-362};
P. Furlan, A.Ch. Ganchev and V.B. Petkova, 
\ijmpa{5}{1990}{2721-2735}. }

\lref\TP{J. Teschner,
{}PRHEP-tmr2000/041  (Proceedings of the TMR network conference
{\it Nonperturbative Quantum Effects 2000}),
 \hepth{0009138}; B. Ponsot and J. Teschner,
{\sl Clebsch-Gordan and Racah-Wigner coefficients for a continuous
series of representations of $U_q(sl(2,R))$}
\mathQA{0007097}.}


\Title{
\vbox{\vglue-8mm\baselineskip12pt
\hbox{ESI 975 (2000)}\hbox{UNN-SCM-M-00-10}\hbox{CERN-TH/2000-355} }}
{
\vbox {
\centerline{The many faces of Ocneanu cells }
}}
\medskip
\centerline{V.B. Petkova
\footnote{${}^*$}{
E-mail: {valentina.petkova@unn.ac.uk}}}
\bigskip
\centerline{\it Institute for Nuclear Research and Nuclear Energy,}
\centerline{\it 72 Tzarigradsko Chaussee,  1784 Sofia, Bulgaria,}
\medskip
\centerline{\it School of Computing and Mathematics,}
\centerline{\it University of Northumbria}
\centerline{\it NE1 8ST Newcastle upon Tyne, UK}
\bigskip
\centerline{and}
\medskip
\centerline{J.-B. Zuber
\footnote{${}^\flat$}{On leave from: SPhT, CEA Saclay, 
F-91191 Gif-sur-Yvette. E-mail:  {zuber@spht.saclay.cea.fr}}}
\centerline{\it TH Division, CERN, CH-1211 Gen\`eve 23}

\vskip .2in

\noindent
{\ninepoint
We define  generalised chiral vertex operators 
covariant under the Ocneanu ``double triangle algebra'' 
 ${\cal A}\,,$  a novel  quantum symmetry intrinsic 
to a given rational 2-d conformal field theory.
This provides a  chiral approach,
which,  unlike the conventional one,  makes explicit
various algebraic structures encountered previously in  
the  study of these theories and of the associated
critical lattice models,  and thus allows their unified treatment.
The triangular Ocneanu cells, the $3j$-symbols of 
the weak Hopf algebra ${\cal A}$, reappear in  several guises.
With ${\cal A}$  and its  dual algebra
${\hA}$ one associates a pair of graphs, $G$ and $\tG$. 
While $G$ are known to encode  complete sets of 
conformal boundary  states, the Ocneanu graphs $\tG$ classify
twisted  torus partition functions.
The fusion algebra of the twist operators provides 
the data determining ${\hA}$.
The study of  bulk field  correlators in the presence of twists 
reveals that the Ocneanu graph quantum symmetry 
 gives also an information on  the   field operator algebra.
}

\bigskip

\Date{}

%
\newsec{Introduction}

\noindent
This paper stems from the desire to understand Ocneanu
recent work on ``quantum groupoids'' \refs{\AO,\AOb}, also called, 
 in a  loose sense, ``finite subgroups of the quantum groups'', 
and to reformulate and 
to exploit it in the context of 2d rational conformal field theories (RCFT).
Our approach is inspired by the study of boundary conditions in CFT,
either on manifolds with boundaries, or on closed manifolds (e.g.
a torus)
where the introduction of defect lines (or twists) is possible.

In Boundary CFT (BCFT), the type of boundary states and the corresponding 
character multiplicities in cylinder partition functions
 are conveniently encoded in a graph (or a set of graphs)  $G$ \BPPZ,
with vertices denoted by $a,b$.  More precisely 
the adjacency matrices of the graphs are given by 
a set of (non negative integer valued) matrices $n_i=\{n_{ia}{}^b\}$
forming a representation of the Verlinde  fusion algebra
\eqn\nim{ n_i\, n_j=\sum_k\, N_{ij}{}^k\, n_k\,,}
and it is usually sufficient to specify only 
a ``fundamental'' subset of them, which generates the other through fusion. 

In accordance with these data we define generalised chiral vertex 
operators (GCVO), covariant under Ocneanu ``double triangle algebra'' (DTA)
$\CA$, a finite dimensional 
``$C^*$ weak Hopf algebra'' (WHA) in the axiomatic setting of \BSz.
They can be looked at as extensions to the complex plane
of the boundary fields and  at the same time
they yield  a precise operator meaning to these fields. 
The fact that the GCVO have 
nontrivial braiding allows to give a global operator definition
 of the half-plane 
bulk fields, described in the traditional approach only through their  small
distance (vanishing imaginary coordinate) expansion. 
 The bulk fields are defined as compositions of  two 
generalised or conventional CVO, which makes the construction of 
their correlators and the derivation of the equations they satisfy  
straightforward.

The $3j$-symbols $\Fo$ of the Ocneanu quantum symmetry, 
also called ``cells'',  reproduce the
boundary field operator product expansion  (OPE) 
coefficients, while the $6j$- symbols $F$
coincide with the fusing matrices, i.e., the OPE coefficients
 of the  conventional CVO.
In the ``diagonal theories'', in which each local field  
 is left-right symmetric, there is a one to one correspondence
 between the set $\CI$ and the spectrum of 
orthonormal boundary states;
 then \nim\ is realised by the Verlinde matrices themselves
 and  the two symbols $F$ and $\Fo$ coincide.  
The $3j$-symbols diagonalise the  braiding matrices of the 
generalised CVO (the $R$ matrix of the quantum symmetry). 
These new braiding matrices are identified  with  the Boltzmann weights
(in the limit $u\to \pm i \infty$ of their spectral parameter)
of the critical $sl(n)$ lattice models
which generalise the Pasquier $ADE$  lattice models and  their fused
versions.  Once again the $3j$- symbols provide
the basic ingredients of these models. In particular their
identification with the Ocneanu intertwining cells gives 
some new solutions for the boundary field  OPE coefficients
in the exceptional $E_r$ cases of $\slh(2)$ theories; for the 
$A$ and $D$-series these constants were computed in \refs{\R,\Rb}. 
\medskip

Through  a discussion parallel to that of boundary states, one may
also study the allowed twists (or defect lines) 
on a torus. The compatibility with conformal invariance and a duality 
argument similar to Cardy's consistency condition \Car\
restrict  the multiplicities $\tV_{ii'}=\{ \tV_{ii';\, x}{}^y\}$ of 
occurrence of representations $(i,i')$ in the presence of twists $x,y$, 
to be now non negative integer valued  matrix representations of 
the {\it squared} Verlinde fusion algebra \PZtw:
\eqn\tosya{ 
\tV_{ii'}\tV_{jj'}= 
\sum_{k,k'}\, N_{i j}{}^{k}\, N_{i' j'}{}^{k'}\,
\tV_{k k'}\,, \ \quad \tV_{ij^*;1}^{{}1}= Z_{ij} \,,
}
where $Z_{ij}$ is the modular invariant matrix.
 Pairs $\tV_{i1}\,, \tV_{1i'}$ of these matrices give rise to 
another graph $\tG$ with vertices  $x,y$ \refs{\AO,\AOb}. 
Combining the concepts of twists and of boundaries, 
i.e. inserting twists in the presence of boundaries, leads
to yet another set of multiplicities, $\tn_x=\{\tn_{ax}{}^b\}$, 
which form a matrix 
realisation of a new, in general non-commutative, fusion algebra:
\eqn\tnrep{\tn_x \tn_y=\sum_z\, \tN_{xy}{}^z\tn_z\ \,,
}
\eqn\tNrep{\tN_x \tN_y=\sum_z\, \tN_{xy}{}^z\tN_z\,.
} 
This algebra admits an interpretation as the algebra
of the twist operators used in the construction of
the partition functions in \PZtw. It is associated with the
Ocneanu graph $\tG$ in the sense of the relation 
\eqn\misys{
\tV_{ij}\, \tN_x=\sum_z\, \tV_{ij;\, x}{}^z\, \tN_z\,,  }
and we shall also refer to it as the $\tG$ {\it  graph algebra}.
In the cases described by a block-diagonal modular invariant
(a diagonal invariant of an extended theory)
it possesses 
subalgebras interpreted as
graph algebras of the chiral graph  $G$, and furthermore a subalgebra 
identified with the extended fusion algebra.  We find, extending the 
analysis in \PZtw\ to correlators in the presence of twists, that
the  representations of \tNrep\ are closely related to the operator
product algebra of the physical local fields of arbitrary spin.
\fig{The simplices}{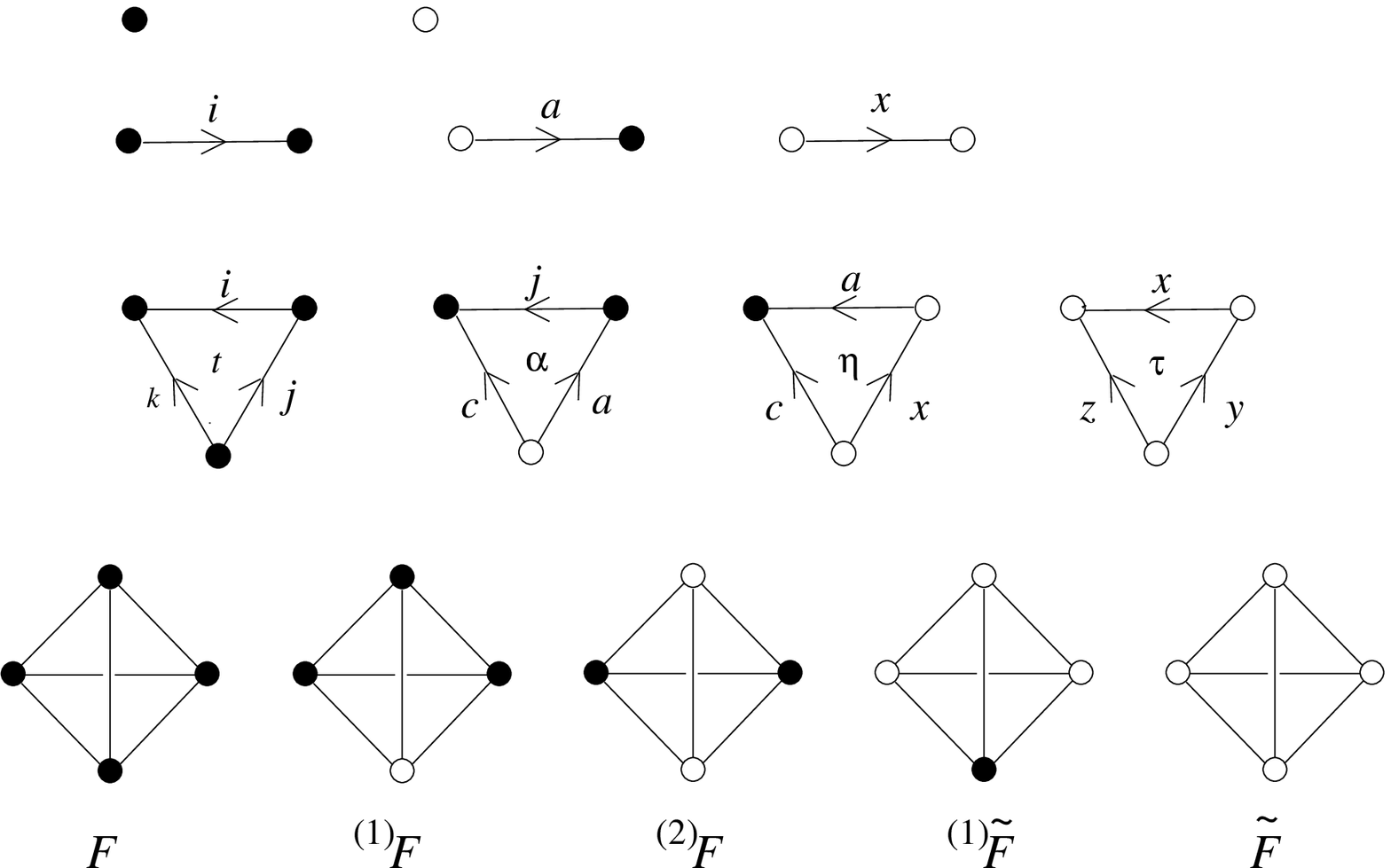}{9cm}\figlabel\simplex
In this approach, we see repeated   manifestations of the 
quantum algebra $\CA$ and of its dual algebra $\hCA$,
both satisfying the axioms of the  WHA  of \BSz. 
The structure as a whole  is maybe most easily described 
in the combinatorial terms of Ocneanu quantum (co)homology 
 \BSz\ (see also  the related notion of ``$2$-category'' in \McL
\foot{We thank A. Wassermann for pointing this out to us.}). 
The latter considers simplicial 3-complexes built
out of the elements depicted on Fig. \simplex. There are three types
of oriented 1-simplices and the triangular 2-simplices 
come with  multiplicities.
Each tetrahedral 3-simplex (arrows omitted in Fig. \simplex{}) 
 is assigned a $\Bbb{C}$-valued 3-chain, subject to  
a set of pentagon relations (the ``Big Pentagon'' of \BSz); the middle
tetrahedron $ \Fot$ appears with its inverse, $\tFot$, 
while  $F\,,\,\Fo\,,\,\tFo\,,\,\tF$ can be chosen unitary. 
These data enable one to construct on an abstract level 
$\CA$ and its  dual $\hA$,  
which are matrix algebras with basis elements represented by two 
sets of ``double triangles'', see Fig. 2., related, 
up to a constant, by $\Fot$.
\fig{The double triangles}{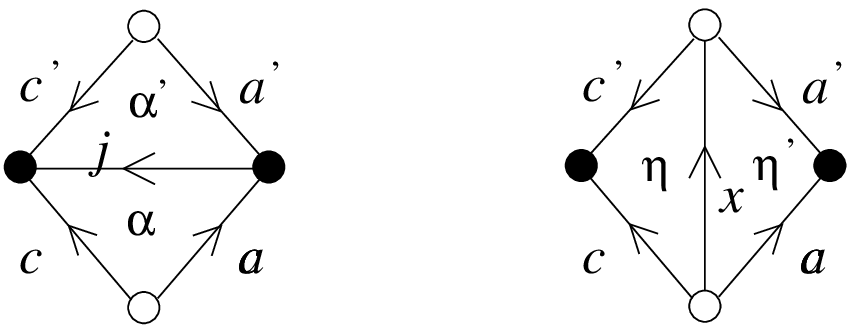}{4cm}\figlabel\dta
\noindent 
In the present context,   the 1-simplices
are labelled by the finite set $\CI$ of representations
of the Verlinde fusion algebra 
and by the sets $\CV$ and $\tCV$ of vertices of graphs $G$ and $\tG$ of
 cardinality  $|\CV|=\tr(Z)\,$ and $|\tCV|=\tr(Z\, ^t\!\!Z)\,$ respectively,
where $Z$ is the modular invariant matrix.  Each 
 triangular 2-simplex comes with a multiplicity label
$t=1,\cdots,N_{ij}{}^k$, $\za=1,\cdots,n_{ia}{}^c$, 
$\eta=1,\cdots, \tn_{ax}{}^c$, $\tau=1,\cdots,\tN_{xy}{}^z$,
and these multiplicities are subject to the relations \nim\ - \misys.
The first two tetrahedra on Fig. \simplex\
 represent the $6j$- and the $3j$-symbols $F\,, \Fo$
discussed above.  

\medskip
Thus Ocneanu's double triangle algebra, which is attached specifically
to each $2d$ CFT and governs many of its aspects 
-- spectrum multiplicities, structure constants, lattice realisations --
appears as its natural quantum symmetry. 
The problem of identifying the underlying quantum symmetry  of
a given CFT is by no means new. Several attempts and partial 
answers were achieved at the end of the 80's and beginning of the 
90's,  see the discussion below in sections 4 and 5.
The previous approaches dealt with the chiral CFT, or equivalently, 
with the diagonal theories.  These are 
also the only examples of CFT  discussed in \BSz, where the relevance 
of the WHA as a quantum symmetry 
was first proposed in the framework of algebraic QFT; in these diagonal
cases the four triangle  multiplicities above coincide with the 
Verlinde fusion multiplicities  $N_{ij}{}^k$ and accordingly all the 
tetrahedra on Fig. \simplex\ reduce to the RCFT fusing matrix $F$.
The development of BCFT on one hand side and the
work of Ocneanu on the other made available new tools and new ideas;
our present considerations yield in particular 
explicit and non trivial examples of the structure of WHA. 
The main novelty of the WHA approach is that it has a coassociative
coproduct consistent with the CFT fusion rules (the Ocneanu 
``horizontal'' product). The presence of boundaries 
provides an extension of the  Hilbert space of the theory consistent
with the fusion rules and basic axioms of the RCFT.
At the same time it should be stressed that the parallel 
with the 
previous discussions on the ``hidden'' quantum group symmetry 
is to some extent superficial,   or deceptive, since this is only
one of the facets of the Ocneanu symmetry;  in contrast
to the former   
the new approach encompasses the full structure of the 2d CFT,
so is much richer in content and applications. 

\medskip
We should not conceal, however, that our understanding is still
fragmentary.  The determination of the cells and of
the remaining tetrahedra
of Fig. \simplex\ from the complicated set of equations they satisfy
poses a difficult technical problem and only partial results
in the $\slh(2)$  related models are known. 
Some of the previous quantities, related in particular to 
the dual structure of the DTA, are still awaiting a better 
field theoretic interpretation. Moreover, several of our results are 
conjectures, tested mainly on the case of $\slh(2)$, but lack 
a general proof. On several of these points, it 
seems that the approach based on the theory of subfactors 
\refs{\AO, \BE, \BEK} is  
more systematic.  Still, our field theoretic approach provides
explicit realisations and exposes some new facts which show 
the consistency of the whole picture.
\medskip

This paper is organised as follows: after a brief summary of notations
(section 2) we introduce  the 
double triangle algebra (section 3),
  then define  the  GCVO and discuss their fusing and 
braiding properties  (section 4).  In section 5  
we show how the  bulk fields may be expressed in terms of 
GCVO and how the equations they satisfy
and the various OPE coefficients may be 
rederived in a more systematic way. Section 6 discusses briefly the
 relation to the
lattice models and the determination of their Boltzmann weights
in terms of the cells. Finally, section 7   deals with the 
construction of solutions of \tosya-\misys\ and of the 
resulting Ocneanu graphs $\tG$  and contains 
a derivation of a formula relating the 
OPE coefficients of arbitrary spin fields to data of the graph. 
Details are relegated to two appendices. 
Sections 5, 6 and 7 may be read independently of one another.
\medskip

Preliminary accounts of this work have been reported at several 
conferences  
(ICMP, London, July 2000; 24th Johns Hopkins Workshop,  Budapest, 
 August 2000; TMR Network Conference, Paris, September 2000 \PZtmr;
Kyoto Workshop on Modular Invariance, 
ADE, Subfactors and Geometry of Moduli Spaces,  
November, 2000)  or have been published separately \PZtw.
It should be stressed that this work was strongly influenced by 
Ocneanu's (unfortunately unpublished) work and that many of the 
concepts and results presented here originate in his work.

\newsec{Notations}

\noindent 
A rational conformal field theory is
conventionally described by data of different nature : \par
\noindent $\bullet$ {\it Chiral data}
specify the chiral algebra $\gA$ and its finite set $\CI$
of irreducible representations $\CV_i$, $i\in\CI$, 
the characters $\chi_i(q)=\tr_{\CV_i} q^{L_0-c/24}$,  
the unitary and symmetric matrix $S_{ij}$ of modular transformations 
of the $\chi$, the fusion coefficients $N_{ij}{}^k$, $i,j,k\in \CI$, 
assumed to be given by Verlinde formula  
\eqn\verl
{N_{ij}{}^k=\sum_{\ell\in\CI}{S_{i\ell}S_{j\ell}S_{k\ell}\over S_{1\ell}}\ .}
Our convention is that the label $i=1$ refers to the 
``vacuum representation'', 
and $\CV_{i^*}$ denotes the representation conjugate to $\CV_i$.
Chiral vertex operators 
$\phi_{ij}^t(z)$ 
and their fusion and
braiding matrices $F_{pt}[{{i\,j\atop k\,l}}]$ and 
$B_{pt}[{{i\,l\atop k\,j}}]$ 
are also part of the set of chiral data.\par
\noindent $\bullet$ {\it Spectral data} specify which representations 
of $\gA\otimes\gA$ appear in the bulk : these 
data are usually conveniently encoded in the partition function
on a torus, with the property of modular invariance
\eqn\toruspf{Z=\sum_{i,j\in\CI}Z_{ij}\chi_i(q)(\chi_j(q))^*\ ;  } 
here, the integer $Z_{ij}$ specifies the multiplicity of occurrence 
of $\CV_i\otimes \overline{\CV_j}$ in the Hilbert space of the theory; 
unicity of the vacuum is expressed by $Z_{11}=1\,$. \par
\noindent $\bullet$ Finally these spectral data must be 
supplemented by data on the structure constants of the
Operator Product Algebra (OPA).  This last set 
of data is the one which is most difficult to determine as
it results from the solution of  a large system of  
non linear equations involving  the  braiding matrices whose 
general form is in general unknown.

It has been recognized  some 
time ago that these spectral and OPA data have to do with graphs. The latter
($ADE$ Dynkin diagrams and their generalizations) (i) encode in the 
spectrum of their adjacency matrices the spectral data
\refs{\CIZ, \DIFZ, \DIFZb}; (ii) contain, through the so-called 
Pasquier algebra, information on the 
OPA structure constants,
see \refs{\Pasq,\PZa,\PZb} and below, section 7. In fact these graphs are 
nothing else than the graphs of adjacency matrices $n_i$ of \nim. 
These matrices $n_i$ 
are diagonalisable in a common orthonormal basis:
\eqn\int{
n_{i a}{}^b =\sum_{j\in {\rm Exp}}\, {S_{ij}\over S_{1j}}\, 
\psi_a^j\, \psi_b^{j*}
}
and obey the identities
\eqn\sint{
n_{i a}{}^b=n_{i^* a^*}{}^{b^*}=n_{i^* b}{}^a
\,.
}
Here and throughout this paper, we make use of the  notation 
$\Exp$ to denote the terms appearing in the diagonal part of the 
modular invariant \toruspf
\eqn\defExp{\Exp=\{(j,\za), \ \za=1,\cdots, Z_{jj}\} \ .}
The two notations $\psi^{(j,\za)}$ and $\psi^j$, $j\in\Exp$ will 
be used interchangeably. 
In the following, $\psi^1$ refers to the Perron-Frobenius eigenvector,
whose components are all positive. Finally, in \sint, 
the conjugation of vertices 
$ a\to a^*$ is defined through  $\psi_{a^*}^j=(\psi_a^j)^*
=\psi_a^{j^*}$. 
\par\noindent
A particular  set of
matrices $n$ is provided by the Verlinde matrices $N$ themselves, 
which form the regular representation of the fusion algebra. 
This is the diagonal case 
for which ${\rm Exp}=\CI$ and the corresponding torus partition function is
simply given by  $Z_{ij} =\delta_{ij}$.


\newsec{ Ocneanu graph quantum algebra}

\noindent  Given a solution of the equation \nim\
 consider an auxiliary  Hilbert space $V^j\cong \IC^{m_j}\,$ 
with basis states $|e^{j,\zg}_{ba}\rangle$, $\zg=1,2,\dots,n_{j a}{}^b$.
It has dimension $m_j= \sum_{a,b}\,
n_{ja}{}^b=\sum_{a,b,\zg}\, 1$, in 
 particular ${\rm dim}\, V^1= \tr(n_1)=|\CV|$.
A scalar product in $\oplus_{j\in \CI}\, V^j$ is defined as
\eqn\sc{
 \langle e^{j,\zg}_{ ba} |
e^{j',\zg'}_{ b'a'}\rangle 
= \delta_{bb'} \delta_{aa'}\,
\delta_{jj'}\,\delta_{\zg' \zg}\,\sqrt{P_a\,P_b\over  d_j}
\,,\qquad d_j={S_{j1}\over S_{11}}\,,\quad P_a={\psi_a^1\over \psi_1^1}\,.
}
We define the 
 tensor product decomposition of states $|e^{i, \za}_{cb}\rangle\otimes
|e^{j,\zg}_{b' a}\rangle$ for coinciding $b'=b$ 
according to 
\eqn\Ima{\eqalign{
&|e^{i, \za}_{cb}\ket \otimes_h |e^{j,\zg}_{b a}\ket=
\, \sum_{k\in \CI }\ \sum_{\zb=1}^{n_{ka}{}^c}\
\sum_{t=1}^{N_{ij}{}^k}\  \Fo_{b k}\left[\matrix{i&j\cr c&a}\right]_{\za\, 
\zg}^{\zb\, t}\ \sqrt{P_b}\
\Big({d_k \over d_i d_j}\Big)^{{1\over 4}}\
|e^{k\,, \zb}_{ca}(ij;t)\ket 
\,.  }}
This is a ``truncated'' tensor product, in the sense that we restrict to a
subspace $V^i \otimes_h\, V^j\,$ of $V^i\otimes V^j\,,$
$(cb)\otimes_h\,(b'a)=\delta_{bb'}\ (cb)\otimes (b'a)\,,$ with dim$(V^i
\otimes_h\, V^j)=\sum_{a,c}\, (n_i n_j)_a{}^c \le m_i m_j$. 
The multiplicity of  $V^k$ in 
$V^i\otimes_h\, V^j$ is identified
with the Verlinde multiplicity $N_{ij}{}^k\,.$
 Then the counting of states in both sides of \Ima\  is consistent,
 taking into account  \nim.  In \Ima\ $e^{k\,, \zb}_{ca}(ij;t)$ give a 
basis, normalised as in \sc, for the space $V^k$ in 
\eqn\fusdec{V^i\otimes_h\, V^j\cong \oplus_{k}\, N_{ij}{}^k\,V^k\ .}
The $\Fo\in \IC$ are Clebsch-Gordan coefficients 
(``$3j$- symbols''), assumed to  satisfy the conditions: 
\item{$-$} if one of the indices $i$ or $j$ is equal to 1, the
tensor product must trivialise  and accordingly
\eqn\trivone{
 \Fo_{b k}\left[\matrix{1&j\cr c&a}\right]_{\za\,
\zg}^{\zb\, t}= \delta_{kj}\, \delta_{b c}\, \delta_{\zb \zg}
\delta_{t1}\delta_{\za 1}\,;}
\item{$-$} the unitarity conditions, expressing 
the  completeness and  orthogonality of the bases in $V^i\otimes_h V^j$
\eqn\unit{\eqalign{
 \sum_{b\,, \alpha\,, \gamma}\,\,
{}\Fo_{b k'}^*\left[\matrix{i&j\cr c&a} \right]_{\za\,\zg}^{\beta'\ t' }
{}\Fo_{b k}\left[\matrix{i&j\cr c&a} \right]_{\za\,\zg}^{\beta\  t }
&= \delta_{k k'}\,
\delta_{\beta\,\beta'}\, \delta_{t t'}\,,\cr
\sum_{k\,, \beta\,, t }\,\,
{}\Fo_{b' k}^*\left[\matrix{i&j\cr c&a} \right]_{\za'\,\zg'}^{\beta\ t }
{}\Fo_{b k}\left[\matrix{i&j\cr c&a} \right]_{\za\,\zg}^{\beta\ t }
&= \delta_{b b'}\, \delta_{\gamma\,\gamma'}\, \delta_{\alpha\, \alpha'}\,,
\cr}}
where $\Fo^*$ is the complex conjugate of $\Fo$.

{In the original (combinatorial) realisation of \AO\ (for the $ADE$
graphs of the case $sl(2)$), $V^j$
is the  linear space of ``essential paths of length $j$'' on the graph $G$.
Then \Ima\ is interpreted as a composition of essential paths,
which is not an essential path in general, but is a linear
combination of such paths. }

The requirement of associativity of the product \Ima\ leads to the  ``mixed''
 pentagon relation 
\eqn\mpone
{ F\  \Fo \ \Fo \, =   \Fo\ \Fo \,,}
or, more explicitly
\eqn\mp 
{\eqalign{ \sum_{m, \, \zb_2, t_3, t_2}\,
F_{m p}\left[\matrix{
i&j\cr l&k} \right]_{t_2\, t_3}^{u_2\, u_3 } &\
{}\Fo_{b l}\left[\matrix{
i&m\cr a&d} \right]_{\za_1\, \zb_2}^{\zg_1\, t_2 }\
{}\Fo_{c m}\left[\matrix{
j&k\cr b&d} \right]_{\za_2\, \za_3}^{\zb_2\, t_3 }\
 \cr
& =\sum_{\zb_1}\,
{}\Fo_{c l}\left[\matrix{
p&k\cr a&d} \right]_{\zb_1\, \za_3}^{\zg_1\, u_2 }\
{}\Fo_{b p}\left[\matrix{
i&j\cr a&c} \right]_{\za_1\,\za_2}^{\zb_1\, u_3 }\,. 
}}
Here $F$ is the matrix  (the ``$6j$-symbols''),
 unitary in the sense of the analogue
 of \unit{},
 relating the two bases in $V^i\otimes_h\, V^j\otimes_h\, V^k$.
{} To  make contact with the standard notation (cf. e.g. \KR{}),
\eqn\sixj
{F_{m p}^*\left[\matrix{i&j\cr l&k}\right]
= \left\{\matrix{i&j&p \cr k&l&m}\right\}\,.
}

\fig{Graphical representation of the $\Fo$ $3j$-symbols, 
of their orthogonality relations,  
of the $6j$-symbols and of the  pentagon identity \mp. Factors
depending on $P_a$ and $d_j$ have been omitted. }
{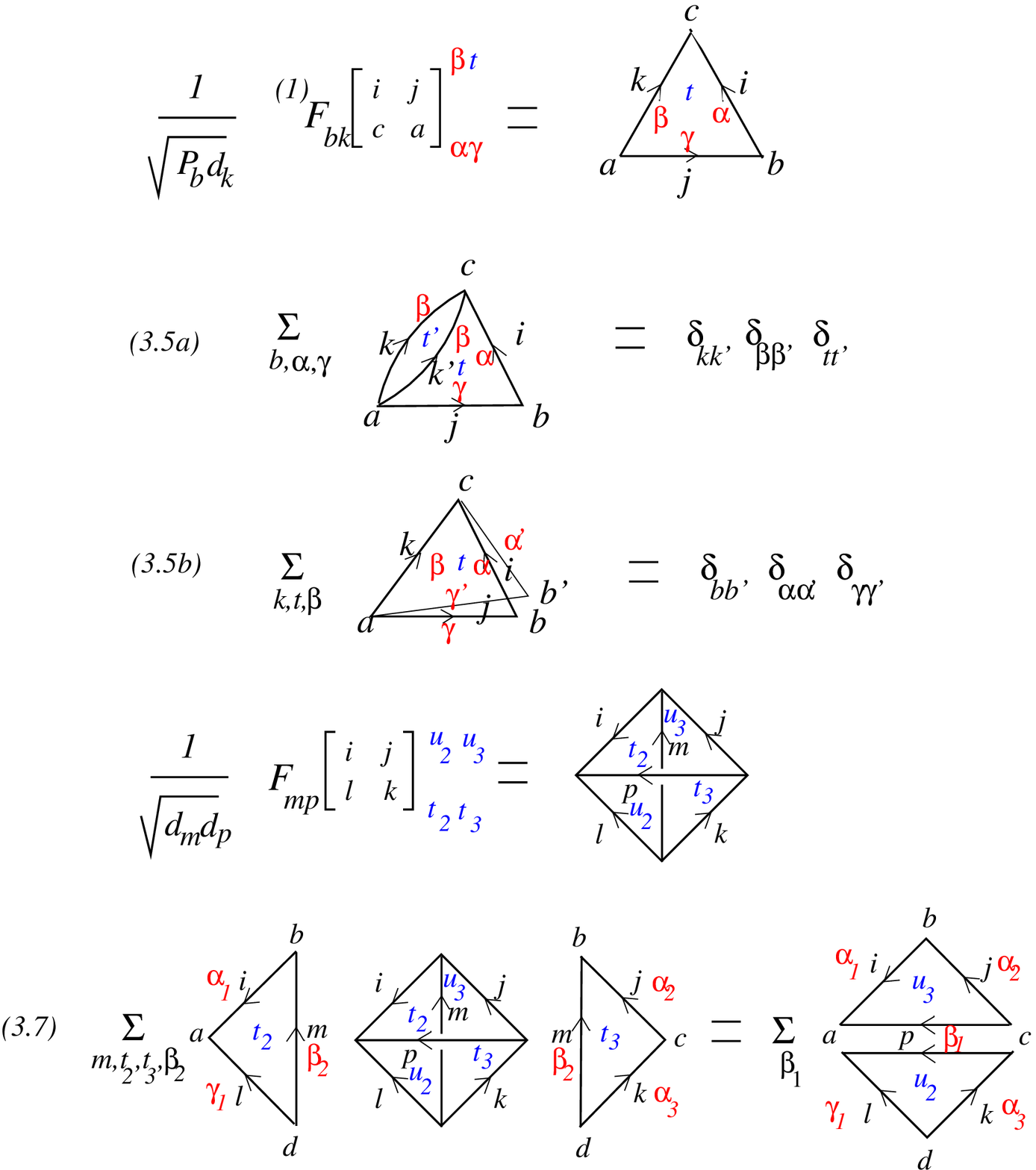}{9.5cm}\figlabel\graphrep

There is a  gauge freedom in ${}\Fo$   
due to the arbitrariness in the choice of basis,
$e^{j,\alpha}_{cb} \to 
\sum_{\alpha} \, U_{cb}^{j, \alpha, \alpha'}\,
e^{j,\alpha}_{cb} $, where $U$ 
is an arbitrary unitary matrix. It is useful to have a 
graphical notation for the $3j$-symbols $\Fo$ by means of triangles, 
and for the $6j$-symbols $F$ by means of tetrahedra (see Fig. \graphrep).
Then relations \trivone-\mp\ are simply depicted
\foot{The reader should not be confused by the multiplicity of graphical
representations used in this paper for the same objects.
It turns out that depending on the question,
a different representation may be clearer or more profitable. The triangles
used here for the ``cells'' may be regarded as obtained from the tetrahedra
of Fig. \simplex\ 
with three $\bullet$ and one $\circ$ by projecting the three edges
$\circ\!{{}\over{\quad}}\!\bullet$ with their label 
$a,b,c$ on the triangle with black vertices.
Likewise, in the representation of Fig. \simplex, the 
pentagon  identity  \mp\ is depicted
by the two  ways of cutting a double tetrahedron into two or
three tetrahedra.}. 
In this 
graphical representation, the gauge freedom consists in changing any 
edge 
$b\buildrel{j, \za}\over \longrightarrow c$ by a unitary matrix
$U_{cb}^{j, \alpha, \alpha'}$.  

The pentagon equation 
 \mp\ can be solved for $F$ given the $3j$-symbols 
$\Fo$ and using the unitarity relation \unit.
Conversely,  the relation  \mp\ can be interpreted, given $F$,
as a (recursive) relation for $\Fo$.  
In fact, the matrix $F$ is taken to be the  
matrix $\Fo$ of the diagonal case ($n_i\equiv N_i$), as we identify 
in that case the $3j$- and the $6j$-symbols and the equation 
\mp\ coincides then with the standard pentagon identity for $F$
 \eqn\pent{
F\  F \ F \, =   F \ F \,.
}
We can also look at \mp\ and its solutions $\Fo$ as providing more
general realisations
of the pentagon identity \pent, corresponding to the 
matrix representations $n_i$ of  the Verlinde fusion algebra \nim.
If we consider along with the states $|e^{j,\zb}_{ca}\rangle$
(triangles with one white and two black vertices), the vector spaces of 
``diagonal'' states $|e^{i, t}_{kj}\rangle$, $t=1,2,\dots N_{ij}{}^k$
(the triangles with three  black vertices in Fig. \simplex), we can identify
 the basis states in the r.h.s. of \Ima\  with the ``mixed'' products 
$|e^{k,\zb}_{ca}\rangle \otimes \ |e^{i, t}_{kj}\rangle$.

A solution of \pent\ is determined by the chiral
 data characterising the CFT.
For instance,   in the theories based on $\slh(2)$, the solution
provides the fusing matrices of the CVO and is
known to be given in terms of the $6j$ -symbols of the quantum algebra
$U_q(sl(2))$, restricted to matrix elements consistent with the
fusion rules. For  given $F$ the solution of  \mp\ is by definition
restricted by the data  in \nim, \int.

In agreement  with the symmetry \sint\ we  introduce
two (commuting) antilinear involutive maps  $ V^j\to V^{j^*}\,,$ 
$(e^{j,\zb}_{ca})^*= e^{j^*,\zb^*}_{c^*a^*}$, 
and $(e^{j,\zb}_{ca})^+= e^{j^*,\zb^+}_{ac}\,.$
 Correspondingly there are two bilinear forms on $V^{j^*}\otimes V^{j}$  
determined by the sesquilinear form \sc, i.e., two dual bases in
 $V^{j^*}$. The first, given by $*$,
 corresponds to the complex conjugation of the components
 of the initial basis in $V^j$ when it is realised through unit
vectors in $\IC^{m_j}$. 
The second basis is determined requiring that
\eqn\bil{
\sqrt{d_j\over P_c}\
\bra e^{1}_{aa}(j^*j)| e^{j^*,\zb^+}_{ac}\otimes_h e^{j,\zb'}_{ca'}\ket
= \bra \,e^{j,\zb}_{ca}
|\,e^{j,\zb'}_{ca'}\ket\,,
}
which implies
\eqn\Ibb{
\Fo_{c 1}\left[\matrix{j^*&j\cr a&a'}\right]_{\beta^+\, \beta' }^{1\,  1}
= \delta_{a a'}\,\delta_{\beta' \beta}\, \sqrt{P_c\over P_a d_j}\,.
}
 This  is a gauge fixing choice
 consistent with the unitarity condition \unit\ and the relation 
\eqn\inta{ d_p\, P_a =\sum_c\, n_{pa}{}^c\, P_c\,, }
derived from \int.  In the diagonal case it coincides  with  the
 standard gauge fixing of the fusing matrices $F$ of the
conformal models based on  $\slh(n)$.  Assuming that on tensor products   
$(x\otimes y)^*=x^*\otimes y^*\,,$ $(x\otimes y)^+ =y^+\otimes x^+
 \,,$ and denoting the dual basis states in the tensor product
 $e^{k^*,\zb^*}_{c^* a^*}(i^*j^*;t^*):=(e^{k,\zb}_{ca}(ij);t)^*$ and
 $e^{k^*,\zb^+}_{ac}(j^*i^*;
\sigma(t^*)):=(e^{k,\zb}_{ca}(ij);t)^+$
(since $N_{j^* i^*}^{k^*}=N_{ij}^k$),
 these maps imply  the symmetry relations for the $3j$-symbols $\Fo$
\eqn\sym{
{}\Fo_{b k}^*\left[\matrix{i&j\cr c&a}\right]_{\za\,\zg}^{\zb\ t}
={}\Fo_{b^* k^*}\left[\matrix{i^*&j^*\cr
c^*&a^*}\right]_{\za^*\,\zg^*}^{\zb^*\ t^*}  
={}\Fo_{b k^*}\left[\matrix{j^*&i^*\cr
a&c}\right]_{\zg^+\,\za^+}^{\zb^+\ \sigma(t)} \,,
}
while from the pentagon relation taken at $l=1$ and \Ibb\ one derives 
\eqn\symb{
{}\Fo_{b k}\left[\matrix{i&j\cr c&a}\right]_{\za\,\zg}^{\zb\ t}
=
\sqrt{P_b\, d_k\over P_c\, d_j}\
{}\Fo_{c j}^*\left[\matrix{i^*&k\cr b&a}
\right]_{\za^+\,\zb}^{\zg\ t^+} \,.
}

\bigskip

The space  $\oplus_{j\in \CI}\,$ End$(V^j)$  is a matrix algebra ${\cal A}
=\oplus_{j\in \CI}\, M_{m_j}$ 
on which a second  product (or a coproduct) is defined   
 via the $3j$-symbols $\Fo$ in \Ima.  This is 
 the Ocneanu double triangle algebra
\AO, 
an example (and presumably a prototype) of
the notion of  weak $C^*$ Hopf algebra
introduced in \BSz; 
 this structure has also received the name
of ``quantum groupoid'' \refs{\AO,\NTV}; 
see also \refs{\BE,\BEK} for recent developments of the original Ocneanu 
approach. Together with its dual algebra, $\CA$ is
interpreted in the present context as the  quantum symmetry of the 
CFT, either diagonal or non-diagonal. We review below briefly
some basic properties of $\CA$ and give further details in  appendix A.
  
 The matrix units in $M_{m_j}$ (block matrices in $\CA$) are
  identified with states in $V^j\otimes V^{j^*}$,
\eqn\mat{  
e_{j;\zb,\zb'}^{(ca)\,,(c'a')}
={\sqrt{d_j}\over (P_c P_a P_{c'} P_{a'})^{1\over 4}}
\ | e^{j\,,\zb}_{ca}\rangle \langle e^{j\,,\zb'}_{c'a'}| \ ,
}
 so that
\eqn\Action{
e_{k;\beta', \beta^{''}}^{(c'a')(c^{''}a^{''})}\ | e^{i\,,\zb}_{ca}\rangle
= \delta_{i k}\,
\delta_{a a^{''}}\, \delta_{c c^{''}}\, \delta_{\beta \beta^{''}}\ 
\Big({P_{a}\, P_{c}\over P_{a'}\, P_{c'} }\Big)^{1\over 4}\
| e^{k\,,\zb'}_{c'a'}\rangle \ .
}

\fig{Two alternative representations of (a) the basis vectors 
$({d_j\over P_a\, P_c})^{1\over 4}\, |e^{j,\zb}_{ca}\rangle$,\qquad \qquad
(b) the matrix units $e_{j;\zb,\zb'}^{(ca),(c'a')}\,$.}
{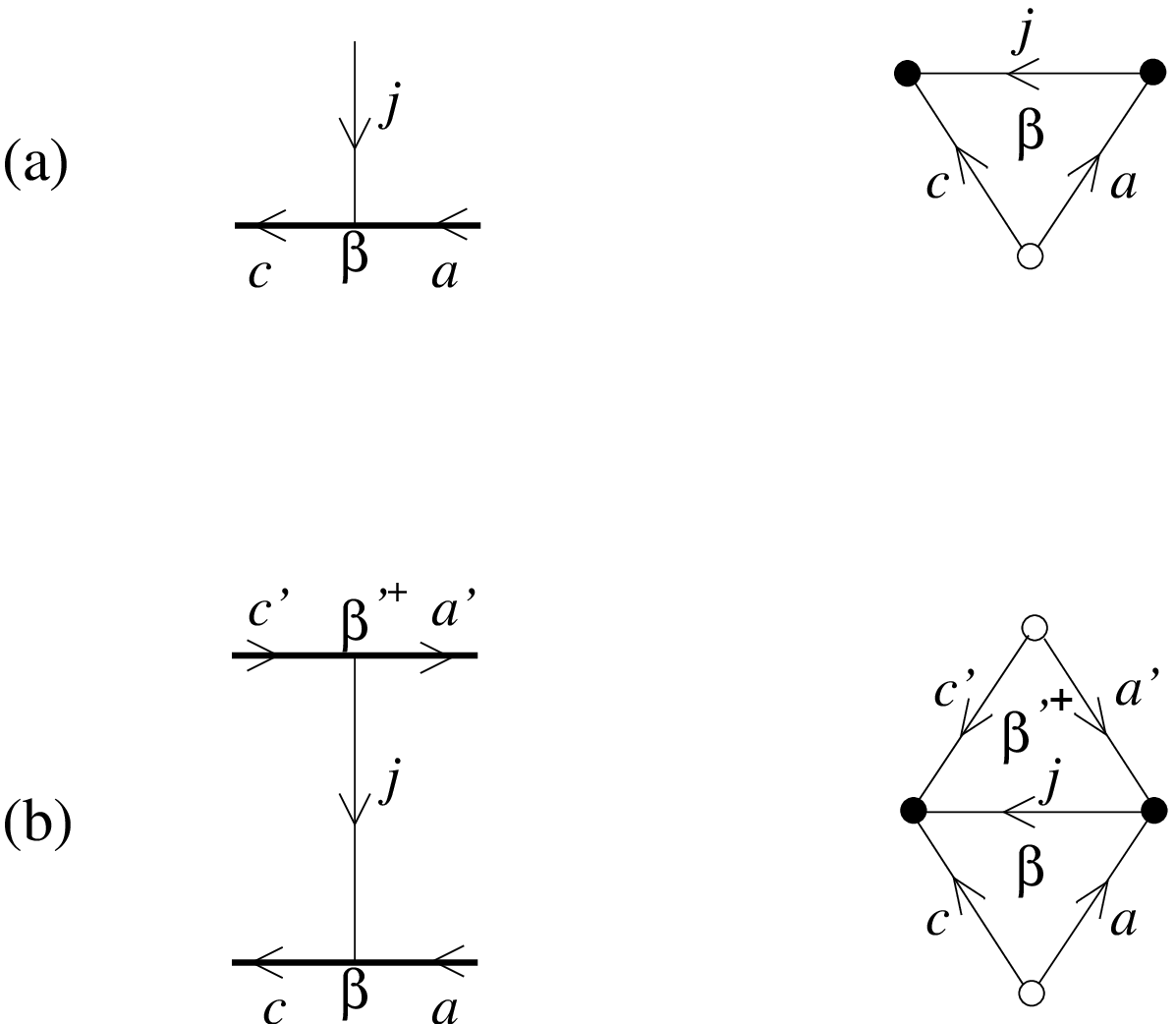}{6cm}\figlabel\cvotri

\medskip
They are depicted as $4$-point blocks in Fig. \cvotri, where  
the states in $V^j$  correspond to 3-point vertices, or, dually,
to triangles, whence the name ``double triangle algebra''
for the algebra $\CA$ spanned by the elements \mat, $j\in \CI$.
Their matrix (``vertical'') multiplication is simply
\eqn\vert{
  e_{j\,,\beta,\beta'}^{(ca)(c'a')} \
e_{i\,,\zg', \zg}^{(d'b')(db)}
=\delta_{ij}\,\delta_{a'b'}\, \delta_{c'd'}\, \delta_{\beta'\zg'}\,
e_{j\,,\beta,\zg}^{(ca)(db)}\,.
}
%
\fig{The vertical product}{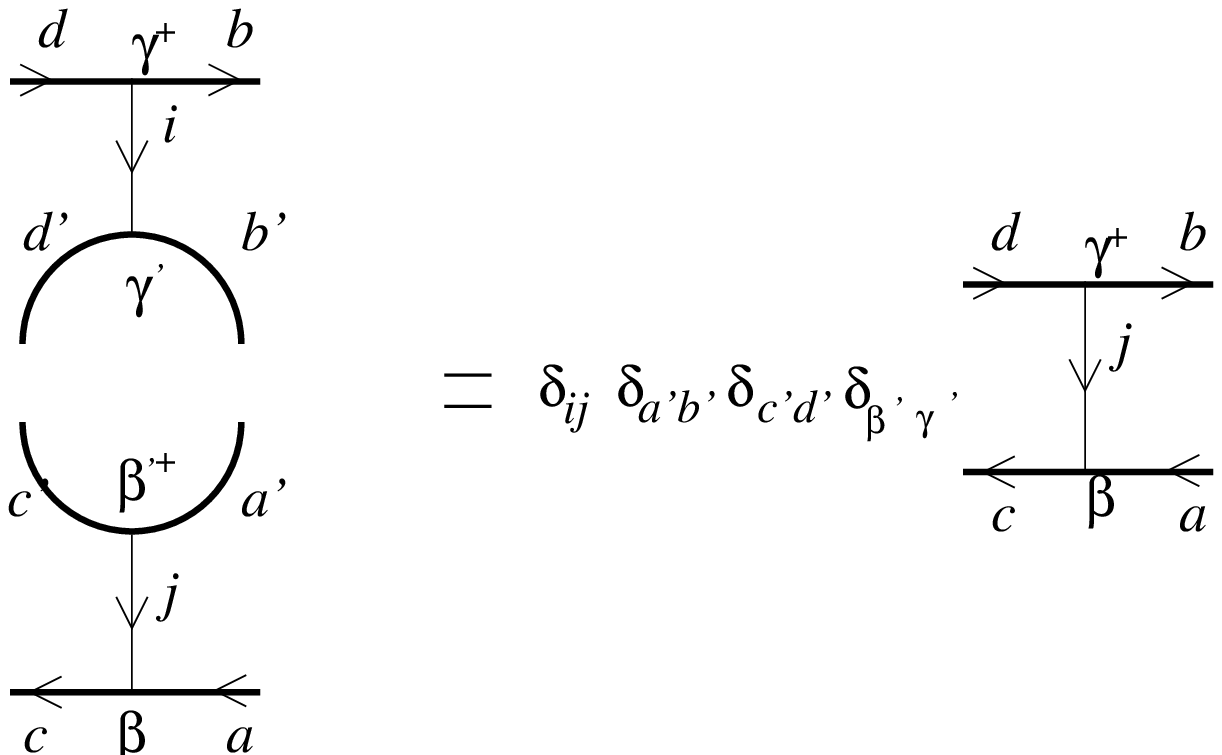}{6cm}\figlabel\dtavp
\smallskip
\noindent
The product \vert\ is illustrated on Fig. \dtavp\ by composing vertically the 
blocks representing the two elements
(the second above the first), and a similar picture represents
\Action. The  identity element $1_v$ in $\cal A$   with 
respect to this multiplication is given by
$1_v=\sum_{i\,,c\,,b\,,\za}\, e_{i\,, \za\,, \za}^{(cb)(cb)}\,. $  
%

\fig{The horizontal product}{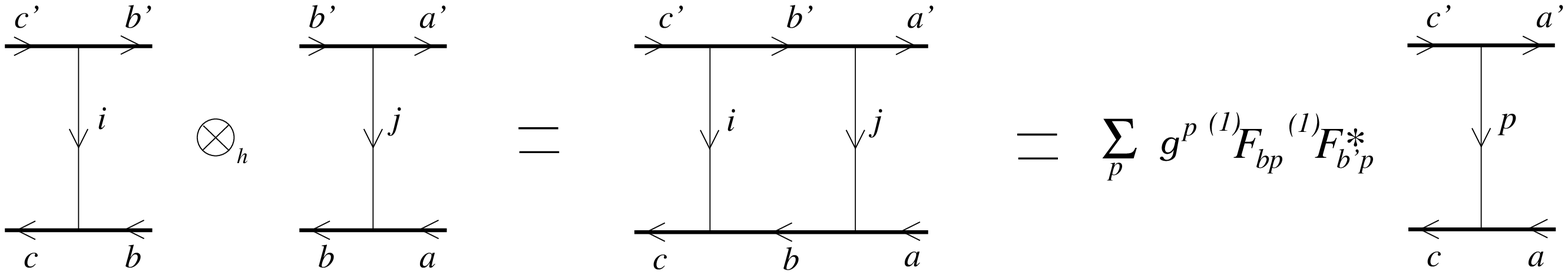}
{12cm}\figlabel\dtahp

\medskip
A second, ``horizontal'',  product is defined
\AO, composing two blocks horizontally, see Fig. \dtahp. Its decomposition
is inherited from the r.h.s. of the product $\otimes_h$ in
\Ima, and thus the r.h.s. in Fig. \dtahp\ involves
 the $3j$-symbols $\Fo$ and $\Fo^*$. The  normalisation constant
%
is chosen for later convenience as given by 
$g_{ij}^{p;b,b'}={c_i^{bc'}\,c_j^{ab'}/c_p^{a c'}}$,
with $\,c_j^{ab'}={d_j\over \sqrt{P_a\, P_{b'}}}\,{S_{11}\over \psi_1^1}\,. $
 Alternatively we can 
define \BSz\ a coproduct  $\triangle:\CA\to \CA\otimes\CA$
\eqn\cop{\eqalign{
&\triangle(e_{k\,,\zb,\zb'}^{(ca)(c'a')}): = \cr
&\sum_{i\,,j\,\atop {t}}\ \sum_{b\,,b'\,\atop{ 
\alpha\,,\alpha'\,, \gamma\,, \gamma'}}\ 
 ^{(1)}F_{b k}^*\left[\matrix{i&j\cr c&a}\right]_{\za\,
\zg}^{\zb\, t}\ \
  ^{(1)}F_{b' k}\left[\matrix{i&j\cr c'&a'}\right]_{\za'\,
\zg'}^{\zb'\, t}\ \
e_{i\,,\za\,,\za'}^{(cb)(c'b')}\otimes
 e_{j\,,\zg\,,\zg'}^{(ba)(b'a')}\,.
}}
The unitarity \unit\ of $\Fo$
 ensures that $\triangle(ab)=\triangle(a)\, \triangle(b)$
while the coproduct is  coassociative,
$(\triangle\otimes Id) \circ \triangle=(Id \otimes \triangle)
\circ \triangle $, whenever there exist a unitary $F$ 
(in the sense of the diagonal analogue of \unit),
satisfying along with $\Fo$  the pentagon identity \mp. 
The ``star'' operation in $\CA$, $(xy)^+=x^+ y^+$, 
is inherited from the map $(+)$ defined above, 
\eqn\star{
\Big(e_{i\,,\za,\za'}^{(cb)(c'b')}\Big)^+=
 e_{i^*\,,\za^+,\za^{'+}}^{(bc)(b'c')}\,.
}
It is a homomorphism of the algebra, i.e. of the vertical
product, and an anti-homomorphism of the horizontal product, 
$(a\otimes_h b)^+ = b^+\otimes_h a^+$,
as well as $(a\otimes b)^+ = b^+\otimes a^+$,
 so that 
$\triangle(a^+)=\triangle(a)^+$.  
The algebra $\CA$ is given a  coalgebra structure defining
a counit $\ze:\CA\to \IC$ according to
\eqn\coun{
\ze(e_{j\,,\zb,\zb'}^{(ca)(c'a')}):=
\delta_{j1}\,\delta_{ac}\,\delta_{a'c'}\,\delta_{\zb
1}\,\delta_{\zb' 1}\,,
}
which satisfies  the compatibility condition
$
(\ze\otimes Id)\, \circ \, \triangle =Id= (Id \otimes \ze)\,
\circ\,\triangle\,.
$

The definitions \cop, \coun\  imply, however,  that $\triangle( 1_v)
\not= 1_v\otimes 1_v$ and that the counit is {\it not} a homomorphism
of the algebra, $\ze(u)\, \ze(w)
\not = \ze(u\,w)$ for general  elements $u,w\in \CA$, i.e., the
DTA is not a Hopf algebra, see the appendix 
 for more details on its structure of ``weak Hopf algebra";
 in particular the antipode is defined according to
\eqn\antipode{
S(e_{i\,,\za,\za'}^{(cb)(c'b')})=\sqrt{P_b P_{c'}\over P_{b'} P_c}\
e_{i^*\,,\za^{'+},\za^+}^{(b'c')(bc)}=\sqrt{P_b P_{c'}\over P_{b'} P_c}\
(e_{i\,,\za',\za}^{(c'b')(cb)})^+
\,.
}

  Using the unitarity \unit\ of the $3j$-symbols $\Fo$,
it is straightforward to show that the elements
\eqn\mcp{
\hat{e}_i=
\, \sum_{c\,,b\,,\za}\ {1\over c_i^{bc}} \ 
e_{i\,, \za\,, \za}^{(cb)(cb)}\,,
}
realise the Verlinde algebra
with respect to the horizontal product in $\cal A$,
\eqn\abV{
\hat{e}_i \otimes_h \hat{e}_j= \sum_k\, N_{ij}{}^k\, \hat{e}_k\,,
}
and $\hat{e}_1$ is
the identity matrix of $\CA$ for that product.

\newsec{Generalised chiral vertex operators}
\noindent
We now return to the field theory. Let $i,j,k\in\CI$
s.t. $N_{ij}{}^k\ne 0$ and let $\hbox{\cyr i}$ 
label  descendent states in $\CV_i$. The chiral vertex operator 
$\phi_{ij,t;\hbox{\cyr i}}^k(z)$, with $t$ -- a basis label, 
$t=1,2,\dots, N_{ij}{}^k$, is an intertwining operator 
$ \CV_j\to \CV_k\, $ \MS. 
We tensor this field  with  an intertwining operator $V^j\to V^k$ 
\eqn\Proj
{P^{k,\alpha;j,\gamma}_{cb,ab}=
\sqrt {d_j\over P_a\,P_b} \
| e^{k,\alpha}_{cb}\ket \bra\,e^{j,\gamma}_{ab} |
\,, }
which corresponds to  a state in $V^k\otimes_h V^{j^*}$. 
This defines a generalised  chiral vertex operator (GCVO)   
$$
\oplus_{j\in \CI}\, \CV_j\otimes V^j \to
\oplus_{k\in \CI}\, \CV_k\otimes V^k\,,
$$
\eqn\gcvo{
^c\Psi_{i, \beta;
\hbox{\cyr i}}^a(z) 
=\sum_{j\,,k\,, t}\ \phi_{ij,t;
\hbox{\cyr i}}^k(z)  \otimes\,
\sum_{ b\,,\alpha\,,\gamma}\
\Fo_{ak}\left[\matrix{i&j\cr c&b}\right]^{ \alpha\, t}_{
\beta\, \gamma}\ P^{k,\alpha;j,\gamma}_{cb,ab}\,. }
The projectors \Proj\ satisfy
\eqn\pro{\eqalign{
 P^{i,\alpha;k,\gamma}_{cb,ab}\ P^{k',\gamma';j,\delta}_{a'b',d b'}&=
\delta_{bb'}\, \delta_{aa'}\,\delta_{kk'}\,\delta_{\gamma \gamma'}\,
 P^{i,\alpha;j,\delta}_{cb,db}\,, \cr
  P^{k,\alpha;j,\gamma}_{cb,ab}\ |e^{j',\gamma'}_{a'b'}  \rangle &= 
\delta_{bb'}\, \delta_{aa'}\,\delta_{jj'}\,\delta_{\gamma \gamma'}\,
  |e^{k,\alpha}_{cb}  \rangle \,,   \cr
\langle e^{1}_{dd} | P^{k,\alpha;j,\gamma}_{cb,ab} |e^{1}_{d'd'}  \rangle
&=\delta_{k1}\,\delta_{j1}\,
\delta_{cd}\, \delta_{bd}\,\delta_{bd'}\, \delta_{ad'}\, P_a\,.
 }}
{}From \gcvo, \pro\ we have in particular
\eqn\Ita{
^c\Psi_{j,\beta}^a(0)\, |0\rangle
 \otimes | e^{1}_{ aa}\rangle
= \phi_{j1}^j(0) |0\rangle  \otimes | e^{j,\beta}_{c a}\rangle
=:|j\,, \beta\ket\,, \quad \beta=1,2\dots, n_{j a}{}^c
}
where $|j\,, \beta\ket $ is the explicit form
of the highest weight state of the chiral algebra representation
$\CV_{j,\,\beta}$,  ``augmented'' with the additional  coupling 
label $\beta$, 
used in the computation of the cylinder partition function in 
the Hilbert space $\CH_{a|c}=\oplus_{j,\zb}\ \CV_{j,\,\zb}\,, $ 
\BPPZ. 
The correlators of the generalised CVO \gcvo\ are  computed 
projecting on ``vacuum'' states $|0\rangle \otimes |e^1_{aa}\rangle$
in the  space ${\cal V}_1\otimes V^1$; recall that $V^1$ has
a nontrivial dimension $|{\cal V}|$.
Since $P^{1,1;1,1}_{ab,db}=\delta_{a b}\, \delta_{b d}\,
P^{1,1;1,1}_{aa,aa}\,,$ the first and the last labels
of any $n$-point correlator coincide, i.e., we can associate with it
a closed path $\{a,a_1,....a_{n-1},a\}$ with elements marked
by the graph  indices and passing through the
coordinate points $z_1\,, \dots, z_n$. E.g.,
the $2$-point correlator reads
\eqn\Itab{
\langle ^a\Psi_{j^*,\beta'}^c(z_1)\ ^c\Psi_{j,\beta}^a(z_2)\rangle_a
=\Fo_{c 1}\left[\matrix{j^*&j\cr a&a}\right]_{ \beta'\, \beta}^{ 
1\, 1} \,  P_a  \ \, \langle 0|\,
\phi_{j^*j}^1(z_1)\ \phi_{j1}^j(z_2)\,  |0\rangle\,.
}
For real arguments one recovers the correlators of the boundary fields.
Note that the normalisation of boundary field correlators following 
from \sc\ differs from that used in \BPPZ, 
(eq. (4.7)) by a factor $\psi_1^1/\sqrt{S_{11}}$, i.e.,
$\bra \un\ket_a= {S_{11}\over (\psi_1^1)^2} \,
 \lim_{L/T\to\infty}\, Z_{1|a}\, e^{-{\pi c\over 6} {L\over T}}=P_a$.

\medskip
The algebra $\CA$ acts on the operators \gcvo\ with the help of the 
antipode \antipode, namely  for 
$\triangle(e_{p;\beta', \beta^{''}}^{(c'a')(c^{''}a^{''})})=
e_{(1)}\otimes e_{(2)}$ we define a representation $\tau(e)$
\eqn\action{\eqalign{
\tau(e_{p;\beta', \beta^{''}}^{(c'a')(c^{''}a^{''})})\,^c\Psi_{i, \beta}^a(z) :
  &= e_{(1)}\, ^c\Psi_{i, \beta}^a(z)\,  S(e_{(2)})\cr 
&= \delta_{i p}\,
\delta_{a a^{''}}\, \delta_{c c^{''}}\, 
\delta_{\beta \beta^{''}} \ 
\Big({P_{a'}\, P_{c}\over P_{a}\, P_{c'} }\Big)^{1\over 4}\
 ^{c'}\Psi_{i, \beta'}^{a'}(z)\,.
}}

Definition \gcvo\ has to be compared with  earlier 
work \refs{\MR,\GS} based on the use of  quantum groups (Hopf algebras), 
or some related versions, e.g. \MSch, obtained by modifying the
standard Hopf algebra axioms (see \BSz\ for a discussion
on the latter and further references).  The papers \refs{\MR,\GS,\MSch} 
deal essentially  with
the diagonal case, and, more importantly, exploit 
the true $U_q(\gg)$ $3j$-symbols at roots of unity
(e.g., for $\gg=sl(2)$) in formulae analogous to \gcvo. 
We stress again that the $3j$-symbols $\Fo$ in \gcvo\ and \Ima\ 
reduce in the diagonal case to the $6j$-symbols of the quantum groups
$U_q(\overline{\gg})$ 
(or products of them), restricted to labels consistent with the
CFT fusion rules.  Thus the decomposition of the Ocneanu
horizontal product fits precisely the CFT fusion,
without the need of additional truncation as in the case
of quantum group representations.  As emphasized in \BSz, 
unlike the alternative  approaches which deviate from 
the standard Hopf algebras, 
the use of a WHA as a quantum symmetry retains coassociativity
reflected in \mp.
A ``price'' to be paid is the multiplicity of vacua, which has, 
however,  a physical interpretation in BCFT, as
providing a complete set of conformal boundary states.

\medskip


{}From the operator representation \gcvo\ one  derives various 
identities. 
In particular
inserting the r.h.s. of \gcvo\ in  the product of two generalised
CVO, then applying the OPE for the standard CVO and finally
using the pentagon identity \mp,\ and once again the
representation \gcvo, we {\it derive} for small $z_{12}$ the OPE
\eqn\bop{\eqalign{
{}^{c}\Psi_{i, \za}^b(z_1)\, {}^{b}\Psi_{j, \zg}^a(z_2)\,
&= \sum_{p\,,\zb\,,t}\
{}\Fo_{b p}\left[\matrix{i&j\cr c&a} \right]_{\za\,
\zg}^{\zb\   \ t }\  \sum_{ \hbox{\cyr p}}\,
\bra  p, \hbox{\cyr p}| \phi_{ij; t}^p(z_{12})|j,0\ket\ {}^{c}\Psi_{p,
\zb; \hbox{\cyr p} }^a(z_2)\cr
&= \sum_{p,\zb, t }\ {}\Fo_{b p}\left[\matrix{
i&j\cr c&a} \right]_{\za\, \zg}^{\zb\  \ t}\
\bra  p,0| \phi_{ij; t}^p(z_{12})|j,0\ket\
{}^{c}\Psi_{p, \zb}^a(z_2) + \  \dots \,.
}}
For arguments restricted to the real line one recovers the
boundary field small distance expansion \CL\ with OPE coefficients
given by the $3j$-symbols of \Ima.
Conversely, the expansion \bop\ was the starting point in \BPPZ\
for the derivation of the  pentagon identity \mp.

Denote by $^c{\cal U}^a_j$ the space of generalised CVO \gcvo.
The  generalised CVO have a nontrivial braiding defined through 
a new braiding matrix with $4+2$ indices of two types, 
\eqn\braidm
{\hat{B}(\epsilon): \oplus_b\, ^c{\cal U}^b_i\,\otimes
{}^b{\cal U}^a_j \to  \oplus_d\, ^c{\cal U}^d_j\, \otimes
{}^d{\cal U}^a_i\,, }
\eqn\Ic{
^c\Psi_{i,\alpha}^b(z_1)\, ^b\Psi_{j,\gamma}^a(z_2)= \sum_{d,
\alpha',\gamma'}\, 
\hat{B}_{bd}\left[\matrix{i&j\cr c&a} \right]_{
\alpha\, \gamma}^{\alpha'\, \gamma'}\!\!\!\!(\epsilon)\
^c\Psi_{j,\alpha'}^d(z_2)\, ^d\Psi_{i,\gamma'}^a(z_1)\,,
\quad
}
\eqn\Idc{
\hat{B}^{12}(\epsilon) \, \hat{B}^{21}(-\epsilon)=1\,,
}
consistently with  the commutativity  of the intertwiners
\eqn\as{
\sum_b n_{ib}{}^c\, n_{ja}{}^b =\sum_d n_{jd}{}^c\, n_{ia}{}^d\,.  }
In \Ic\ 
$z_{12} \notin \IR_-\,,$ and  $\epsilon$ stands for $\epsilon_{12}=
 {\rm sign(Im}(z_{12}))$ 
and for $i=1\,$ or $j=1$ the matrix $\hat{B}$ is trivial. 
The braiding matrices $\hat{B}$ satisfy the ``Yang-Baxter (YB) equation'' 
\eqn\ybe{
\hat{B}^{12}(\epsilon_{12})\, \hat{B}^{23}(\epsilon_{13})\,
  \hat{B}^{12}(\epsilon_{23})\,
=  \hat{B}^{23}(\epsilon_{23}) \, \hat{B}^{12}(\epsilon_{13})\,
 \hat{B}^{23}(\epsilon_{12}) \,.
}
Combining \Ic\ with the definition \gcvo\ of the generalised CVO,
using then the braiding of the standard CVO and projecting 
on the state $|0\ket$, we obtain the relation
\eqn\Ibbb{
\sum_{d\,, \alpha'\, \gamma'} \hat{B}_{bd}\left[\matrix{i&j\cr c&a} \right]_{
\alpha\, \gamma}^{\alpha'\, \gamma'}\!\!\!\!\!\!(\epsilon)\ 
{}\Fo_{d k}\left[\matrix{j&i\cr c&a} \right]_{\alpha'\,
\gamma'}^{\beta\  t }= e^{-i \pi\epsilon\, \triangle_{ij}^k}\
{}\Fo_{b k}\left[\matrix{i&j\cr c&a} \right]_{\alpha\,
\gamma}^{\beta\, t }\,,
}
where
the phase in the r.h.s., depending on the scaling dimensions
$\triangle_{ij}^k=\triangle_i+\triangle_j-\triangle_k\,,$  
comes  from the standard CVO braiding matrix $B$.
 In the diagonal case,
where we can identify ${}\Fo$ and $\hat{B}$ with
the standard fusing and braiding matrices, $F$ and $B\,,$
 this relation is nothing else than the
simplest hexagon relation (the $q$-Racah identity).
 Inverting \Ibbb\ we get a bilinear representation of $\hat{B}$ 
in terms of $\Fo$   
\eqn\Ie{
 \hat{B}_{bd}\left[\matrix{i&j\cr c&a} \right]_{
\alpha\, \gamma}^{\alpha'\, \gamma'}\!\!\!\!\!\!(\epsilon)\, 
= \sum_{k\,, \beta\,, t }\,
{}\Fo_{b k}\left[\matrix{i&j\cr c&a} \right]_{\alpha\,
\gamma}^{\beta\  t }\ 
\, e^{-i \pi\epsilon\, \triangle_{ij}^k}\
{}\Fo_{d k}^*\left[\matrix{j&i\cr c&a} \right]_{\alpha'\,
\gamma'}^{\beta\  t }\,.  }
This formula determines   $\hat{B}$ whenever we know ${}\Fo$
and the scaling dimensions $\triangle_{j}$, i.e., 
the $3j$-symbols ${}\Fo_{bi}$ diagonalise the 
matrix $\hat{B}_{bd}$. It also implies the symmetries 
\eqn\Bsymm{
\hat{B}_{bd}\left[\matrix{j&k\cr c&a} \right](\epsilon)=
\hat{B}_{b^*d^*}\left[\matrix{k&j\cr a^*&c^*} \right](\epsilon)=
\hat{B}_{db}\left[\matrix{j^*&k^*\cr a&c} \right](\epsilon)=
\hat{B}_{b^*d^*}^*\left[\matrix{j^*&k^*\cr c^*&a^*} \right](-\epsilon)\,.}

The relation \Ibbb\ is a particular case of the more
general identity derived 
by inserting \gcvo\ in \Ic\
and using the analog of \Ic\ for the standard CVO
$$
\Fo\, \Fo\,B= \hat{B}\, \Fo\, \Fo\,, 
$$ 
or, more explicitly
\eqn\Ii{\eqalign{
\sum_{n\in {\cal I}}\,
{}\Fo_{a n}\left[\matrix{j&m\cr c&b} \right]\
{}\Fo_{ck}\left[\matrix{i&n\cr d&b} \right]\
B_{n l}\left[\matrix{i&j\cr k&m} \right]\ \cr
=\sum_{c'\in \CV}\ 
\hat{B}_{ c c'}\left[\matrix{i&j\cr d&a} \right]\
{}\Fo_{a l}\left[\matrix{i&m\cr c'&b} \right]\
{}\Fo_{c' k}\left[\matrix{j&l\cr d&b} \right]\,,
}}
\eqnn\Il
\fig{Equations \Ii-\Il.  }{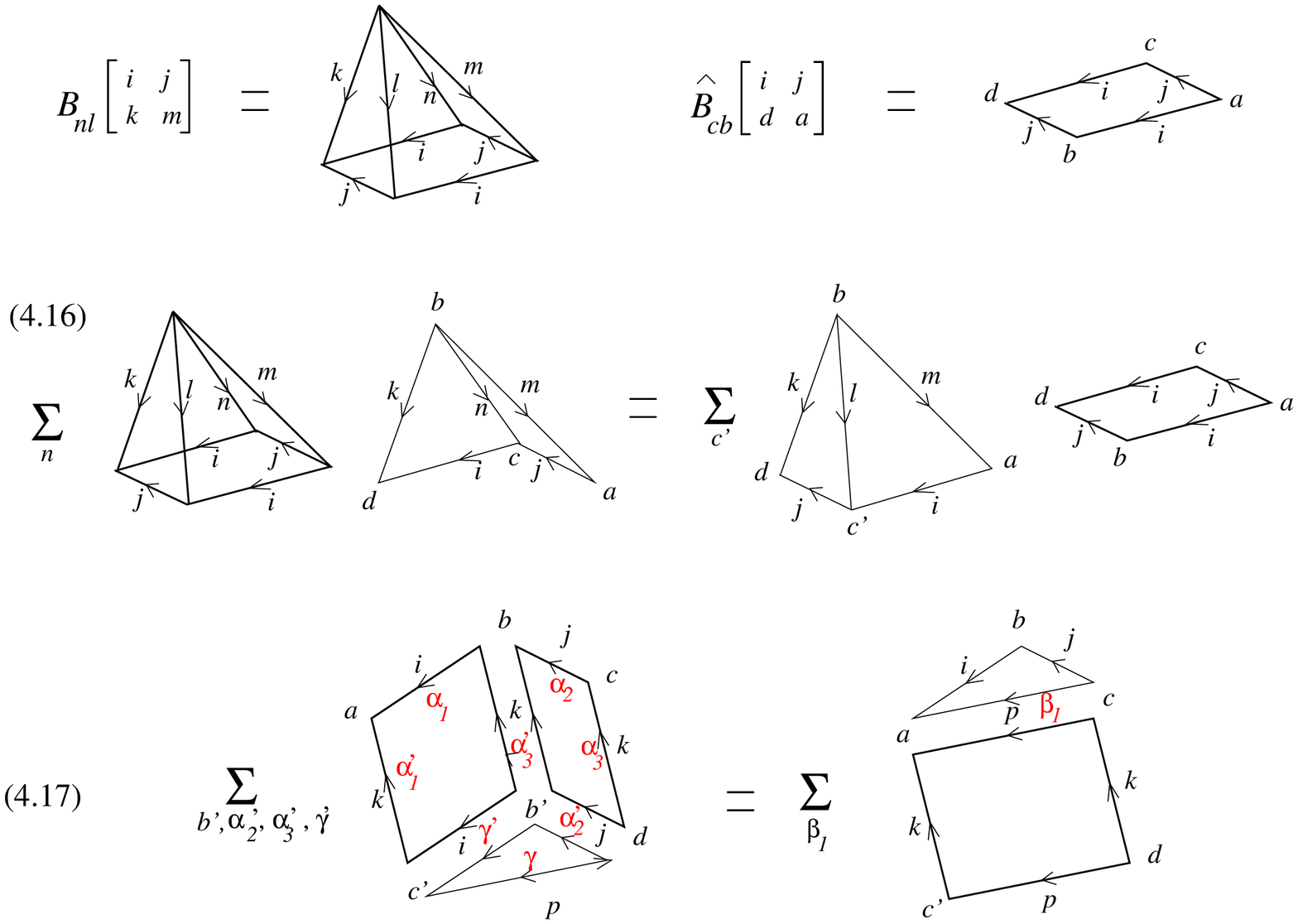}{12cm}\figlabel\triang
\noindent as illustrated on Fig. \triang.
For $m=1$ we recover \Ibbb. 
Eq. \Ii\ implies that (products of two) $3j$ symbols $\Fo$ 
intertwine the two 
representations $B$ and $\hat B$ of the braiding group. This is
to be compared with the  ``cells'' introduced in lattice models, 
 \refs{\Vpa,\Roch,\DIFZ,\PZh}, see section 6.

Another relation derived from the product of three GCVO $\Psi$
gives a generalisation of the braiding-fusing identity
of Moore-Seiberg \MS\ 
$$
 \hat{B} \ \Fo =  \hat{B}\  \hat{B}\ \Fo\,,
$$
or,
$$\eqalignno{
&\sum_{\beta_1}\,
\hat{B}_{c c'}\left[\matrix{p&k\cr a&d}\right]_{\beta_1\, \alpha_3
}^{\alpha_1'\, \gamma}\,
\Fo_{b p}\left[\matrix{i&j\cr a&c}\right]_{\alpha_1\,
\alpha_2}^{\beta_1 \,t}\,  & \Il \cr
&=\sum_{b'\,, \alpha_2'\,, \alpha_3'\,,\gamma'}
\, 
\Fo_{b' p}\left[\matrix{i&j\cr c'&d}\right]_{\gamma'\,\alpha_2'}^{
\gamma \ t}\,
\hat{B}_{b c'}\left[\matrix{i&k\cr a&b'}\right]_{\alpha_1\,
\alpha_3'}^{\alpha_1'\, \gamma'}\,
 \hat{B}_{c b' }\left[\matrix{j&k\cr b&d}\right]_{\alpha_2\,
\alpha_3}^{\alpha_3'\,\alpha_2'}\,
}$$
In the diagonal case this is the equation from which one
obtains (taking $d=1$) the relation between the braiding $B$ and
fusing $F$ matrices; inserting this relation back in \Il\ reproduces 
the standard pentagon identity for $F$.
In general  \Il\ provides a
recursive construction of $\hB$ in the spirit of \KRS. Namely,
solving it for  the $\hB$ matrix in the l.h.s., i.e., writing it as 
$\hat B= \sum_{b,b'}\, \Fo\, \hat{B}\, \hat{B}\ \Fo^*\ , $
we get an  equation which 
determines $\hat B$ recursively, given the subset of $3j$-symbols
with one of the labels $i,j,p$ fixed to the fundamental representation(s).
Using \Ie\ the r.h.s of \Il\ can be completed to 
$\hB_{23}\, \hB_{12}\, \hB_{23}$, i.e., to the r.h.s. of the YB equation \ybe.
Similarly one derives a second braiding - fusing identity so that its
r.h.s. is completed to the l.h.s. of \ybe. Comparing the two identities
 and using twice in the l.h.s. of one of them
the  pentagon identity \mp\ and the unitarity of $F$, 
 one recovers the YB relation.

\medskip

{} Together with the interpretation of $\Fo$ as $3j$-symbols, 
the braiding matrix $\hat B$ can be interpreted 
as the $\CR$ -- matrix of a quasitriangular WHA, 
see Appendix A.  The relation \Ie\ is an analogue
of the representation of the $\CR$ matrix in terms of the 
$3j$-symbols, while \Ii\ 
 is an analogue of the relation between vertex representation and
 path representation of the quantum group $\CR$ matrix 
(Vertex -IRF correspondence), see \VP. 

There is an important difference  in the analogy with the r\^ole
of  quantum groups in CFT.  Namely in the present approach 
 the summations in all
identities, like e.g., over $k$ in \Ie, or  over $n$ in \Ii, run
according to the fusion rules, while in their  analogues, where the
true quantum group $3j$-symbols appear, these summations run within 
the standard classical tensor product bounds.  
When interpreted in the CFT framework the analogue of the braiding
relation \Ic\ is then required  to hold only on a ``physical''
subspace, or alternatively, the   conformal Hilbert spaces 
(and the conventional CVO in the definition
of the covariant CVO of \MR) have to be  extended to accomodate
``unphysical'' intermediate states incompatible with the fusion rules,
\refs{\MSchb,\GS, \FGP}, see also the recent work \HT\ for a
related discussion and further references. 

%
\newsec{Bulk fields -- chiral representation}
\noindent
Let now the pairs $I=(i,\bi)$, $i\,,\bi\in \CI$, label the ``physical'' 
spectrum,
corresponding to nonzero  matrix elements of the modular mass matrix 
$Z_{i\bi}$, or, in a more precise notation, which we will  
for simplicity skip in this section, $(i,\bi;\alpha)\,, \za=1,2\dots
 Z_{i\bi}$.
We define  (upper) half-plane bulk fields as compositions of two
GCVO \gcvo\ 
\eqnn\hp
$$\eqalignno{
\Phi_{(i,\bi)}^H(z, \bar z)&= \sum_{a, b\,,\beta'\,,\beta}\,
\Big(\sum_{j, \za, u}\, R_{a,\za}^{(i,\bi^*, u)}(j)\,
\Fo_{bj}^*\left[\matrix{i&\bi^*\cr a&a}\right]^{ \alpha\, u}_{
\beta\, \beta'}\Big)\,
^a\Psi_{i, \beta}^b(z)\ ^b\Psi_{\bi^*, \beta'}^a(\bar{z})
\cr 
&=\sum_{n\,, k\,, l\,,t\,,t'}\ 
\phi_{ik;t'}^n(z)\, \phi_{\bi^* l;t}^k(\bar z)\otimes \,
\sum_{a\,,b'\,,\zg\,,\zg'}\,
C^{n,k,l;t',t}_{(i,\bi)\,a,b',a;\zg,\zg'}\
P^{n,\zg;l,\zg'}_{a b',a b'} \,.  & \hp\cr
}$$
Here $\bar{z}\in H_-$ is the complex conjugate of $z\in H_+$
 and the field $\Phi_{(i,\bi)}^H(z, \bar z)$
 transforms under a tensor product representation  
of one copy of the chiral algebra $\gA$
labelled by the pairs $(i,\bi^*)$,
see \refs{\RSch,\FS,\BPPZ}
for discussions of more general gluing conditions.
\foot{
For convenience we keep the same notation for
the half-plane field $ \Phi_{(i,\bi)}^H$ (with $(i,\bi^*)$
appearing in the CVO product in the r.h.s. of \hp) as for its
full-plane counterpart $ \Phi_{(i,\bi)}^P\,,$
with the second label in $(i,\bi)$ corresponding to
a representation of a second copy of the chiral algebra;
in our convention the diagonal torus modular invariants
correspond to the fields $ \Phi_{(i,i)}^H$, $ i\in \CI$.
}
The choice of the constants in \hp,  related  according to  
\eqn\bcc{
C^{n,k,l;t',t}_{(i,\bi)\,a,b',a;\gamma,\alpha}
= \sum_{j\,,u\,,u'\,, \alpha'}\, \ R_{a,\za'}^{(i,\bi^*, u)}(j)
\ 
{}\Fo_{an}\left[\matrix{j&l\cr a&b'}\right]^{\gamma\,u'}_{\alpha'\,\alpha}\,
F_{kj}^*\left[\matrix{i&\bi^*\cr n&l}\right]^{u'\,u}_{ t'\, t}\,,
}
is such that when applying for small $z-\bar{z}=2i\,y$ the OPE \bop\ for the
two CVO in \hp\ (and projecting on $|e^1_{aa}\ket$)
we recover in the leading order the boundary field $^a\Psi_{j, \za}^a(x)$ 
contributing with the {\it bulk-boundary reflection coefficient}
$R_{a,\za}^{(i,\bi^*, u)}(j)=
C^{j,\bi^*,1;u,1}_{(i,\bi)\,a,a,a;\alpha, 1}
$ of \CL. 
(We denote here $R_{a,\za}^{(i,\bi, u)}(j)$ what was denoted
$^{a;\za}B_{(i,\bi)}^{j;u}$ in \BPPZ.)
For  $j=1$ it is expressed
in terms of the graph eigenvector matrices $\psi_a^i$
\eqn\IVb{
R_a^{(i,i^*)}(1)={\psi_a^i\over  \psi_a^1 }\ {e^{i\pi \triangle_i}
\over   \sqrt{d_i}}\,.
}

{}From the operator representations \gcvo\ and \hp,
which involve the two  sets of constants, $\Fo$ and 
 $R$  (or,  $C$),
 one recovers all correlators of the fields $\Psi$ and $\Phi^H$;
 they are expressed as linear combinations
of standard CVO correlators. E.g., the  
 $2$-point function projected on the state $|e^1_{aa}\ket$, is
\eqn\bba{
\bra\, ^c\Psi_{j, \alpha}^b(z_2)
\Phi_I^H(z, \bar z)\ket_{a}\
=\delta_{ab}\,
\delta_{ac}\, {P_a\over \sqrt{d_j}}\,\sum_{t}\,
R_{a,\za^+}^{(i,\bi^*, t)}(j^*)
\, \bra 0|\phi_{jj^*}^{1}(z_2)\,
\phi_{i\bi^*; t}^{j^*}(z)\, \phi_{\bi^* 1}^{\bi^*}(\bar z)\,  |0\ket\,.
}
In \bba\  we have adopted an 
ordering corresponding to real parts increasing from right to left,
i.e., Re $(z_2-z) >0$. 
The inverse order would give a function which
differs by a phase (due to the nontrivial braiding 
of products of CVOs), even if the difference of scaling dimensions,
 the spin $s_I=\triangle_i-\triangle_{\bi}$, is (half)integer,
as required from the physical spectrum. The phase 
 vanishes if we furthermore restrict the argument $z_2$
of the generalised CVO to the real axis  boundary of $H_+$
and thus the bulk and the boundary fields commute. 

Let us now briefly review the derivation of the  equations
resulting from the sewing constraints of Cardy--Lewellen \refs{\CL,\Lew}
in the BCFT. The operator
representations introduced here both for the boundary and the bulk fields 
make these derivations straightforward (in fact also slightly more general)
 and reduce them to the use of the  fusing and
braiding relations for the conventional CVO. 
First requiring locality (commutativity)
of a bulk and boundary operators, 
$\Phi_I^H(z,\bar z)\,
^a\Psi_j^b(x_2)= {{}^a\Psi}_j^b(x_2)\, \Phi_I^H(z,\bar z)\,, $
has further implications, leading to an equation for
the unknown constant $C$  in the operator representation
\hp.  It reads, omitting for simplicity the multiplicity indices
\eqn\fle{\eqalign{&
\sum_l\, C_{(i,\bar{i})\, a,b',a}^{n,k,l}\ \ 
\Fo_{b l}\left[\matrix{j&g\cr a&b'}\right]\ 
B_{ll'}\left[\matrix{\bi^*&j\cr k&g}\right] (-)
=
\cr &
\sum_{k'}\,  C_{(i,\bar{i})\, b,b',b}^{k',l',g}\ \ 
\Fo_{b n}\left[\matrix{j&k'\cr a&b'}\right]\ 
B_{k'k}\left[\matrix{j&i\cr n&l'}\right] (-)\,.
}}
 Projecting the product of two fields on $|0\ket$, or on $\bra0|$, i.e.,
setting $g=1$, or $k=1$ in \fle, one 
recovers the (first) bulk-boundary Cardy-Lewellen equation
\refs{\CL,\Lew}; \fle\ is a slightly more general version of it,
corresponding to a $5$-point chiral block.  This equation provides 
a closed expression for the scalar reflection coefficients
$R_{a,\za}^{(i,i^*, t)}(j)$ in terms of the $3j$-symbols $\Fo$ and the
modular matrix $S(j)$ of  $1$-point torus correlators 
\eqn\refl{
{P_a\over \sqrt{d_j}}\, {R_{a,\alpha}^{(i,i^*)}(j^*)\over R_1^{(i,i^*)}(1)}
= S_{i1}\ \sum_{k,b,\beta}\, {\psi_b^i\over \psi_1^i}\ 
\Fo_{ak}\left[\matrix{k&j\cr b&a}\right]_{\beta\, \alpha}^{\beta}
\ S_{ki^*}(j)\,. 
 }
%
In the diagonal case the l.h.s. reproduces $S_{ai}(j)/S_{1i}$
\refs{\R,\BPPZ}.
%

 With \hp\ at hand one also {\it derives}
the OPE of two bulk fields $\Phi^H_{K}\,\Phi^H_{L}$
first expressing their product as a product of four standard CVO, 
 exchanging then the second and third fields and fusing 
 each of the two pairs labelled by
$(k,l)$ and $(\bk^*, \bl^*)$ (this can be depicted by a 
$6$-point chiral block diagram slightly more general than 
Fig. 10 of \BPPZ{}). In the process one finds an expression for
the OPE coefficients, to be denoted $d_{K\,L}^{J;t,t'}$.
It reads symbolically, ordering the constants in the l.h.s.
 in the sequence they appear in the above steps, 
\eqn\slea{
 F\, F\, B(-)\,  C \,  C = d\,  C \,,
}
or   
\eqn\sle{\eqalign{ 
&C_{(k,\bar{k})\, a,b,a}^{m,n,g'}\
  C_{(l,\bar{l})\, a,b,a}^{g',\bar{n},i}=\cr
 &
\sum_{j,\bj,g}\, d_{K\,L}^{J}\,
 B_{gg'}\left[\matrix{l&\bk^*\cr n&\bar{n}}\right](+)\
F_{nj}^*\left[\matrix{k&l\cr m&g}\right]\
 F_{\bar{n} \bj^*}^*\left[\matrix{\bk^*&\bl^*\cr g&i}\right]\ 
 C_{(j,\bar{j})\, a,b,a}^{m,g,i}\,.
}}
Setting $i=1$ and substituting  the constants $C$
with the reflection coefficients $R $ as in \bcc, this 
 can be  also rewritten, introducing a new constant $M$, as
\eqn\sleb{
R_{a,\za_1}^{(k,\bar{k};u_1)}(p_1)\
R_{a,\za_2}^{(l,\bar{l};u_2)}(p_2)=\sum_{j,\bj, p_3, u_3,\za_3 }\ 
M_{(k,\bar{k},u_1;p_1,\za_1))
(l,\bar{l},u_2;p_2,\za_2)}^{\qquad(j,\bar{j},u_3;p_3,\za_3); a}\ 
R_{a,\za_3}^{(j,\bar{j};u_3)}(p_3)\,
}
with $u_1=1,\dots, N_{k\bk}{}^{p_1}\,, \, \za_1=1,\dots, 
n_{p_1 a}{}^a$, etc.
This is the second  of the two basic bulk-boundary Lewellen equations \Lew. 
In the diagonal case $K=(k,k)$  the OPE coefficients $d=d^{(H)}$
coincide with their full-plane counterparts
$d^{(P)}$ and   in the unitary gauge used here are simply
 $d_{K\,L}^{(P)\, J;t,t'}=\delta_{t t'}$ for $N_{kl}{}^j\not =0$. 

The equation \sleb, taken at $p_1=p_2=1$,
 allows to derive and generalise to higher
rank cases (see \refs{\PSS,\BPPZ}) the empirical $sl(2)$
result of \PZa\ on the coincidence of the relative scalar OPE coefficients
and the structure constants of the Pasquier algebra \Pasq.
The latter algebra has  
$1$-dimensional representations (characters) given by
${\psi_a^i\over\psi_a^1}= e^{-\pi i \triangle_i}\,
\sqrt{d_i}\, R_{a}^{(i,i^*)}(1)$, cf. \IVb.
A generalisation of this result will be discussed in section 7 below.
 
The  reflection coefficients  satisfy 
\eqn\conj{
(R_{a}^{(i,\bi)}(j))^* =R_{a^*}^{(i,\bi)}(j^*)\,
e^{-i \pi \triangle_{i\bi}^j}=
R_a^{(i^*,\bi^*)}(j^*)
\,e^{-i \pi \triangle_{i\bi}^j}
\,,
}
and furthermore \sle\ implies, choosing (the positive) constant 
$d_{K\, K^* }^{1}=1$.
\eqn\orta{
\sum_{a\,,\za}\,
R_{a,\za}^{(k,\bar{k};u)}(j)\ R_{a,\za}^{(l,\bar{l};u')\,*}(j)\
 (\psi_a^1)^2\
 {d_k\over d_j }
=\delta_{lk}\, \delta_{\bar{l}\bar{k}}\,  \delta_{u u'}\,
\zd_{\bk k^*}
\,.
}
The identity
\orta\ reduces for $j=1$ to the orthonormality property of $\psi_a^l$ 
(expressing the completeness of the set of boundary states)
 and in the diagonal cases to the unitarity relation for the 
 modular matrices $S(j)$.
In general $d^{(H)}$ and $d^{(P)}$
differ by phases depending on the spins $s_K=\triangle_k-\triangle_{\bk}$,
\eqn\hafu{
d_{K\, L}^J =e^{-i{\pi\over 2}(s_K+s_L-s_\J)}\, 
d_{K\, L}^{(P)\, J} \,,
}
reproducing in particular the spin-dependent full-plane $2$-point 
function normalisation, $d_{K\, K^*}^{(P)\, 1}=(-1)^{s_K}$,
proved to be  consistent with the locality
and reflection positivity requirements \VBPb.

\newsec{Relation to integrable lattice models}
\noindent
Some of the identities  in sections 3 and 4, most  notably the YB
equation, coincide with the basic identities of the related
IRF integrable  lattice models.  The lattice Boltzmann weights, 
however, depend on a spectral parameter $u$, 
which does not appear in the CFT, and to compare 
 the two discussions, a proper limit of this parameter has to be
taken. This correspondence has been established in the diagonal
cases, \Reh, and in this section we show how it generalises
to all models built on graphs related to $\slh(n)_{h-n}$ CFT.

The data required to define the generalised $sl(n)$-IRF models that we 
consider are a graph $G$ -- we postulate  that one of the 
graphs met in the CFT discussion is appropriate-- and a pair 
of representations $j_1$ and $j_2$ for $sl(n)$.
Then to each vertex of the square lattice is assigned a vertex $a$ 
of the graph. 
 The Boltzmann weights $W_{j_1 j_2}\left({c\atop b}{d\atop a}\right)(u)$
are functions of the four vertices $a,b,c,d$
around a square face and of a spectral parameter $u$. 
It is conventional to tilt the lattice by 45 degrees and to
represent the Boltzmann weights as in Fig. 8.
Representation $j_1$ is  assigned to the SW-NE bonds, and $j_2$
to the SE-NW   ones \JKMO. Intuitively, one goes from
vertex $a$ to vertex $b$ through the action of  $j_2$, and from
$b$ to $c$ through $j_1$, and accordingly,
the weights  depend also  in general on bond labels
$\alpha,\gamma,\cdots$, which specify which path from $a$ to $b$,
from $b$ to $c$ etc is chosen, : $\alpha=1,\cdots, n_{j_1\, b}{}^c$,
$\zg=1,\cdots,n_{j_2\,a}{}^b$, etc.

The Boltzmann weights are solutions of the spectral parameter dependent 
YB and inversion (unitarity) equations.  Knowing them  
for the fundamental representation(s) enables one to construct 
the other weights by a fusion procedure \refs{\DJKMO,\WDA,\PZhf}.  


\def\youngone{{ \hbox{\epsfxsize=1.5mm\epsfbox{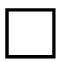}} }}
\def\youngtwo{ \hbox{\raise -0.5mm\hbox{\epsfxsize=1.5mm\epsfbox{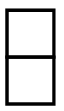}} }}
\def\youngtwoh{ \hbox{\hbox{\epsfxsize=3mm\epsfbox{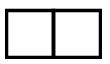}} }}

In the simplest case  where implicitly all the bonds carry 
the fundamental representation  $\youngone$ of $ sl(n)$, 
the Boltzmann weights have the general form
\eqn\bolw{
W\left(\matrix{c&d\cr b&a}\right)_{
\za\, \zg}^{\za'\, \zg'}(u)=
\sin({\pi\over h}-u) \delta_{bd} +\sin(u)\, [2]_{q}\, 
U\Big[{c\atop a}\Big]_{b\,\za\,\zg}^{d\,\za'\zg'}
\,, }
where 
$ [2]_{q}=2\cos({\pi\over h})$ for $q=e^{-2\pi i {h-1\over h}}\,,$
($h$ the Coxeter number of the graph $G$), 
and $[2]_{q}\,U$ are  Hecke algebra  generators satisfying $U^2=U$ etc.  
Choosing the  labels $j=k=\youngone$ in the  bilinear representation \Ie\
for the braiding matrix
$\hat{B}\,$, we can cast it into a form similar to \bolw\ 
\eqn\bow{
\hat{B}_{bd}(\ze)=\delta_{bd}\, q^{a\ze}\  - q^{b\ze}
\ C \, U_{bd}\,, 
}
with (cf. Fig. 8) 
\eqn\bowe{
U\Big[{c\atop a}\Big]_{b\,\za\zg}^{d\,\za'\zg'}
= \sum_{\zb}\, 
\Fo_{b\, \youngtwo}
\left[\matrix{\youngone&\youngone\cr c&a}\right]_{\za \, \zg}^{\zb\,1}\,
\Fo_{d\, \youngtwo}^*
\left[\matrix{\youngone&\youngone\cr c&a}\right]_{\za' \, \zg'}^{\zb\,1}
\ .}
%
\fig{(a): the Boltzmann weight $W$ for $j_1=j_2=\youngone$; 
(b) $U$ as a product of two cells}{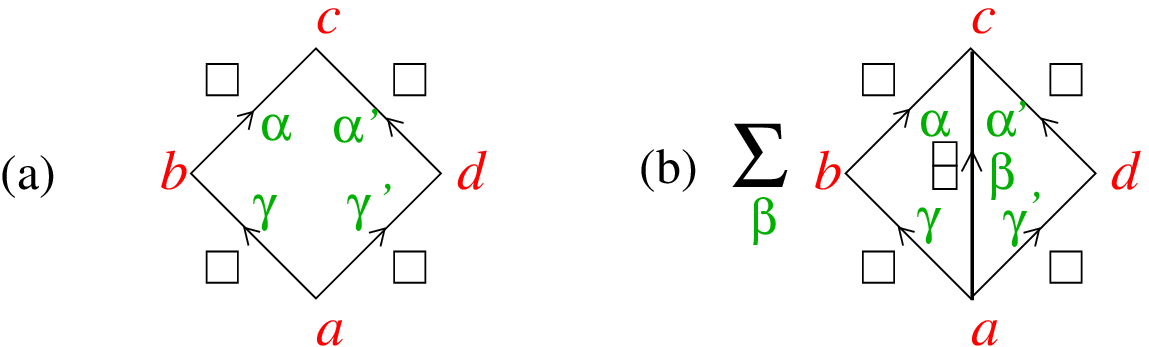}{6cm}\figlabel\diamond

\noindent
The constants  $a,b,C$ are determined from \Idc\ and \Ibbb; 
 from \Idc\ we get $C=q^{a-b}+q^{b-a}$, and  from \Ibbb\ we get 
$a=-h(\triangle_{\youngone}^{\rm Su}-{1\over 2 }\triangle^{\rm
Su}_{\youngtwoh })\,, 
2b-a =-h(\triangle^{\rm Su}_{\youngone}-{1\over 2 }\triangle^{\rm
Su}_{\youngtwo })\,$, hence $C=[2]_q$.
Here  $\triangle_{\zl}^{\rm Su}$ are  Sugawara conformal dimensions,
while   in \Ibbb\ enter the 
dimensions
$\triangle_{\zl}=\triangle_{\zl}^{\rm Su} -<\lambda,\rho> $
 of the minimal $W_n$ model of central charge 
$c=(n-1)[1+2n(n+1) -n (n+1)({h-1\over h} +{h\over h-1})]$;
 this shift of the dimensions is
 accounted for by the sign in front of the second term in \bow{}.
One obtains $a={n-1\over 2n}\,, b=-{1\over 2n}$.

When \bow\ is inserted in the YB equation \ybe, the latter reduces
to the Hecke algebra relation for the operators $[2]_q\,U$ in \bowe, which
can be identified with the operators in 
the r.h.s. of
 \bolw. Thus
 the Hecke generators are expressed in terms
of the $3j$-symbols $\Fo$, recovering a formula  in \AOb.
Furthermore comparing \bolw\ and \bow\ we obtain
\eqn\bw{
\hat{B}_{bd}\left[\matrix{\youngone&\youngone\cr c&a} \right]_{
\za\, \zg}^{\za'\, \zg'}\!\!\!\!(\epsilon)\, 
= 2i\, q^{-\Ge{1\over 2 n}}\, \lim_{u\to -i\Ge\infty}\,
e^{-i\pi\Ge u}\, W\left(\matrix{c&d\cr b&a}\right)_{
\za\, \zg}^{\za'\, \zg'}(u)\,.
}
In other words, we can look at the correlators of the generalised
 CVO with all representation labels fixed to the fundamental ones as
realising a representation of the corresponding Hecke algebra  -- in 
parallel to the path representations of the lattice theory.
In the $sl(2)$ case \bow, \bowe\ reproduce, inserting \Ibb,   the
Boltzmann weight of the $ADE$ Pasquier models \Pas\
\eqn\paslat{
q^{{1\over 4}}\,\hat{B}_{bd}\left[\matrix{f&f\cr a&c}\right]
=q^{1\over 2} \delta_{bd}
- \delta_{ac}\, {\sqrt{P_b P_d}\over P_a}\,.
}
There is no general information on $\Fo$ in the higher rank cases,
however the particular (fundamental) matrix elements in \bowe\
are recovered from  the $sl(n)$ examples of Boltzmann weights
found in the literature, \refs{\JMO,\Wen,\Fen, \DIFZ,\Soch}. 
Recently  exhaustive  results were obtained \AOb\ for
 all $sl(3)$ graphs but one. 
A general existence theorem for $\Fo$ and $W$ for a  subclass of 
graphs corresponding to conformal embeddings appears in \Xu.
On the other hand, as the counter-example of \AOb\ shows, 
some solutions of \nim\ do not support a representation of the
Hecke algebra, i.e., a system of $3j$-symbols $\Fo$.

In the $sl(2)$ and $sl(3)$ cases one can formulate \AOb\
a quartic relation directly for the cells $\Fo$ which in turn implies
the Hecke algebra (or YB) relations; in our notation it reads
\eqn\Pita
{ \eqalign{
&\sum_{b',\atop{ \za_1,\za_2,\atop{ \za_3\, \za_4}}}\,
\Fo_{b' \youngtwo}^*\!\!\!\left[{\youngone\atop a}{\youngone\atop c}\right]_{\za_1  \za_2}^{\zg_1}\, 
\Fo_{c\, \youngtwo}\!\!\left[{\youngone\atop b'}{\youngone\atop d}\right]_{\za_2  \zg_2}^{\za_3}\, 
\Fo_{c' \youngtwo}^*\!\!\left[{\youngone\atop b'}{\youngone\atop d}\right]_{\za_4 \zg_3}^{\za_3}\, 
\Fo_{b' \youngtwo}\!\!\left[{\youngone\atop a}{\youngone\atop c'}\right]_{\za_1  \za_4}^{\zg_4}\,  \cr
&=\sum_{b',\atop{ {\za_1,\za_2}\atop \za_3,
\za_4}}\!{\sqrt{P_c\,P_{c'}}\over P_{b'}}\, 
\Fo_{b' \youngtwo}^*\!\!\left[{\youngone\atop a}{\youngone\atop c}
\right]_{\za_1 \za_2}^{\zg_1\,1}\!
\Fo_{b' \youngone^*}\!\left[{\youngtwo^{\!*}\atop\!\!\! d}{\youngone\atop c}
\right]_{\za_3 \za_2}^{\zg_2^+\,1}\!
\Fo_{b' \youngone^*}^*\!\left[{\youngtwo^{\!*}\atop \!\!\!d}{\youngone\atop c'}
\right]_{\za_3 \za_4}^{\zg_3^+\,1}\!  
\Fo_{b' \youngtwo}\!\!\left[{\youngone\atop a}{\youngone\atop c'}\right]_{\za_1 \za_4}^{\zg_4\,1}\,\cr 
&= {1\over [2]^2}\, \Big({ \sqrt{P_c P_{c'}}\over P_a }\ 
 \delta_{ad}\,  \delta_{ \zg_1\,\zg_2}\, \delta_{\zg_3\,\zg_4}
+  \delta_{cc'}\, \delta_{ \zg_1\,\zg_4}\, \delta_{\zg_2\,\zg_3}  \Big)
}}
The first delta-term  here is present only for the $sl(3)$ case where
$\youngtwo^*=\youngone$ and the two 3-point couplings corresponding to the
$\delta$ function are identical; for $n=2$, where by convention
$\youngtwo$ refers to the identity representation,  the first term is
zero; accordingly we recover the  TLJ algebra relation. 

The fused Boltzmann weights are similarly expected to 
be related to more  general braiding matrix elements.
The recursive construction of the general $\hat B$ elements
using the  fusing-braiding relation \Il\
is analogous to the fusion procedure of the lattice models
yielding the fused Boltzmann weights. 
The ``inversion equation'' for the Boltzmann weights 
in the lattice models  turns into
 the unitarity identity  \Idc. The relation \Il\ taken
for $p=1$ leads to the (crossing) identity 
\eqn\tetrah{
\sum_{b'}\ \hat{B}_{c b' }\left[\matrix{i^*&k\cr b&d}\right]
\!\! (\epsilon)\, \hat{B}_{b d}\left[\matrix{i&k\cr a&b'}\right]
(\epsilon)\, \sqrt{P_{b'}\, P_a\over P_b\, P_d} =\delta_{a c}\,, 
}
while \Ibbb\ with $i=1$ reads
\eqn\ppty
{\sum_{d}\, \hat{B}_{bd}\left[\matrix{j&j^*\cr a&a} \right] \!\!
(\epsilon)\ \sqrt{P_d} 
=\sqrt{P_b}\, e^{-2 \pi i \epsilon \,\triangle_j }\,,
}
a property analogous to one satisfied by the full ($u$ - dependent)
 Boltzmann weights.

We now  turn to the relation \Ii, which has the form of the 
intertwining relation for the square  Ocneanu cells,  
studied in \refs{\Vpa,\DIFZ,\Roch,\PZh}.  To make contact with 
the  notation in \DIFZ, 
$\Fo_{a j}\left[{\youngone\atop c}{i\atop b }\right]$ 
is identified with $Y\left[{a\atop c}{i\atop j}\right]$ 
with $b$ fixed and $i,j,c,a$
 restricted by $n_{ib}{}^a\,, n_{jb}{}^c\ne 0$.  
The  data found in those papers provide thus a partial
information on the $3j$-symbols (i.e., on the boundary 
field OPE coefficients), namely determine those
matrix elements in which one of the representation labels is 
fixed to the fundamental weight $\youngone$ and $b$ fixed to $1$.
On the other hand, knowledge of the cells 
$\Fo_{a j}\left[{\youngone\atop c}{i\atop b }\right]$ for  
arbitrary $a,b,c$ and $i,j$ is sufficient to determine all the cells
using the pentagon equation, in a way similar to the discussion at 
the end of section 4. A general solution for $\Fo$ for the 
$\slh(2)$ D-series has been found in \Rb. 

\medskip

We conclude with the remark that it would be
interesting to extend the correspondences discussed in this section
to the boundary lattice theories,  see \refs{\BPZ,\BPa},  and in 
particular to clarify the r\^ole of the reflection 
equations \Ch\ in the present setting.

\bigskip


\newsec{Ocneanu graphs and the associated algebras}

\def\M{{\bf T}} 

\noindent
In the following we shall
motivate on physical grounds and by analogy with a situation 
already encountered in BCFT  the construction of 
new sets of (non-negative integer valued) matrices and their
associated graphs. On a mathematical level, this construction 
has been justified in the subfactor approach \refs{\AO,\AOb,\BE,\BEK},
but the field theoretical approach provides new insight. 

In BCFT we know that three sets of matrices play an
interlaced r\^ole, generalising the fusion matrices $N_i$. 
The first is the set of $|\CV|\times|\CV|$ matrices $n_i$ defined 
in \nim, which form a 
representation of the fusion algebra and define the graph $G$.
As recalled in section 2, their diagonalisation introduces a set of 
orthonormal eigenvectors 
$\psi_a^j$, $a\in \CV$, $j\in\Exp $. 

The second set of matrices, also of size $|\CV|\times|\CV|$, 
denoted 
$\hN_a=\{\hN_{ba}{}^c\}$ in \BPPZ, forms the 
regular representation of an associative algebra, 
\eqn\hNalg{\hN_a\hN_b=\hN_{ab}{}^c\hN_c}
(the Ocneanu algebra) \refs{\AOold,\Vpa}. 
It is attached to the graph in the sense that 
\eqn\hatN{n_i \hN_a=\sum_b n_{ia}{}^b \, \hN_b\ ,}
i.e., if the matrix $\hN_a$ is assigned to vertex $a$ of the graph $G$, 
the action of $n_i$ on $\hN_a$ gives a sum over the neighbouring 
matrices $\hN_b$ (neighbouring in the sense of the adjacency matrix 
$n_{ia}{}^b$).

In general, these matrices $\hN_a$ have entries that are integers, 
but in general of indefinite sign
\foot{A case where this integrality property 
of the $\hN_{ab}{}^c$ seemed invalid was pointed out in \PZb, but later
it was shown by Xu that integrality could be restored at the expense of 
commutativity \Xu, see below sect 7.2.}. 
At this point, we recall that RCFT and the associated graphs $G$ 
come in two types. Those for which the  modular invariant partition 
function is block-diagonal and expressible in terms of the $n$ matrices as
\eqn\typeI{Z_{ij}=\sum_{a\in T}\, n_{i1}{}^a n_{j1}{}^a }
for a certain subset $T$ of vertices are called of type I. 
They are interpreted as diagonal theories in the sense of some extended 
chiral algebra $\gAe$. The set $T$ is in one-to-one correspondence 
with the set of ordinary representations of that algebra $\gAe$
and the integer $n_{i1}{}^a $ is the multiplicity mult$_a(i)$
 of representation
$\CV_i$ of $\gA$ in the representation of $\gAe$ labelled by $a$. 
Then all matrices 
$\hN_a\,,\, a\in \CV$ have non negative integer entries and
the subset $\{\hN_a\}_{a\in T}$ forms a subalgebra isomorphic
to the fusion algebra of $\gAe$, \DIFZb
\foot{
These statements are for us empirical facts, of which we know
no general proof. They seem to have been established for 
a variety of cases in the subfactor approach or are taken as
assumptions. 
Note that our definition of type I in \typeI\ 
above is slightly more restrictive than the one used previously
\refs{\PZb,\BPPZ}. It rules out one of the graphs ($\CE^{(12)}_3$
in the Table of \BPPZ). See also 
\BEc\ for cases which go beyond this simple classification.}

An interpretation 
of the whole set of $\hN_{ab}{}^c$ as fusion coefficients 
of a class of ``twisted'' representations of $\gAe$ broader 
than considered in section 2 has been 
proposed in \refs{\BPPZ,\FSoo}, see also \Hon\ and sect 7.6 below.  
In contrast, a theory of type II cannot be written as in \typeI\ and 
is obtained from some type I one --its ``parent theory''-- through 
an  automorphism of its fusion rules
acting on  its right sector with respect to the left one \refs{\MS,\DV}. 
We thus expect many of their properties to be more
simply expressed in terms of data pertaining to the parent 
theory.  For example, their torus partition function reads
\eqn\typeII{Z_{ij}
 =\sum_{a\in T}\, n_{i1}{}^a n_{j1}{}^{\zeta(a) }\ , }
where the $n$'s are those of the parent type I theory.

We can then define the dual (in the sense of \BI) 
of the $\hN$ algebra by the algebra of linear maps $\hN \to \IC$. 
This algebra,  also called the Pasquier algebra, is realised by 
matrices $M_{(i,\za)}$ labelled by the elements of $\Exp$ 
and as mentioned in section 5 relates to the scalar OPE coefficients. 

As a side remark, we recall that in the $sl(2)$ case 
it is this $M$ algebra which also 
appears as the perturbed chiral ring of $N=2$ superconformal CFTs 
perturbed by their least relevant operator (or of their topological 
counterparts) \refs{\many}, 
hence as a specialisation of the Frobenius
algebra \Dub. 
We shall return to these algebras and their CFT interpretation
in the next sections. 

\bigskip
In the following, we are going to introduce four sets of matrices, which 
generalise the previous three, define again graph(s) $\tG$, and satisfy
analogous relations. The matrices $n$ gives rise to two sets, 
denoted $\tV$ and $\tn$, while the dual pair $(\hN,M)$ generalises to
a pair $(\tN,\tM)$.


\subsec{The $\tV$ matrices and Ocneanu graphs}
\noindent
We  first consider the integral, nonnegative matrix  solutions of a  system
of equations for commuting matrices $\tV_{ii';\, x}{}^y$ with $i,i'\in \CI$.
It  generalises \nim, with the Verlinde fusion
 multiplicities $N_{ij}{}^k$ replaced by the product
 $N_{ij}{}^k \,N_{i'j'}{}^{k'}$
\eqn\tosya{
\sum_y\, \tV_{ij;\, x}{}^y\,  \tV_{i'j';\, y}{}^{z} = \sum_{i'',j''}\, 
N_{i i'}{}^{i''}\, N_{j j'}{}^{j''}\,  \tV_{i''j'';\, x}{}^z \,.
}
The labels $x,y,\cdots$ of these matrices take their values in a finite
set denoted $\tCV$, whose cardinality equals $|\tCV|=\sum_{j\bj}(Z_{j\bj})^2$
in terms of the modular invariant matrix $Z$.

This property, and more generally the physical interpretation of \tosya, 
follow from the discussion of torus partition functions in the presence 
of twist operators (physically defect lines) 
denoted $X_x$, see \PZtw\ for details. The discussion 
is parallel to the way equation \nim\ appears in the study of cylinder 
partition functions and involves the consistency between two alternative 
pictures. In one picture, two twist operators $X_x^\dagger$ and $X_y$, 
attached to  
homology cycles of type {\tt a} of the torus,  act in the Hilbert 
space of the ordinary bulk theory, $\CH=\oplus Z_{i\bi}\ \CV_i\otimes 
\bar\CV_{\bi}$, and are assumed to commute with the generators of the 
two copies of the chiral algebra $\gA$. This forces them to be linear 
combinations of operators $P^{(k,\bk; \zg,\zg')}$
intertwining the different copies of equivalent 
representations of $\gA\times\gA$
\eqn\tpro{
X_x=\sum_{i,\bi\atop \za,\za'=1,\cdots Z_{i\bi}}\
{\Psi_x^{(i,\bi;\za,\za')}\over  \sqrt{S_{1i}S_{1\bi}}}\ 
P^{(i,\bi; \za,\za')}
\ ,}
with 
\eqn\projP{P^{(i,\bi; \za,\za')} P^{(j,\bj; \zb,\zb')}
=\delta_{ij}\delta_{\bi\bj}\delta_{\za'\zb}\, P^{(i,\bi; \za,\zb')}\ .
}
The other picture  makes use of a Hilbert space $\CH_{x|y}$
associated with the  homology cycles of type {\tt b};  
the non-negative integer $\tV_{ij^*;\, x}{}^y$ 
describes the multiplicity of representation $\CV_i\otimes\overline{\CV_j}$
in $\CH_{x|y}$. 
The equality of the twisted partition functions computed in these two
alternative ways  leads to a consistency condition of the form
\eqn\Cardy
{  \tV_{i\bi;\, x}{}^y=
\sum_{\J\atop \za,\za'=1,\cdots Z_{j\bj}}\,
{S_{ij}S_{\bi\bj} \over S_{1j}S_{1\bj}}\
\Psi_x^{(j,\bj; \za,\za')}\ \Psi_y^{(j,\bj;\za,\za')\, *}
\,,\qquad i,\bi\in\CI \,,
}
where $\Psi_y^{(j,\bj;\za,\za')\, *}$ is the complex conjugate
of $ \Psi_y^{(j,\bj;\za,\za')\,}$.
Then $\Psi=\{\Psi_x^{(\J;\za,\zb)}\}$ is assumed to be 
a square, unitary matrix, labelled 
by the  $x\in\tCV$ and by the pairs $\J=(j,\bj)$ of labels  
supplemented by  their multiplicities in the spectrum 
$\za,\zb=1,\cdots, Z_{j\bj}$
\eqn\Ortho
{\eqalign{
\sum_{x\in \tCV} \Psi_x^{(\J; \za,\zb)} 
\Psi_x^{(\Jp; \za',\zb')\, *}
&= \delta_{jj'}\delta_{\bj\bj'}\delta_{\za\za'}\delta_{\zb\zb'}\cr
\sum_{\J\atop \za,\zb=1,\cdots Z_{j\bj}} \Psi_x^{(\J; \za,\zb)}
\Psi_{x'}^{(\J; \za,\zb)\, *} 
&=\delta_{xx'}\ .\cr }}
Following a standard argument, in \Cardy\ the $\Psi^{(\J;\za,\zb)}$ 
appear as the eigenvectors and the ratios ${S_{ij}S_{\bi\bj}/S_{1j}S_{1\bj}}$
as the eigenvalues of the $\tV_{i\bi}$ matrices. As the latter
  satisfy the double fusion algebra \tosya,  so do the matrices $\tV$. 

In fact the integer numbers $\tV_{ij;\,x}{}^y$ may be regarded 
not only as the entries of $|\tCV|\times |\tCV|$ matrices 
$\tV_{ij}$, $i,j\in\CI$, as we just did, but also as those 
of $|\CI|\times|\CI|$ matrices $\tV_x{}^y$, $x,y\in\tCV$. 

By convention, the label $1$ refers to the trivial (neutral) twist, and 
it is thus natural to impose the further constraint that for 
$y=z=1$,  $\tV$ reduces to the modular invariant matrix, up to a conjugation 
\eqn\trivial{ \tV_{i\bi^*;\, 1}{}^1=Z_{i\bi}\ .}
This is consistent with \Cardy\ if 
\eqn\Psiid{\Psi_1^{(\J;\za,\za')}= \sqrt{S_{1j} S_{1\bj}}\,\delta_{\alpha
\alpha'}=: \Psi_1^{(\J)}\, \delta_{\alpha \alpha'}  \ .}
In particular $\Psi_1^{(1)}=S_{11}$ and 
denoting $\tilde{d}_x={\Psi_x^{(1)}\over
\Psi_1^{(1)}}$, this implies, using the unitarity of $\Psi$
and of the modular matrix $S$, the ``completeness'' relation 
\eqn\eqdim{
\sum_{x\in \tilde{\CV}}\, \tilde{d}_x^2 ={1\over (S_{11})^2}=
\sum_{i\in \cal I}\, d_i^2\,,
}
(see also \BE).
It also follows from \Cardy\ that
$\tV_{ij;\, x}{}^y=\tV_{i^*j^*;\,y}{}^x$ and conversely, 
this latter property, together with \tosya, suffices to guarantee the
diagonalisability of the $\tV$ in an orthonormal basis, as in \Cardy.

Then, if we define the matrices $\M^x_{i,j}:=\tV_{ij^*;\, 1}{}^x\,,$
 thus $\M^1=Z$,  taking the $x=z=1$ matrix element of \tosya\ yields
\eqn\tosy{
\sum_x\, \M^x_{i,j} \ \M^x_{i'^*,j'^*}       
= \sum_{i'',j''}\, N_{i i'}{}^{i''}\, N_{j j'}{}^{j''}\,  Z_{i''j''}
\,,  }
which is the way the matrices $\M^x$ appeared originally in the 
work of Ocneanu, under the name of  ``modular 
splitting method''.  

\medskip
The set of matrices $\tV_{ij}$ may be regarded as the adjacency matrices 
of a set of graphs with a common set of vertices $\tCV$. In
any RCFT, the fusion ring is generated by a finite number of 
representations $f$ of $\CI$ called fundamental, and because of
\tosya, it is sufficient to give the graphs of $\tV_{f1}$ and of 
$\tV_{1f}$ for these  representations
to generate the whole set by fusion. 
(For example, for the $\slh(2)$ theories, $\tV_{21}$ and $\tV_{12}$
suffice.) 
Following Ocneanu \AO, it is convenient to represent these graphs
simultaneously on the same chart, with edges of different colours.
We shall refer to this multiple graph as the Ocneanu graph $\tG$
associated with the graph $G$ of the original theory.
Examples are given in Fig. 10 of  Appendix B for $\slh(2)$ 
theories, and additional ones may be found in \refs{\AOb,\BEK}.
If one attaches the matrix $\M^x$ to vertex $x$, the 
two kinds of edges of the graph
describe the action  of the fusion matrices on the left and right indices
 of the $\M^x$. For example the edges of the first colour (red full lines
on Fig. 10) encode the $\tV_{f1;\, y}{}^z  $ in
\eqn\fusact{N_{fi}{}^{i'}\M_{i'j}{}^z= 
\sum_y\, \tV_{f1;\, y}{}^z\  \M_{ij}{}^y  }
or in short, $ (N_f\otimes I) \M^z=\sum_y\, \tV_{f1;\, y}{}^z\,  \M^y$, 
and likewise, those of the second colour (blue, broken) describe
$ (I\otimes N_f) \M^z =\sum_y\, \M^y\, \tV_{1f;\, y}{}^z\,.$ 

\subsec{The $\tN$ algebra}
\noindent
In turn, this Ocneanu graph $\tG$ may be used to define a new algebra
in the same way as $\hN$ was attached to the graph $G$. 
{} To each vertex $x$ of the graph we attach a matrix 
$\tN_x=\{\tN_{yx}{}^z\}$ of size $|\tCV|\times |\tCV|$.  
For the special vertex $1$,  $\tN_1=I $.  
The matrices $\tN$ are  assumed to satisfy the algebra \misys: 
$ \tV_{ij}\, \tN_x=\sum_z\, \tV_{ij;\, x}{}^z\, \tN_z\ , $
(compare with \hatN).
Using the spectral decomposition \Cardy\ of the $\tV$, one 
may construct an explicit solution for these matrices $\tN_x$
\eqn\Ibl{
\tN_{y x}{}^z=\sum_{\J;\za}\,\sum_{\zb,\zg }\, \Psi_y^{(\J;\,\za,\zb)}\,
{\Psi_x^{(\J;\, \zb,\zg)}\over \Psi_1^{(\J)}}\,
\Psi_z^{(\J;\, \za,\zg)\, *}\,.  }
 Taking into account the orthonormality of the $\Psi$, one finds 
that $\tN_{1x}{}^z=\tN_{x1}{}^z=\delta_{xz}$, 
and
that  $\tN_x$  form a matrix representation of an algebra 
\def\hE{{\hat E}}
\eqn\tNrepb{
    \hE_x \  \hE_y=\sum_z \tN_{xy}{}^z\hE_z\,, }
with an identity and a finite basis.
The algebra is associative, but in general non-commutative 
if some $Z_{ij}>1$. Indeed, if all  $Z_{ij}=1$, the summation over 
$\za,\zb,\zg $ in \Ibl\ is trivial, and this equation is, once again, 
nothing else than the spectral decomposition of the matrices $\tN$ 
in terms of the one-dimensional representations $\Psi_x/\Psi_1$ 
of the algebra. If, however, some  $Z_{ij}>1$, 
the matrices $\tN$ are not simultaneously diagonalisable, but 
rather {\it block-diagonalisable} with blocks 
 $\gamma_x^{(\J;\zb,\zg)} ={\Psi_x^{(\J;\zb,\zg)}/ \Psi_1^{(\J)}} $  
forming a $Z_{j\bj}$-dimensional representation of the algebra
\eqn\Ibla{
\sum_{\zb}\, \gamma_{x}^{(\J;\za,\zb)}\, \gamma_{y}^{(\J;\zb,\zg)} =\sum_z\,
 \tN_{x y}{}^z\, \gamma_{z}^{(\J;\za,\zg)} \ .  }
(see also \BE\ (Lemma 5.2) for a similar although somewhat less explicit
variant of \Ibl, with $\tN_{xy}^z=\bra \beta_z,  \beta_x\circ \beta_y, \ket$ 
being the    ``sector product matrices'').
By inspection, one checks, at least in all $ADE$ cases 
(see below and Appendix B),  
that these $\tN$ matrices have non negative integer entries.
They may indeed be viewed as multiplicities (of  dual triangles
with three white vertices), 
and accordingly  the algebra \tNrepb\ appears as the algebra  of the 
center of $\hA$,  with the product in the l.h.s. of \tNrepb\ given by  
the vertical product of \AO, compare with \abV\  and  
see Appendix A. 
These matrices are recovered also directly from \tpro, \projP,
\eqn\trtw{
\tN_{yx}{}^z={\rm Tr}( X_y\ X_x \ X_z^{\dagger})
}
using that Tr$P^{(\J;\, \za,\zb)}:= \delta_{\za\zb}\, S_{1j}\, S_{1\bj}$ 
(this definition of the trace may be justified in unitary CFT's 
in exactly the same way as the norm of the Ishibashi states,
 via the $\tau \to \infty$ asymptotics of the characters $\chi_j(\tau)$, 
see \refs{\Ishi,\BPPZ} and (4.2) of \PZtw). Equivalently, we have
\eqn\fustw{X_x X_y=\tN_{xy}{}^z X_z\ ,}
thus justifying the name of twist fusion algebra that we give to
the $\tN$ algebra. In this latter context, the non-commutativity 
of this $\tN$ algebra may be viewed as coming from the inpenetrability 
and the resulting lack of commutativity 
of the defect lines to which the twists $X_x$ and $X_y$ are attached.

If a conjugation in the set $\tCV$ is defined through 
\eqn\conjPsi{\Psi_{x^*}^{(\J;\,\za,\zb)}=
\big(\Psi_x^{(\J;\,\zb,\za)}\big)^*}
(note the reversal of the indices $\za$ and $\zb\,$!), it follows that
\eqn\defxst{\tN_{yx}{}^1=\delta_{xy^*} \ ,}
and the noncommutativity of $\tN$ modifies the analogue of the symmetry
relations \sint\ according to   
\eqn\stN{
\tN_{y x }{}^z=\tN_{x^* y^*}{}^{z^*}=\tN_{z x^*}{}^y
\,.  }

Equation \Ibl\ may be rewritten
\def\e{{\bf e}}
as a sum of $\sum_{j\bj; \alpha} 1=\sum_{j\bj} Z_{j\bj}$
 (matrix) idempotents $\e^{\J;\za}_{yz; \zb,\zg}={1\over Z_{j\bj}}\,
\Psi_y^{(\J;\za,\zb)}\, \Psi_z^{(\J;\za,\zg)\, *}\,,$
\eqn\idem{
 (\e^{\J;\za}\, \e^{\J;\za})_{xz;\zb,\zg}=\sum_{y,\zg'}\,
 \e^{\J;\za}_{xy; \zb,\zg'}\  \e^{\J;\za}_{yz;\zg',\zb}=
 \e^{\J;\za}_{x z; \za,\zb}\, , \qquad  
\sum_{\J,\za,\zb}\,  Z_{j\bj}\, \e^{\J;\za}_{\zb,\zb}=\tN_1
\,,}
\eqn\decom{
\tilde{N}_{x}=\sum_{\J,\za}\,
\sum_{\zb,\zg}\,  Z_{j\bj}\, \gamma_{x}^{\J;\zb,\zg}\, \e^{\J;\za}_{\zb,\zg}\,=
\sum_{\J}\, Z_{j\bj}\, \gamma_{x}^{\J}\,\sum_{\za}\, \e^{\J;\za}\,,
}
where, suppressing the matrix indices, the sum runs over the physical spectrum
$(\J,\za):=(\J,\za,\za)$.
These are the labels of the  representations of the $\tN$  algebra, which
  are $Z_{j\bj}$ -dimensional and
 given according to \Ibla\  by the matrices $\gamma^\J_x$, i.e.,
\eqn\Nrep{\eqalign{
&\triangle_{\J;\za, \zb}: \tilde{N}_{x} \to
\triangle_{\J;\za,\zb}(\tilde{N}_{x})=
 \gamma_{x}^{\J;\za,\zb}\,,\cr
& \triangle_{\J;\za,\zg}(\tilde{N}_{x}\, \tN_y)
=\sum_{\beta}\, \triangle_{\J;\za,\zb}(\tilde{N}_{x})
\triangle_{\J;\zb,\zg}(\tilde{N}_{y})=\sum_z\,  \tN_{xy}^z\,
\triangle_{\J;\za,\zg}(\tilde{N}_{z})\,.
}}
The formula \decom\ is then interpreted as a decomposition of the
regular representation of the $\tN$-algebra into a sum of representations
$\triangle_\J$
each appearing with multiplicity $Z_{j\bj}$ so the dimension is 
$\sum_{j\bj}\, Z_{j\bj}\,$ dim($\triangle_\J$)=$\sum_{j\bj}\,
Z_{j\bj}^2\,=|\tCV|$. In \BEK\  a formula analogous to \Nrep,
or \Ibla\ appears directly for the elements $\hat{E}_x$ in 
$\hA$ spanning (with respect to the vertical product)
 the algebra \tNrep, see Appendix A.

\bigskip
We now return to the graph  algebra $\hN$ of the chiral graph $G$
mentioned in the introduction to 
this section.  In fact, we shall restrict our attention to type I cases, 
which are the only ones for which all the matrix elements of
the $\hN_a$ are non negative integers. Because in this case, equation
\typeI\ applies, each exponent appears $(n_{j1}{}^a)^2$ times 
for each  representation $a$ of the extended algebra, identified 
with a vertex $a\in T$.
It is advantageous to denote the corresponding eigenvectors of 
the $n$ matrices as $\psi^{(j,a;\za,\zb)}$, with $\za,\zb=1,\cdots,
n_{j1}{}^a$. In \PZb, various formulae have been established 
for the components  
$\psi_b^{(j,a)}$, $b\in T$. It is 
easy to extend them  to
\def\Se{S^{{\rm ext}}}
\eqn\psiinT{
\eqalign{\psi_1^{(j,a;\za,\zb)}=\delta_{\za\zb}\, \psi_1^{(j,a)}\,,
\quad  \psi_1^{(j,a)}
=\sqrt{S_{1j}\Se_{1a}}\,,\quad 
{\rm for}\ n_{j1}^a\ne 0\,,
\cr
\psi_b^{(j,a;\za,\zb)}= \delta_{\za\zb}\, \psi_1^{(j,a)}\,
\,{\Se_{ba}\over \Se_{1a}}\,, \qquad {\rm for\ }a, b\in T\ , \cr} }
using the modular invariance identity
\eqn\mipr{
\sum_{i\in \CI}\ S_{ij}\  n_{i1}{}^b   = 
\sum_{a\in T}\   S_{ba}^{\rm ext}\ n_{j1}{}^a\,, \qquad b\in T\,.
}
The similarity with the case of the $\tN$ algebra (of which the 
$\hN$ algebra turns out to be a subalgebra in these type I cases)
suggests a formula which encompasses and generalises all known cases
\eqn\hNform{
\hN_{c b }{}^d=\sum_{a\in T\,, j\in \CI
\atop \za,\zb,\zg=1,\cdots, n_{j1}{}^a }\,
 \psi_c^{(j,a;\,\za,\zb)}\,
{\psi_b^{(j,a;\, \zb,\zg)}\over \psi_1^{(j,a)}}\,
\psi_d^{(j,a;\,\za,\zg)\, *}\,. }
It is an easy matter to check that the relations \hNalg\ and 
\hatN\ are indeed satisfied. 
We have checked 
in the simplest case $n=2$ of $\slh(2n)_{2n}\subset
\widehat{so}(4n^2-1)_1$, for which multiplicities $n_{i1}{}^a >1$ are known 
to occur, and we conjecture in general, that this
formula always yields non negative integers, and gives an  
explicit realisation of the considerations of \refs{\Xu,\BEb} 
\foot{It is understood that in cases 
where exponents come with a non trivial multiplicity, 
the remaining  arbitrariness  in the choice of the $\psi$
is used to make the $\hN$ nonnegative integers, and this seems always
 possible in type I cases.}.


\subsec{The $\tn$ matrices}
\noindent
We then introduce a new set of matrices $\tn_x=\{\tn_{ax}{}^b\}$, 
$a,b\in \CV$,  which form a non negative integer valued representation
(nimrep) of this $\tN$ algebra, see \tnrep, in clear analogy with \nim.
\foot{
In the subfactor approach, given  an inclusion of subfactors $N\subset M$,
the  equality \tnrep\ is interpreted as an associativity condition
for the $M-M$, $M-N$ sectors, similarly as the analogous
identity \nim\ 
for the  $N-N$, $N-M$ sectors \BEK.}  Like the $\tN$, these matrices 
are non-commuting in general, if some $Z_{j\bj}>1$, and they 
admit a block decomposition like \Ibl
\eqn\tnblo{\tn_{ax}{}^b=
\sum_{j}\,\sum_{\za,\zb=1,\cdots,Z_{jj} }\, \psi_a^{j,\za}\,
{\Psi_x^{(j,j;\, \za,\zb)}\over \Psi_1^{(j,j)}}\,
\psi_b^{j,\zb\, *}= \tn_{bx^*}{}^a\,.  }
One also checks, using the orthonormality and conjugation properties of 
the $\psi$ and $\Psi$,  
that
\eqn\onemore{
\sum_{x\in\tCV} \tn_{ax}{}^{a'}\tn_{b'x^*}{}^b =\sum_{i\in\CI}
 n_{ia}{}^b n_{i^* b'}{}^{a'}\ . }
These matrices are again interpreted as multiplicities: namely
$\tn_{ax}{}^b$  
describes  the dimension of the space $\hV_{a x}^b$ of dual triangles
with fixed markings $x,a,b$ (one black, two white vertices). Varying $a,b$, 
 they form a basis of the dual vector space $\hV_x$. 
Then \tnrep\ serves as a consistency condition needed to give sense
to the dual (vertical) product $\hV_x\otimes_v\, \hV_y$,
in which $\hV_z$ appears with multiplicity $\tN_{xy}{}^z$,
the latter replacing the Verlinde multiplicities in a formula 
analogous to \fusdec.
On the other hand \onemore\ is interpreted as the equality between 
the dimensions of the space of double triangles
and that of  dual double triangles with $a,a',b,b'$ fixed
 and justifies a change of basis considered in Appendix A (see Fig. 9)
Recalling that in 
section 3, $m_j=\sum_{a,b\in\CV}\,n_{ja}{}^b$ stands for the dimension 
of the space $V^j$ of triangles (or CVOs), we  now denote
$\tm_x=\sum_{a,b\in \CV} \tn_{xa}{}^b$ the dimension of the 
 space $\hV_x$.
The equality of the dimensions of the double triangle  algebra $\CA$  
and of its dual $\hA$ amounts to the identity
\eqn\equdim{\sum_{j\in\CI} m_j^2=\sum_{x\in\tCV} \tm_x^2}
which results indeed from the summation over $a,a',b,b'$ in \onemore.
On the other hand a less trivial equality holds, 
checked case by case in all $sl(2)$ cases,
\eqn\eqdi{
\sum_{j\in\CI}\, m_{j } 
=\sum_{C_x\,,\, x\in \tCV}\, \tilde{m}_{ x}\,
}
where the sum in the r.h.s. runs over the ``classes'' $C_x$ in $\tCV$
(or classes in the $\tN$ algebra), determined by 
$x \sim y\,,$ iff $
\forall \J,\ {\rm tr}(\triangle_\J(\tN_x))
=\sum_{\za,\za}\,\zg^{\J, \za,\za}_x ={\rm tr}(\triangle_\J(\tN_y)) $, 
i.e., the characters of the representations of the $\tN$ algebra
are constant on the class $C_x$. 
For cases with trivial multiplicities
$Z_{j\bj}=0,1$ the summation in the r.h.s. runs over the set $\tCV$
and  \eqdi\ expresses the equality of dimensions of the 
regular representations of $\CA$ and $\hA$. 
In the $sl(2)$ $D_{\rm even}$ cases there are two nontrivial 
classes $C_x$, each containing the fork vertices in the chiral subgraphs 
of $\tG$, see Appendix B.

\medskip

 The physical interpretation of the matrices \tnblo\
 is obtained by looking 
at the effect of a twist in the presence of boundaries. One 
consider the RCFT on  a finite cylinder with boundary states 
$|a\rangle$ and $\langle b|$ at the ends and a twist operator
$X_x^\dagger$ in between. Repeating the calculations of partition functions
carried out in \refs{\BPPZ,\PZtw}, one finds that the ``open string channel''
is described by a Hilbert space with representation $\CV_i$
occurring with multiplicity $(n_i\tn_x)_{a^*}{}^{b^*}$, i.e. the matrix
element of a (commuting) product of the matrices $n_i$ and
$\tn_x$. 
Thus  $(\tn_x)_{a^*}{}^{b^*}$ is the multiplicity of the
identity character in this ``twisted'' cylinder partition function.


\subsec{The $\tM$ matrices}
\noindent
The last set of matrices that we may associate with the Ocneanu graph 
generalises the Pasquier algebra.   
We can define a dual (in the sense of \BI) 
of the $\tN$ algebra by the algebra of linear maps 
\eqn\dualch{\eqalign{
\triangle^+_{\J;\zb, \zb'}:\  
\tN_{x} \to \triangle^+_{\J;\zb,\zb'}(\tilde{N}_{x}) &=
{\Psi_1^{(\J;\zb,\zb')}\over \Psi_x^1}\,\triangle_{\J;\zb, \zb'}(\tilde{N}_{x})
={\Psi_x^{(\J;\zb,\zb')}\over \Psi_x^1}\ ,\cr
(\triangle^+_{\I;\za, \za'}\, \triangle^+_{\J;\zb, \zb'})(\tN_{x})
&= \triangle^+_{\I;\za,\za'}(\tilde{N}_{x})\,
\triangle^+_{\J;\zb,\zb'}(\tilde{N}_{x})\cr 
&= \sum_{\K;\zg,\zg'}\, 
\tM_{(\I;\za,\za')\,(\J;\zb,\zb') }{}^{\!\!(\K;\zg,\zg')}\
\triangle^+_{\K;\zg,\zg'}(\tilde{N}_{x})
}}
with structure constants
\eqn\Mtilde
{\tM_{(\I;\za,\za')\,(\J;\zb,\zb') }{}^{\!\!(\K;\zg,\zg')}=\sum_x\,
{\Psi_x^{(\I;\za,\za')}\over  \Psi_x^{(1)}}\,
 \Psi_x^{(\J;\zb,\zb')}\, \Psi_x^{(\K;\zg,\zg')\, *}\,.
}
This algebra is abelian and its $1$-dimensional representations, or characters,
are given by \dualch. An involution $(*)$ in the set $\{(\I;\za,\za')\}$
is  defined   by 
the complex conjugation $\Psi_x^{(\I;\za,\za')^*}=
\Psi_x^{(\I;\za,\za')\,*}$ so that 
$M_{(\I;\za,\za')^*}={}^t\!M_{(\I;\za,\za')}$.
The subset of the numbers formed by the $\tM_{(\I;\,\za,\za)
(\J;\,\zb,\zb)}{}^{(\K;\,\zg,\zg)}$, i.e. diagonal in the 
multiplicity indices, 
plays a 
physical r\^ole. Their explicit computation
(in the $ADE$ cases)
 shows  that (i) they are non negative algebraic 
numbers; (ii) they give the modulus squares of the relative structure 
constants of the OPA of the corresponding CFT
\eqn\OPA{
|d_{(\I;\, \za)(\J;\,\zb)}{}^{(\K;\,\zg)}
|^2=
\tM_{(\I;\,\za,\za)\, (\J;\,\zb,\zb)}{}^{(\K;\,\zg,\zg)}
\ .  }
We recalled in section 5  that the Pasquier 
algebra gives access to the relative structure constants of 
spinless fields.
The OPA structure constants of non left-right symmetric fields, 
however, were escaping in general
this determination in terms of graph-related data
\foot{The equation \slea\ represents the
constants $d$ in terms of the $3j$- and $6j$-symbols and  the 
general  nonscalar reflection coefficients.}.
The empirical result in \PZa\ only states that in the cases
of conformal embeddings $D_4, E_6, E_8$
the l.h.s. of \OPA\ factorises into a product of scalar constants
(and hence is expressed by the Pasquier algebra structure constants)
and that for the $D_{\rm even}$ series this factorisation holds in a 
somewhat weaker sense; this factorisation
 is confirmed (see Appendix B) by what is computed for the r.h.s. 

 In fact \OPA\ can be derived extending the consideration 
of \PZtw\ to $4$-point functions of physical fields in the presence 
of twists; it is sufficient to look at the functions on the plane,
which can be interpreted as the $L/T\to \infty$  limit of the torus
correlators, $\lim_{L/T\to \infty}\ $Tr$ (e^{-2L H}\, ......)$, 
when we map it to the plane through $w\to z=^{-2\pi i w\over T}$.
 Let us sketch the argument which is a generalisation of the
derivation of the locality equations; we shall use the convention
of notation in \PZb. 
We consider a $4$-point function with insertion of two twist
operators \tpro\ (omitting the labels $(P)$ on the fields and
the OPE coefficients)
\eqn\vone {\eqalign{
\bra 0|&\Phi_{(J^*; \zb^*)}(z_1, \bz_1)\,
 \Phi_{(I^*; \za^*)}(z_2, \bz_2)\, 
 X_x\,
\Phi_{(I;\za')}(z_3, \bz_3)\,
\Phi_{(J;\zb')}(z_4, \bz_4)\, X_x^\dagger |0\ket \cr 
&= \sum_{k,\bk,\zg,\zg'}\,
d_{(J^*; \zb^*) (J; \zb)}^{(1) }\
d_{(I^*; \za^*)(K; \zg,\zg') }^{(J; \zb)}\
{\Psi_x^{(k,\bk;\zg,\zg')}\over \Psi_1^{(k,\bk)}}\
d_{(I; \za') (J; \zb')}^{(K; \zg, \zg') }\
{\Psi_x^{(1)}\over \Psi_1^{(1)}}\cr
&\quad \bra 0| \phi_{j^* j}^1(z_1)\, \phi_{i^* k}^j(z_2)\, \phi_{i j}^k(z_3)\, 
\phi_{i 1}^i(z_4)|0\ket \times ({\rm right\ chiral\ block})\,,
\cr}}
taking into account that $d_{(J; \zb')(1) }^{(J; \zb')}=1$.
The limit $z_{21}, z_{34} \to \infty$  of
this correlator is alternatively represented by the
identity contribution in the correlator
\eqn\vtwo{\eqalign{
\bra 0&| \Phi_{(I^*; \za^*)}(z_2, \bz_2)\
 X_x\, \Phi_{(I;\za')}(z_3, \bz_3)\,
\Phi_{(J;\zb')}(z_4, \bz_4)\, X_x^\dagger\, 
\Phi_{(J^*; \zb^*)}(z_1, \bz_1)|0\ket  \cr
&= \sum_{p, \bp, \zd,\zd'}\,
d_{(I^*; \za^*) (I; \za)}^{(1) }\
{\Psi_x^{(i,\bi;\za,\za)}\over \Psi_1^{(i,\bi)}}\
d_{(I; \za')(P; \zd, \zd') }^{(I; \za)}\,
d_{(J; \zb') (J^*; \zb^*)}^{(P; \zd,\zd') }\
{\Psi_{x}^{(j,\bj;\zb,\zb)}\over \Psi_1^{(j,\bj)}}\cr
&\quad \bra 0| \phi_{i^* i}^1(z_2)\, \phi_{i p}^i(z_3)\, \phi_{jj^*}^p(z_4)\, 
\phi_{j^* 1}^{j^*}(z_1)|0\ket \times ({\rm right\ chiral\ block})\,,
}}
%
i.e., by the first term $p=1=\bar{p}$.
Next we use the braiding relations for the chiral blocks to
identify the two products of chiral correlators , i.e., move $j^*$
and $\bj^*$ to the very  right-- 
this brings about the product of  fusing matrices $F_{kp}\left[
{j^*\atop i}{j\atop i}\right]\,    
F_{\bk \bp}\left[{\bj^*\atop \bi}{\bj\atop\bi}\right]$ 
taken at $p=1=\bar{p}$.
This implies $\za=\za'$ and $\zb=\zb'$ and also 
 trivialises the fusion matrices to the ones in the diagonal
counterpart of \Ibb, i.e., we get ratios of square roots of q-dimensions,
which precisely match the factors $\Psi_1$ 
\psiinT\ coming from the twists.
Equating the coefficients 
and taking also into account the symmetries of the OPE
coefficients (this produces the same sign
$(-1)^{s_I+s_J}$ in both sides), see \PZb, we finally obtain 
\eqn\tcor{
\sum_{k,\bk, \zg, \zg'}\,
|d_{(I; \za) (J; \zb)}^{(K; \zg,\zg') }|^2\
{\Psi_x^{(K;\zg,\zg')}\over \Psi_x^{(1)}}
={\Psi_{x}^{(I;\za,\za)}\over \Psi_x^{(1)}}
\, {\Psi_x^{(J;\zb, \zb)}\over \Psi_x^{(1)}}\,,
}
from which \OPA\ follows.
In deriving \tcor\ we have assumed that the decomposition of the physical
 fields  involves several copies of each product
of  left and right chiral blocks,
i.e., $\Phi_{I;\za}(z,\bar z)= \sum_{j,\bj,k,\bk,\zb, \zb',\zg,\zg'}
\ d_{(I; \za) (J; \zb,\zb')}^{(K; \zg,\zg') }\
\Big(\phi_{ij}^k(z)\otimes
\phi_{\bi \bj}^{\bk}(\bar z)\Big)_{(\za,\za) (\zb,\zb')}^{(\zg,\zg')}\,.$
 These copies are  labelled by the  pairs $(\zb,\zb')\,,\,(\zg,\zg')$ and 
 they correspond to the multiplicity of  states in the
 projectors $P_x^{(k,\bk; \zg,\zg')}$ in \tpro; 
in the only nontrivial $sl(2)$ case, the $D_{\rm even}$ series, 
$\tM_{(I; \za,\za) (J; \zb,\zb)}^{(K; \zg,\zg') }$, and hence
  $d_{(I; \za) (J; \zb)}^{(K; \zg,\zg') }$,
  are identically zero for  $\zg \not = \zg'$.
In the  previous discussion
we have suppressed for simplicity
the multiplicity indices $ t=1,2,\dots, N_{ij}{}^k\,$ and 
$ \bt=1,2,\dots, N_{\bi\bj}{}^{\bk}$ appearing in the
 higher rank cases; when restored the modulus square
in the l.h.s. of \tcor\ and \OPA\ is replaced by $\sum_{t,\bar{t}}\,
|d_{(I; \za) (J; \zb)}^{(K; \zg,\zg;\, t, \bar{t}) }|^2\,.$
Note that in the presence of a twist operator the identity
$1$-point function appears normalised as
$\bra 0|\Phi_{(1)}\,X_x |0 \ket = {\Psi_{x}^{(1)}\over \Psi_1^{(1)}}
=\tilde{d}_x$. 

An  intriguing issue
is the fact that from a mathematical point of view,
 the indices $x$ play a r\^ole dual to that of
representation labels $i\in\CI$, (dual in the algebraic sense, going from a
linear space to the space of its linear functionals, see Appendix A and also
equation \eqdim), while from a physical point of view,
they play a r\^ole dual to that of the 
labels of bulk fields:   this is apparent in equation \Ibl\ where
there is a (Fourier-like) duality between the set $\tCV$ of $x$ and
the set $\tExp$ of pairs $(\J)$
counted with a multiplicity $(Z_{j\bj})^2$, that is between the vertices
and the ``exponents'' of the graph $\tG$.

We conclude this subsection with the remark that some correlators 
including twist operators may be interpreted as generalised 
order-disorder field correlators, compare
 with \VBPb, where such functions matching the operator
content of the $\IZ_2$-twisted torus partition functions of \refs{\Ca, \Z} 
were constructed 
and their OPE coefficients computed. We recall \PZtw\
that the partition functions of \refs{\Ca, \Z} provide the
simplest examples of solutions of \tosy.

%

\subsec{Constructing the $\tG$ graphs}
\noindent
Let us see now how the matrices $\tV_1{}^x$ and  $\Psi$, 
from which the graph $\tG$ and 
all the other matrices $\tV,\ \tN,\ \tn$ and $\tM$ may be constructed, 
can be determined in a given CFT, i.e. starting from a given 
modular invariant $Z$ and the associated graph $G$.  
(see \Coq\ for a detailed discussion of the particular $E_6$ case 
of $\slh(2)$ following a different approach.)

First, in the case of a diagonal theory, $Z_{ij}=\delta_{ij}$, 
it is natural to identify the set $\tCV$ with the set $\CI$ of 
representations, since their cardinality agrees, and to take
\eqn\diag{\tV_{ij}=N_i\ N_j}
understood as a matrix product, i.e.
$\tV_{ij;\, x}{}^y=\sum_{k\in \CI} N_{ix}{}^k N_{jk}{}^y $, 
in particular $\tV_{ij;\, 1}{}^k=N_{ij}{}^k$.
The corresponding $\Psi_x^{(j,j)}$ are just the modular matrix elements
$S_{xj}$ and the Ocneanu graph $\tG=\tA$, which is generated by 
the ``fundamental'' $\tV_{f1}$ and $\tV_{1f}$, 
both equal to $N_f$,  is identical to the ordinary graph $G=A$.

 As a second case, consider a non-diagonal theory
with a matrix $Z_{ij} = \delta_{i\zeta(j)}$, where $\zeta$ is
the conjugation of representations  or some other
automorphism of the fusion rules
(like the $Z_2$ automorphism in the $D_{2\ell+1}$ cases of $\slh(2)$
theories).  Then $\tCV=\CI$, and 
$ \tV_{ij} =\tV^{({\rm diag})}_{i\zeta(j)} =N_iN_{\zeta(j)}$.
The graph is generated by $\tV_{f1}=N_f$ and $\tV_{1f}=N_{\zeta(f)}$, 
each one giving a subgraph isomorphic to $A$.  (see Fig.  10 \ for
the case of $D_{{\rm odd}}$).
The $\Psi_x^{(\J)}=S_{xj}\delta_{j\zeta(\bj)}$, and one finds that 
the $\tN$ matrices reduce to those of the diagonal
case, i.e.  $\tN_{xy}{}^z=N_{xy}{}^z$, $x,y,z\in \CI$, 
while $\tn_{ax}{}^b=n_{xa}{}^b$ and $\tM_{(i,\zeta(i))\,(j,\zeta(j))}{}^
{\!\!(k,\zeta(k))}=N_{ij}{}^k$.

General expressions may be obtained for type I theories \typeI.
The algebra \hNalg\ enables one to define a partition 
of the set $\CV$ into equivalence classes 
$T_\kappa$:\  $a\sim a'$ if $\exists\, b\in T: \hN_{ab}{}^{a'}\ne 0$
\refs{ \BI,\DIFZ}. 
The number of such classes equals the number of representations of $\gA$
coupled to the identity in the modular invariant, i.e. of $i\in \CI$
such that $Z_{1i}\ne 0$ 
\foot{This may be established in cases where the 
$\hN$ algebra is commutative, and where the structure constants
of both $\hN$ and  
its dual $M$ are non negative, following the work of \BI. It seems to
extend to non-commutative cases as well, as we checked on the aforementioned
case of $\slh(4)_4$, where some entries of $M$ are negative or
even imaginary.}.
Since \typeI\ applies to the matrix $\M^1=Z$, 
it suggests to look for similar expressions for the other $\M^x$. 
We find that 
in all known type I cases, in particular for $\slh(2)$ theories, 
the labels $x$ may be taken of the form
$(a,b,\kappa)$, $a,b\in \CV$, $\kappa$ a class label,  and
\eqn\Vtilde{\M^x_{ij}=\tV_{ij^*;\, 1}{}^{(a,b, \kappa)}=
P^{(\zk)}_{ab}:= {\sum_{c\in T_\kappa}} n_{ic}{}^a n_{jc}{}^b}
with $c$ running over a certain subset $T_\kappa$ of vertices,
or equivalently
\eqn\Vtilde{\tV_{ij;\,1}{}^{(a,b, \kappa)}=
{\sum_{c\in T_\kappa}} n_{ic}{}^a n_{jb}{}^c\ .}
One checks that indeed $\M^1=Z$, the modular invariant matrix 
as given in \typeI.
As the matrices $n_i$ form a representation of the fusion algebra,
$n_i n_j =N_{ij}{}^k n_k$, one finds that upon left multiplication by any
$N_f$, 
\eqn\leftN{
N_f.P^{(\zk)}_{ab}=\sum_{a'}\,  n_{fa'}{}^{a}\  P^{(\zk)}_{a'b} }
and likewise, upon right multiplication $P^{(\zk)}_{ab}.N_{f^*}=
\sum_d P^{(\zk)}_{ad}  n_{f d}{}^b$ 
by repeated use of $n_f^T=n_{f^*}$. 
\medskip

In theories like $\Bbb{Z}_n$ orbifolds of $\slh(n)$ theories, there is a 
partition of the set of vertices into classes $T_\alpha$ such that 
\eqn\cond
{\forall a,a'\in T_\alpha\ ,\ \forall b\in T_\beta\ne T, 
\qquad {\hN}_{ba}{}^{a'}=0 \ , }
because the $\hN$ algebra respects the $\Bbb{Z}_n$ grading of the vertices. 
Then let us prove that \tosy\ follows from the Ansatz \Vtilde\
with $x=(a,1,\zk)$
\eqn\valya
{\eqalign{
\sum_a \sum_\zk (P^{(\zk}_{a1})_{ij}(P^{(\zk)}_{a1})_{i'^*j'^*}
&= \sum_{c\sim c'}\sum_{i''\in \CI}\,N_{ii'}{}^{i''} \sum_d
n_{i''1}{}^d\, \hN_{dc'}{}^c\,n_{j1}{}^c\, n_{j'1}{}^{c'^*}\cr
&= \sum_{c,c'}\sum_{i''\in \CI} N_{ii'}{}^{i''} 
\sum_{d\in T} n_{i''1}{}^d\, \hN_{dc'}{}^c\,n_{j1}{}^c\, n_{j'1}{}^{c'^*}\cr
&= \sum_{i'',j''} N_{ii'}{}^{i''}\,N_{jj'}{}^{j''} Z_{i''j''}\,
}}
where we have repeatedly used  
\hatN\ and \nim\ and on the second line, we have used \cond\ to restrict the 
summation over $d$ to the set $T$; the constraint $c\sim c'$ is 
then automatically enforced, which enables us 
to sum over independent $c$ and $c'$.


\def\gh{{\goth h}}
For the case of a conformal embedding $\hat\gh_k\subset \hat\gg_1$, 
we checked in all $\slh(2)$ cases and conjecture in general
 that the label $\kappa$ may be dropped, and $x$ represented by a 
pair of vertices $(a,b)$,
$a\in\CV$, and $b$ running over a subset of vertices. 
Then  we  make use of  formula  \psiinT\ 
to express the eigenvectors $\psi_c^{j,d;\,\za,\zb}\,,$  $c\in T$  
in terms of the modular $S^{{\rm ext}}$ matrix of the 
extended algebra (i.e. of the $\hat\gg_1$ current algebra).
In that way we find, multiplying \Vtilde\ with $S_{i^*j}S_{\bi^*\bj}$, 
\eqn\eigenvem{
\sum_{\zg }\ \Psi_x^{(J;\zg,\zg)\, }=\sum_{d,\za,\bar{\za}} 
{\psi_a^{(j,d;\,\za,\za)\, }\,
\psi_b^{(\bj,d;\,\bar{\za},\bar{\za})\,*}\over S_{1 d}^{{\rm ext}}}\,,
\qquad x=(a,b)
}
where the sum in the l.h.s. runs according to
$\zg=1,\cdots, Z_{j\bj}=\sum_{d\in T}\, n_{j1}{}^d\, n_{\bj1}{}^d\,,$
and that in the r.h.s.  runs on $\za=1,\cdots, n_{j1}{}^d$,
$\bar\za=1,\cdots, n_{\bj1}{}^d$ .
If there is only one $d\in T$ in the sum we can identify 
$\zg =(\za,\bar{\za})\,,$
if there are more, first $\zg$ has to be split into a multiple index
and then each identified
with a  pair  $(\za,\bar{\za})$ depending on $d$.
For $a\in T$  $\Psi_a^{(J;\zg,\zg')}\, =\delta_{\zg\zg'}
\, \Psi_1^{(J)}\
 {S_{a d}^{{\rm ext}}\over S_{1 d}^{{\rm ext}}}$ is consistent with
\eigenvem\ and implies that
$\tN_{ab}{}^c =^{\rm ext}\!\!N_{ab}{}^c$ for $a,b,c \in T\,,$ using
$\sum_{j,\bj\in d } \, Z_{j\bj}\, S_{1j}\,S_{1\bj}=
(S_{1 d}^{{\rm ext}})^2$,
and hence, that $\tN_a\,, \, a\in T$ form a subalgebra isomorphic
to the extended fusion algebra.

\medskip

\noindent
In particular in the cases with commutative $\tN$ algebra one computes
\eqn\tilint{\eqalign{
\tV_{ij;\, (a_1,b_1)}{}^{(a_2,b_2)}&=
\sum_{c\in T}\, (n_i\, \hat{N}_{a_1})_c{}^{a_2}\,
(n_{j^*}\, \hat{N}_{b_1})_{c}{}^{b_2}\,,\cr
\tilde{N}_{(a_1,b_1)\,(a_2,b_2)}{}^{\!\!(a_3,b_3)}
&=
\sum_{c\in T}\,
(\hat{N}_{a_1}\, \hat{N}_{a_2})_{c}{}^{a_3}\,
 (\hat{N}_{b_1}\, \hat{N}_{b_2})_c{}^{b_3}\,\cr
\tilde{n}_{(a,b)}&= \hat{N}_a\, \hat{N}_b\,, 
}}
which ensures that these matrices are integral, nonnegative valued.
In particular
$$
\tV_{i1;\, (a_1 1)}{}^{(a_2,1)}=
n_{i a_1}{}{}^{a_2}\,, \quad \tV_{1j;\, (1,b_1)}{}^{(1,b_2)}=
n_{j b_1}{}^{b_2} \ .
$$

\bigskip

        The description of $\M^x$ as a bilinear form in the 
	$n$ matrices does not seem to restrict to type I theories
 	like in \Vtilde. Indeed, this is what happens 
 	in the $D_{{\rm odd}}$ and $E_7$ cases of $\slh(2)$ theories. 
Knowledge of the $\M^x$ matrices
(and in general of $\tV_{ij;(1,1,\kappa)}^x$ for any $\kappa$)
 determines the whole structure.  
It is easy to invert the (block-)diagonalisation formula of the  
 $\M^x$ and to get,  using also  \Psiid,
\eqn\getPsi{ \sum_{\zg}\, \Psi_x^{(\J;\zg,\zg)\, *}= 
\sqrt{S_{1j}S_{1\bj}}\,
\sum_{i,\bi\in\CI} \M^x_{i\bi^*} S_{i^*j}S_{\bi^*\bj} \ ,}
This determines $\Psi_x$ completely for cases with $Z_{j\bj}=1$,
while higher multiplicities $Z_{j\bj}>1$ require a little 
more work and care, see Appendix B for an illustration on the 
$D_{2\ell}$ case of $\slh(2)$.
Once  $\Psi$  is known, it is a simple matter to obtain 
all $\tn,\tN,\tV$ and $\tM$ matrices.  

\newsec{Conclusions and perspectives}
\noindent
The reader who has followed us that far
should by now be convinced of the relevance and utility 
of Ocneanu's DTA $\CA$ in the detailed study of rational CFT. 
In our view, two new concepts developed in this paper in connection 
with this quantum algebra have proved particularly useful:
\item{$-$} the generalised CVO, which are covariant under the 
action of $\CA$, unify the treatment of bulk and boundary fields
and permit a more direct discussion of their operator algebras;
\item{$-$} the twist operators, whose r\^ole manifests itself in 
several ways, give a physical interpretation to the abstract 
labels $x$ of the dual algebra $\hat\CA$ and to the coefficients
$\tN_{xy}{}^z$ and also, through their interplay with bulk fields, 
provide a new way to determine the general OPA structure constants 
in the bulk.

Several points deserve further investigation. First, as 
already pointed out in the Introduction, many of our statements which
rely on the explicit examination of particular cases, mainly based
on $sl(2)$ and $sl(3)$, and which are presented as conjectures in general, 
should be extended in a systematic way to all RCFT. The case of orbifold
theories, in which the relevant graphs would be {\it affine} Dynkin
diagrams and their generalisations, should be quite instructive.
Other directions of generalisations include irrational CFTs
(generic $c\ge1$ CFTs or $N=2$ superconformal CFTs)
 or non compact theories like Liouville \TP.

Secondly, among the five types of 3-chains ${{}^*F}$ attached to the 
tetrahedra of Fig. \simplex, only two, namely $F$ and $\Fo$ have
received a physical interpretation,
as they  underlie both the CFT and the related integrable critical
lattice models. Understanding the meaning of the 
others, which all involve one or several twist labels $x$, presumably 
requires a deeper discussion  of the interplay of twist operators
with bulk and/or boundary fields. 

\def\hchi{\hat\chi}
In fact the general properties of twists and  
their relations with ``twisted'' representations of the underlying chiral
algebra $\gA$ await a good discussion. We regard as quite significant
that {\it all} partition functions either on a torus or on a cylinder
with or without defect lines (twists) are expressible 
as linear or bilinear forms with non-negative integer coefficients of
the $|\CV|$ linear combinations of characters 
\eqn\hachi{
\hchi_a:=\sum n_{i1}{}^a \chi_i= Z_{a|1}\,.
}
This follows from eqs \typeI,\typeII\
and from our Ansatz \Vtilde\ that in type I
the matrices $\M^x$ are bilinear in the $n$'s. (In type II theories, 
we recall that 
the $n$'s that appear here are those of the parent type I theory).
The $\hchi_a$ thus appear as the building blocks of all 
partition functions. Their natural interpretation, as alluded above, is that
they are the characters of a class of more general 
representations of the extended 
algebra $\gAe$. Among them, the subset $a\in T$ represents the ordinary, 
untwisted, representations. The other have been called twisted \BPPZ, 
or solitonic \FSoo.  
The induction/restriction method \refs{\Xu,\BEb} of constructing 
 these ``twisted sectors'' essentially amounts to the recursive solution
 of the system of equations \nim,  \hatN, \hNalg.
On the other hand, the direct definition of (some) of 
these twisted representations, closer in spirit to
the concept of twist as developed in section 7, has
been achieved only in a limited number of cases, see e.g. \Hon.
%
\bigskip

\noindent {\bf Acknowledgements}

\noindent 
We want to thank Gabriella B\"ohm, Robert Coquereaux, Patrick
Dorey, David Evans, C\'esar G\'omez, Liudmil Hadjiivanov, 
Paul Pearce, Andreas Recknagel, Christoph Schweigert, 
Korn\'el Szlach\'anyi, Raymond Stora, Ivan Todorov, Peter
Vecserny\'es, Antony  Wassermann, Gerard Watts for useful discussions,  
and especially, Adrian Ocneanu, for the inspiration for this work.
We thank Ivan Todorov for inviting us to the programme 
``Number Theory and Physics'' at ESI, Vienna.
Hospitality of CERN where this long paper was completed is also
gratefully acknowledged. 
V.B.P. acknowledges the  support and hospitality of ICTP and INFN, Trieste 
and  partial support of the Bulgarian National Research Foundation 
(Contract $\Phi-643$).



\appendix{A}{Ocneanu DTA -- dual structure}

\noindent 
This appendix contains some more details on the
Ocneanu DTA  \refs{\AO, \BEK} and its WHA interpretation \BSz.

The coproduct \cop\   does not preserve the
identity, i.e., 
\eqn\forgot{
\triangle(1_v)= 
 \sum_b\,\sum_{i,j,a,c,\za,\zg}\,
e_{i\,,\za,\za}^{(cb)(cb)}\otimes
e_{j\,,\zg,\zg}^{(ba)(ba)}\, \Big(=: 1_{v\,(1)}\otimes 1_{v\,(2)}\,\Big)
\not=
 1_v\otimes 1_v
}
while $\triangle(1) =1\otimes 1$ is one of  the axioms of a Hopf algebra.

For $u\,,w\in \CA$ and $uw$ -- the matrix (vertical) product, one
has
\eqn\oubli{
\ze(u\,w) =
\sum_{i,j,a,b,c,\za,\zg}\,
\ze( u\, e_{i\,,\za,\za}^{(cb)(cb)})\
\ze(e_{j\,,\zg,\zg}^{(ba)(ba)}\,w)\, \Big(= :\ze(u\, 1_{v\,(1)})\
\ze(1_{v\,(2)}\,w)\Big)\,,
}
e.g. for $u=\sum_{a,b} C_{a,b} e_1^{aa,bb}$,
$w=\sum_{a,b} C_{a,b}^{'} e_1^{aa,bb}$ --
one gets $\ze(u\,w)=$ tr$(CC^{'})$,  and
$\ze(u)\, \ze(w)=\sum_{a,b}C_{ab}\sum_{a',b'}C_{a'b'}^{'}\not =\ze(u\,w)
$ in general,  while  the counit of a Hopf algebra is an algebra homomorphism.

The antipode is a linear anti-homomorphism $S(uw)=S(w)\,S(u)$,
 defined according to \antipode, and 
so that $S^{-1}(u)=(S(u^*))^*$. It is also an anti-cohomomorphism
i.e., inverts the coproduct, in the sense that 
\eqn\ant{
\triangle \circ
S= (S\otimes S)\circ \triangle^{op}\,,
}
Here $\triangle^{op}(u)=u_{(2)} \otimes u_{(1)}$ for 
$\triangle(u)=u_{(1)} \otimes u_{(2)}$. Furthermore
instead of the Hopf algebra postulate
 $S(u_{(1)})\,u_{(2)} =1_v \epsilon(u)$ the antipode of a WHA satisfies
\eqn\antb{
S(u_{(1)})\,u_{(2)} \otimes u_{(3)}= (1_v\otimes u)\triangle(1_v)\
\big(= 1_{v\,(1)}\otimes u\, 1_{v\,(2)} \big)\,,
}
The relations \ant, \antb\ are
 checked using both unitarity relations \unit, as well as \sym,\symb;
the choice of the coefficient in \antipode\ is essential.

One turns  $\cal A$ into a {\it quasitriangular WHA }
by defining an
 $\CR$ - matrix, i.e.  an element $\CR\in \triangle^{op}(1_v)\ \CA\otimes
\CA\ \triangle(1_v)$, which intertwines the two coproducts,
\eqn\commut{
\triangle^{op}(u)\,\CR= \CR\, \triangle(u)\,,
  }
subject to the constraints,
\eqn\fubr{
(\triangle\otimes Id )\, \CR 
 = \CR_{13}\, \CR_{23}\,, \qquad
(Id \otimes  \triangle )\, \CR 
= \CR_{13}\, \CR_{12}\,. \ 
}
Namely 
\eqn\rmat{
\CR= \ \sum_{i,j, p \atop{a,a',b,c,c',d
\atop{\za,\za',\zg,\zg', \zb, \zb', t}}}\, 
\Fo_{b p}\left[\matrix{i&j\cr c'&a'}\right]_{\za'\,
\zg'}^{\zb'\, t}\ 
 w_{a,a';c,c';\zb,\zb'}^{i,j,p}\ 
\Fo_{d p}^*\left[\matrix{j&i\cr c&a}\right]_{\za\,
\zg}^{\zb\, t}\ 
 e_{j\,,\zg'\,,\za}^{(ba')\,(cd)}\otimes
 e_{i\,,\za'\,,\zg}^{(c'b)\,(da)}
}
Here $ w_{a,a';c,c';\zb,\zb'}^{i,j,p}$ is a unitary matrix
and to make contact with the CFT under consideration we choose 
\eqn\phase{
 w_{a,a';c,c';\zb,\zb'}^{i,j,p}\ =\delta_{\zb \zb'}\,
 \delta_{a a'}\, \delta_{c c'}\ e^{-i \pi \triangle_{ij}^p}\,,
}
so that  the coefficient
 in \rmat\ reproduces the braiding matrix $\hat{B}(+)$ in \Ie.
 Similarly one defines  $ \CR^* \in \triangle(1_v)\ \CA\otimes
\CA\ \triangle^{op}(1_v)$,  corresponding to $\hat{B}(-)$;
the  inversion relation \Idc\  is equivalent to
$ \CR^*\, \CR =\triangle(1_v)\, $, $ \CR\, \CR^* =\triangle^{op}(1_v)$.
The relations \fubr\ are equivalent to the fusing-braiding 
relation \Il\ and its counterpart discussed in section 4.
Denoting by $P$ the permutation operator in $V^i\otimes V^j$,
 the definition \rmat\  with the choice \phase\ implies 
\eqn\rmatr{
P\,\CR\ {1\over \sqrt{P_b}}\, e_{cb}^{i,\za'}\otimes_h
e_{ba}^{j,\zg'}=\sum_{d, \atop{\za,\zg}}\,
\hat{B}_{bd}^*\left[\matrix{i&j\cr c&a} \right]_{
\za'\,  \zg'}^{\za\, \zg}\!\!\!\!\!(-)\ {1\over \sqrt{P_d}}\,
e_{cd}^{j,\za}\otimes_h e_{da}^{i,\zg}\,.
}
\medskip

The horizontal product in $\CA$ depicted on Fig. \dtahp\ reads
more explicitly
\eqn\Io{
e_{i\,, \za\,,\za'}^{(cb)(c'b')}\otimes_h 
e_{j\,,\zg\,,\zg'}^{(da)(d'a')}=\delta_{bd}\, \delta_{b'd'}
 \sum_{p \atop{\zb, \zb',t}}\,
g_{ij}^{p;b,b'}\,
\Fo_{b p}\left[\matrix{i&j\cr c&a}\right]_{\za\,
\zg}^{\zb\, t}\
  \Fo_{b' p}^*\left[\matrix{i&j\cr c'&a'}\right]_{\za'\,
\zg'}^{\zb'\, t}\ 
 e_{p\,, \zb\,, \zb'}^{(ca)(c'a')} \,.
}

  The dual algebra $\hA$ of $\CA$ is the space of
linear functionals on $\CA$. It is a matrix algebra
$\hA=\oplus_{x\in \tCV}\, Mat_{\tm_x}$  with matrix unit 
basis $\{E_x^{(a'a;\eta)(d'd;\zeta)}\,,\, x\in \tCV\}$,
 depicted  by double triangles, or blocks,  with an intermediate index $x$
 see Figs.  \dta, 9.  The indices $(a,a';\eta)\,,$ $ a,a'\in \CV\,,\,
\eta=1,2,\dots \tilde{n}_{a x}{}^{a'}$ label the states in a linear vector 
space $\hat{V}_x$  of dimension dim$(\hat{V}_x)=\sum_{a,a'}\,
\tilde{n}_{a x}{}^{a'}=\tilde{m}_x$. They are
depicted in Fig. \simplex\ as  triangles with two white and 1 black vertices. 
The vertical and horizontal products are exchanged in the dual algebra,
 i.e., the horizontal product is the matrix product in $\hA$
 and the vertical product for the basis elements $E_x$ is given
 by a  dual analogue of \Io, with the convention that the second element
appears above the first. In this product  the  r\^ole 
of the multiplicities $N_j$ and $n_j$ is taken over by $\tN_x$ and $\tn_x$,
with the relation \tnrep\ serving now as a consistency relation
 replacing \nim. The dual  $3j$- and $6j$-symbols  $\tFo$
 and $\tF$, the last two of the tetrahedra in Fig. \simplex,
satisfy unitarity relations analogous to \unit\ and
two more pentagon relations parallel to \pent\ and \mp\ respectively.
 The matrix  $\tF$ dual to the   fusing matrix   $F$ has  all
indices of type $x$, while  $\tFo_{bz}\left[{y\atop c}{x\atop a}\right]$ 
 is a matrix with 3+3 indices of type $a,b,c\in \CV$ and $x,y,z\in \tCV$. 
All the steps of section 2 can be repeated, in particular we can
choose a gauge fixing for $\tFo$ analogous to \Ibb, 
using that $\td_x\, P_a = \sum_b\, \tn_{a x}{}^b\, P_b$.

\fig{Relating the two bases}{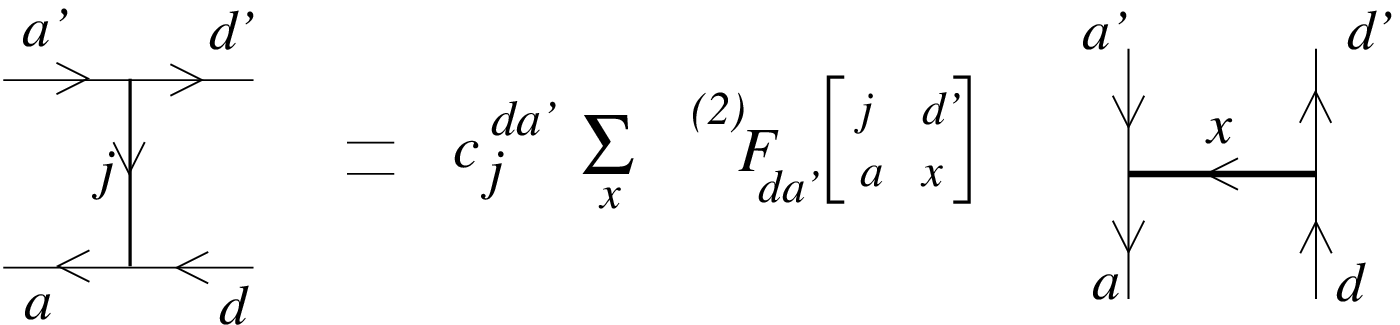}{8cm}\figlabel\dualdta

The finite dimensional algebras $\CA$ and $\hA$ can be identified, looking at  
$\{e_j^{\za,\za'}\}$ and $\{E_x^{\eta,\zeta}\}$ as 
providing different bases,  see Fig. \dualdta.
This  introduces a new ``fusing'' matrix $\Fot$,  given,
up to a  constant, by the numerical value of the linear functional
$E^{\eta\,,\zeta}_x(e_j^{\za\,, \za'})\in \IC$.
$\Fot$ is the third tetrahedron on Fig. \simplex,  supported by
two black and two white vertices, and two types of triangles  
of multiplicities  $n_j$ and $\tn_x$. More explicitly we have
\eqn\Ir{
E^{\eta\, \zeta}_x( e_{j}^{\za\, \za'})=
E_{x;\eta\,,\zeta}^{(a'a)(d'd)} \Big(e^{(cb) (c'b')}_{j;\za\,\za'} \Big) =
\delta_{ac}\ \delta_{bd}\delta_{a'c'}\ \delta_{b'd'}\
c_j^{da'}\, \tilde{c}_x^{da'}\
\Fot_{b\, c'}\left[\matrix{j&b'\cr
c&x}\right]_{\za\, \zeta}^{\eta\, \za'}\,
}
with
\eqn\norma{
 c_j^{bc'}\ \tilde{c}_x^{bc'} ={\td_x\, d_j\over P_b\, P_{c'}} \,
 ({S_{11} \over \psi_1^1 })^2\,.
}
The equality \onemore\ ensures that the number of elements
on both sides of Fig. \dualdta\ for fixed $a,a',d,d'$ and varying $j$ and $x$
is the same, so 
the linear transformation $\Fot$ is invertible, the inverse denoted  
$\tFot^*$, in the sense of the relations
\eqn\inverse{\eqalign{
\sum_{x, \eta,\zeta}\, c_j^{bc'}\ \tilde{c}_x^{bc'}\
\Fot_{b\, c'}\left[\matrix{j&b'\cr
c&x}\right]_{\za\, \zeta}^{\eta\, \zb}\
\tFot_{b\, c'}^*\left[\matrix{j'&b'\cr
c&x}\right]_{\za'\, \zeta}^{\eta\, \zb'}
&=\delta_{jj'}\, \delta_{\za\, \za'}\, \delta_{\zb \zb'}\cr
\sum_{j, \za,\zb}\,  c_j^{bc'}\ \tilde{c}_x^{bc'}\
\Fot_{b\, c'}\left[\matrix{j&b'\cr
c&x'}\right]_{\za\, \zeta}^{\eta\, \zb}\
\tFot_{b\, c'}^*\left[\matrix{j&b'\cr
c&x}\right]_{\za\, \zeta'}^{\eta'\, \zb}
&=\delta_{xx'}\, \delta_{\zeta\, \zeta'}\, \delta_{\eta\, \eta'}\,.
}}
We shall require that $\Fot$ and  $\tFot$
are trivial for $x=1$ and $j=1\,,$
analogously to \trivone. This is consistent with the inverse relations
\inverse, inserting \norma\ and using that 
\eqn\dx{
\sum_x\, \tn_{a x}{}^b\
\td_x =({\psi_1^1\over S_{11} })^2\,
P_a\, P_{b}=\sum_j\, n_{j a}{}^b\ d_j\,.
}
We recall that the
ratio of constants $ c_j^{bc'}$ 
appears in the normalisation of the horizontal product \Io,
and similarly  a ratio of the constants $\tilde{c}_x^{bc'}$ 
determines the constant
in the vertical product of the dual basis elements.
Inserting  the relation in Fig. \dualdta\ in both sides of the horizontal
product \Io\ and using furthermore that the horizontal
product acts trivially on the dual basis by a formula analogous to
\mcp,   one  gets the pentagon relation \AO\
%
%
\eqn\Is
{\eqalign{\sum_{b', \za', \zg', \zeta}  
&\Fo_{b' p}\left[\matrix{i&j\cr a'&c'}\right]_{\zg'\,
\za'}^{\zb'\, t}\
 \Fot_{ba'}\left[\matrix{i&b'\cr
a&x}\right]_{\zg\,\zeta}^{\eta\, \zg'}\
\Fot_{cb'}\left[\matrix{j&c'\cr
b&x}\right]_{\za\,\eta'}^{\zeta\, \za'}\  \cr
&\qquad=
\sum_{\zb}  
\Fo_{b p}\left[\matrix{i&j\cr
a&c}\right]_{\zg\, 
\za}^{\zb\, t}\
\Fot_{ca'}\left[\matrix{p&c'\cr
a&x}\right]_{\zb\,\eta'}^{\eta\, \zb'}\ .
}}
In terms of the functional values \Ir\ the identity  \Is\ reads
\AO\
\eqn\Iss{
\sum_{\zeta}\, 
E^{\eta\, \zeta}_x( e_{i}^{\zg\, \za})\ E^{ \zeta\, \eta'}_{x'}(
e_{j}^{\zg'\, \za'}) =\delta_{xx'}\ E^{\eta\, \eta'}_x( e_{i}^{\zg\,
\za}\otimes_h\, e_{j}^{\zg' \,\za'}) \,.
}
Similarly starting from the vertical product  analogue
of \Io\  for the dual basis we obtain the dual analogue
of \Is, with $\Fo$ and $\Fot$ replaced by $\tFo$ and $\tFot$.
The relation in Fig. \dualdta\ allows to define a sesquilinear
form in the algebra determined by the pairing  \Ir\ on $\CA \otimes
\hA$, s.t.  $\bra E_x\,,\,E_{x'} \ket=\delta_{x x'}\,\tilde{c}_x$. 
Assuming furthermore that  $\bra e_j\,,\,e_{j'} \ket=\delta_{j j'}\,c_j$  
leads to the identification $\Fot=\tFot$. 
Then the above two dual pentagon identities are equivalent
to the identities relating, via the pairing, 
 the coproduct in each of the two algebras to the product in
its dual \BSz,
\eqn\codp{\eqalign{
\bra E_x  E_y \,,\,  e_p\ket & =
\bra E_x \otimes E_y\,,\, \triangle(e_p) \ket \cr
\bra E_z\,,\, e_i  e_j  \ket & =
\bra \triangle(E_z)\,,\, e_i \otimes e_j  \ket\,,
}}
where the products in the l.h.s. stand for the
algebra multiplications in $\hA$ and $\CA$ (i.e., the horizontal and
vertical products respectively).
\foot{
The above identification $\Fot=\tFot$ appears in \BSz\ (up to
different notation) as a solution in the diagonal cases, where
the r.h.s. of \norma\ simplifies to a ratio of q--dimensions.  In
general the matrix defining the pairing on $\CA \otimes
\hA$ and its inverse are left unrelated  and the equalities
\codp\ lead to dual (with respect to the $3j$-symbols) pentagons
in both of which  only the inverse matrix enters.  We are
indebted to Gabriella B\"ohm and Korn\'el Szlach\'anyi for a
clarifying e-mail correspondence on this point.}
The coproduct and the horizontal product are related via the
scalar product in $\CA$ defined above 
\eqn\cohor{
\bra e_i\otimes_h\, e_j\,,\, e_p  \ket =
\bra e_i\otimes e_j\,,\, \triangle(e_p)  \ket \,.
}
 We shall furthermore assume the analogues of the  symmetry
relations \sym, \symb{} 
(compatible with the form of $\Fot$ and  with the relation
$\tn_{ax}{}^b=\tn_{b x^*}{}^a$) 
\eqn\symd{
\Fot_{b\, c'}\left[\matrix{j&b'\cr
c&x}\right]= \sqrt{P_b\, P_{c'}\over P_{b'}\, P_{c} }\ 
\Fot^*_{c\,b'}\left[\matrix{j^*&c'\cr
b&x}\right]= \Fot_{c'\, b}\left[\matrix{j^*&c\cr
b'&x^*}\right]\,,
}
Inserting the first equality of \symd\ in the relation obtained from \Is\
for $p=1$, one obtains using \Ibb\
\eqn\unitb{
\sum_{b', \zeta, \zg'}\ \Fot^*_{a\, b'}\left[\matrix{j&a'\cr
b&x}\right]_{\zg\, \eta}^{\zeta \,\zg'} \ \Fot_{c\, b'}\left[\matrix{j&a'\cr
b&x}\right]_{\za\, \zeta'}^{\zeta \,\eta'}=
\delta_{ac}\, \delta_{\za \zg}\, \delta_{\eta \eta'}\,\,.
}
Using \symd\ one also checks that the conjugation operation
${}^{+}$ computed directly from
\Ir\  coincides with $ E^+(e)=\overline{E(S^{-1}(e)^+)}\,.$
To make contact with the basis for the dual triangles
 exploited in  \BSz\ one has to
introduce 
$\phi_x^{(cc')(bb')}:=\sqrt{P_b\, P_{c'}\over P_{b'}\,P_c}\,
E_{x}^{(cc')(bb')}= S(E_{x^*}^{(b'b)(c'c)})$,
so that $\bra\phi^+, e\ket=\overline{\bra \phi, S(e)^+\ket}$.

The identity \Is\ and its dual complete the set of pentagon type
relations called ``the Big Pentagon'' in \BSz. In the diagonal
case $Z_{j\bj}=\delta_{j\bj}$ where all multiplicities $N_i\,,
n_i\,, \tilde{N}_{x}$ and $\tilde{n}_{x}$ coincide with the
Verlinde one, one can identify $\Fo=F=\Fot=\tFo=\tF$ since
all pentagon relations involved coincide with \pent\ and the
unitarity relations \unit\ and their dual counterparts, as well
as \inverse,  reduce to the unitarity of $F$. The next simple
cases are the permutation modular invariants $Z_{j\bj}=Z^{{\rm
diag}}_{j\zeta(\bj)}$, where $\zeta$ is an automorphism of the
fusion rules. For any of these cases $\tCV$ is identified with
$\CI$, $\tN=N\,, \tn=n$, and accordingly $\tF=F\,,\ \tFo=\Fo\,.$
We notice that in these cases the pentagon identity
\Iss\ looks like the fusing-braiding identity \Ii\
and this suggests that given $\Fo$, and hence by \Ie, given $\hat{B}$, the
latter matrix may provide, up to some constant, a solution for
$\Fot$. In the simplest example of the $D_{\rm odd}$ $sl(2)$ series
the matrices $\Fo$ were computed in \Rb.

Defining  the dual counterparts of \mcp\
%
\eqn\mcpd{ 
\hat{E}_x=\sum_{c\,,b\,,\eta}\ {1\over \tilde{c}_x^{bc}}
\ E_{x\,, \eta\,, \eta}^{(cb)(cb)}\,,
}
and using the analogues of \Io\ and \unit\ with $\Fo$ replaced by $\tFo$
one obtains the algebra \tNrepb\ with the multiplication identified with
the vertical product 
\eqn\Imm{
\hat{E}_x \otimes_v\, \hat{E}_y= \sum_z\, \tilde{N}_{x y}{}^z\, \hat{E}_z\,.
}
The identity in $\hA$ is given by 
$\un_h=\sum_{x\,, c\,,b\,,\eta}\ E_{x\,, \eta\,, \eta}^{(cb)(cb)}\,$
 and \Ir,
\inverse\ ensure that $\un_h$  coincides with the counit ${1\over |\CI|}\,
\epsilon$. 
\foot{
The factor $1/P_b P_{c'}$ in
\norma,  dictated by the requirement of
 consistency of the full set of pentagon 
and inversion equations,
can be assigned  entirely to one of the
constants $c_j$ or $\hat{c}_x$. Then one of the formulae \mcp\ or
\mcpd\   gives elements in the center of the corresponding algebra,
the so called ``minimal central projections''. 
However there seems to exist no consistent renormalisation
of the two products and of the relation on Fig. \dualdta\ making
central the basis elements of both algebras \abV\ and   \Imm.
}
The dual algebra  $\hA$ 
cannot be  turned in general into a quasitriangular WHA.

The relations  \Imm, \tosya, \misys, imply that the  ``chiral generators''  
\eqn\cgen{
p_j^{+}=\sum_x\, \tV_{j1;\, 1}{}^x \, \hat{E}_x\,,\qquad
p_j^{-}=\sum_x\, \tV_{1j^*;\, 1}{}^x \, \hat{E}_x\,,
}
%
satisfy 
the Verlinde algebra
\eqn\cgenb{p_i^{\pm}\otimes_v\, p_j^{\pm}=\sum_k\, N_{ij}{}^k\ p_k^{\pm}\,,
}
%
while 
\eqn\cgenc{
p_i^{+}\otimes_v\, p_j^{-}=\sum_z\, \M^z_{ij}\, \hat{E}_z
\,. 
}
Since ${1\over |\tCV|}\,
\tilde{\ze}(\hat{E}_z)= \delta_{z1}$, applying 
${1\over |\tCV|}\, \tilde{\ze}$
to \cgenc\ reproduces in the r.h.s. the modular invariant matrix 
$Z_{ij}=\M^1_{ij}$  
\AO; in  \BEK\  the analogous relation  reads 
$\bra \za^+_i, \za^-_j \ket=Z_{ij}$.

\bigskip


\appendix{B}{The $sl(2)$ theories}

\def\tP{\tilde P}

\def\hN{{\widehat {N}}}


\noindent 
In this appendix, we illustrate the construction of Ocneanu graphs 
and of the associated matrix algebras on 
the $\slh(2)$ theories and modular invariants of $ADE$ type.
 
\ifx\answ\bigans \fig{The Ocneanu graphs of $ADE$ type: each vertex $x$ is assigned
its matrix $\tV_1^x$, written as a $P$ or $\tP$ matrix as in \Vtilde\ or
in (B.9). Edges of $\tV_{21}$, resp $\tV_{12}$ are shown in red full lines,
resp blue broken ones, and the vertices of the
different ``cosets'' for the action
of $\tV_{21}$ are depicted in different colours.}
{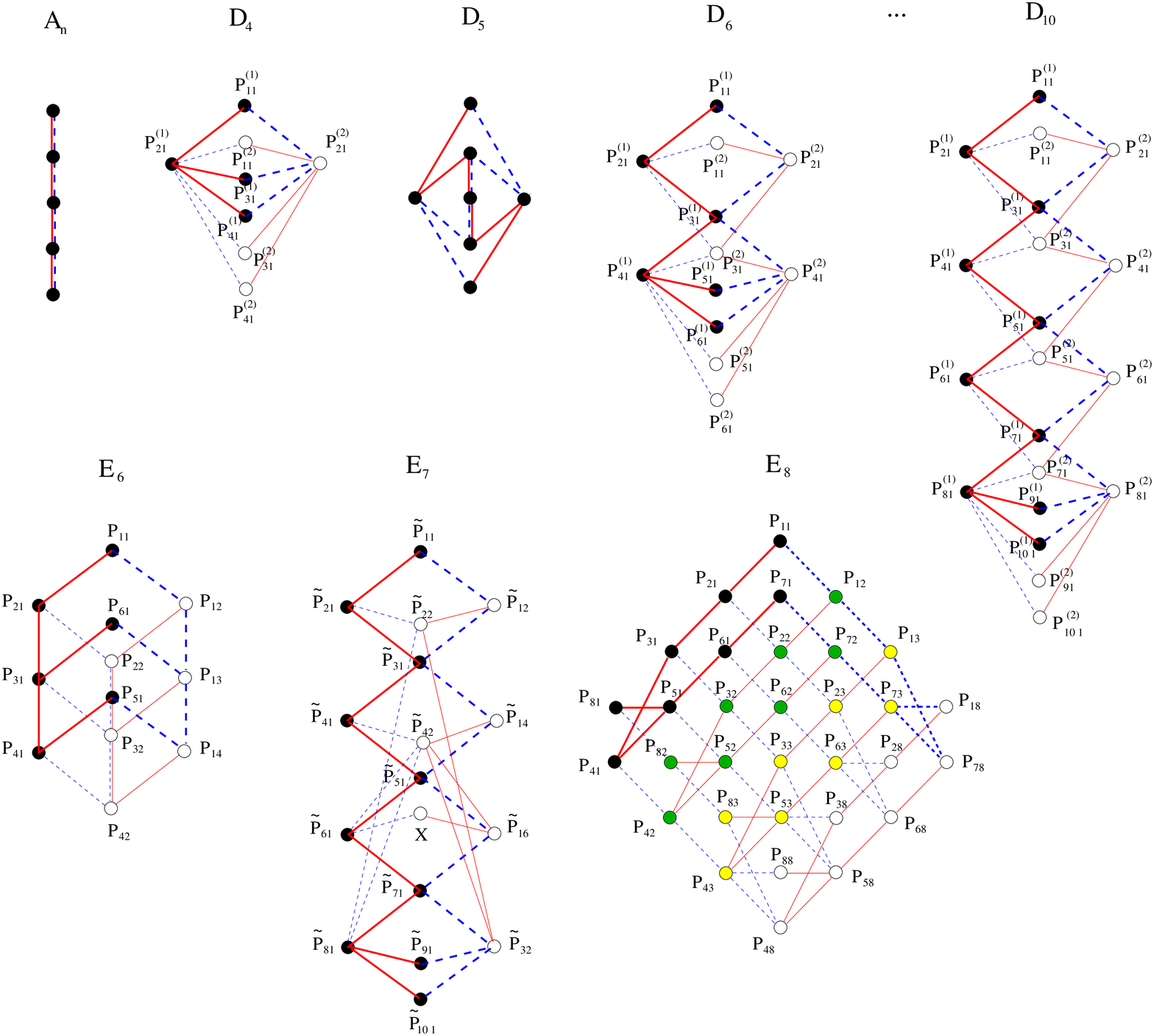}{15cm}\figlabel\ocngraph
\else 
\fig{The Ocneanu graphs of $ADE$ type: each vertex $x$ is assigned
its matrix $\tV_1^x$, written as a $P$ or $\tP$ matrix as in \Vtilde\ or
in (B.9). Edges of $\tV_{21}$, resp $\tV_{12}$ are shown in red full lines,
resp blue broken ones, and the vertices of the
different ``cosets'' for the action
of $\tV_{21}$ are depicted in different 
colours.}{ocngraph.eps}{12cm}\figlabel\ocngraph\fi

The cases of $A_{n}$ and $D_{2\ell+1}$ have been covered by the discussion
in  section 7.5: the $A$ cases are diagonal theories and the  $D_{2\ell+1}$
 case is obtained from the diagonal case $A_{4\ell-1}$ with the same
Coxeter number $h=4\ell$ 
by the $\Bbb{Z}_2$  automorphism $\zeta(j)=h-j$ of the fusion rules. 
The $D_{2\ell}$, $E_6$ and $E_8$ cases have also been implicitly 
covered there. But we shall collect here additional data on them 
and present the $E_7$ case which does not follow from the previous formulae.
 Throughout this appendix, we follow the notations of \BPPZ\
on the vertices and on the eigenvectors of ordinary Dynkin diagrams.

\def\sqt{\sqrt{2}}
For the $D_{2\ell}$ theories, in which condition \cond\ is satisfied, 
formula \Vtilde\ applies with $b=1$ and $\zk=0,1$. 
Diagonalising the matrices $\tV_1{}^x$ as explained in section 7.5, and with a 
little extra insight to  find the appropriate combinations of eigenvectors
$\Psi$ with exponent $(\J)=(j,\bj)=(2\ell-1,2\ell-1)$ of multiplicity 
$(Z_{j\bj})^2=4$, we find that 
\ifx\answ\bigans
\eqn\PsiDev{       
\{\Psi_x^{(\J;\za,\zb)}\}=\Bigg(  
\underbrace{
\matrix{
{\psi_a^j\over\sqt}&{\psi_a^j\over\sqt}&{\psi_a^{h-j}\over\sqt}
&{\psi_a^{h-j}\over\sqt}
\cr
{\psi_a^j\over\sqt}&-{\psi_a^j\over\sqt}&-{\psi_a^{h-j}\over\sqt}
&{\psi_a^{h-j}\over\sqt}
\cr}
}_{j=1,3,\cdots,2\ell-3} 
\matrix{
{\hbox{\raise -3 mm \hbox{}}\psi_a^{(2\ell-1,+)}} &{\psi_a^{(2\ell-1,-)}}&0&0\cr
0&0&{\psi_a^{(2\ell-1,+)}}&{\psi_a^{(2\ell-1,-)}}\cr
}
\Bigg)
}
\else
\eqn\PsiDev{
\{\Psi_x^{(\J;\za,\zb)}\}\!=\!\!\Bigg(
\underbrace{
\matrix{
{\psi_a^j\over\sqt}&{\psi_a^j\over\sqt}&{\psi_a^{h-j}\over\sqt}
&{\psi_a^{h-j}\over\sqt}
\cr
{\psi_a^j\over\sqt}&-{\psi_a^j\over\sqt}&-{\psi_a^{h-j}\over\sqt}
&{\psi_a^{h-j}\over\sqt}
\cr}
}_{j=1,3,\cdots,2\ell-3}
\matrix{
{\hbox{\raise -3 mm \hbox{}}
\!\!\psi_a^{(2\ell-1,+)}} &\!\!\!\!{\psi_a^{(2\ell-1,-)}}&0&0\cr
0&0&\!\!\!\!{\psi_a^{(2\ell-1,+)}}&\!\!\!\!{\psi_a^{(2\ell-1,-)}}\cr
}
\Bigg)
}
\fi
\noindent which should be understood as follows: 
the exponent of $\psi$  in the  first four columns 
 run over $j=1,3,\cdots,2\ell-3$, ($h=4\ell-2$ is the 
Coxeter number), and for each $j$ the corresponding value of the pair 
$(\J)=(j,\bj)$ is successively $(j,j),\,(j,h-j),(h-j,j),\,(h-j,h-j)$.  
In the last four columns, the exponent of $\Psi$ is 
$({h\over 2},{h\over 2};\,\za,\zb)$
with successively $(\za,\zb)=(1,1),(2,2,),(1,2),(2,1)$.
The row index $x$ is of the form $x=(a,\kappa)$, as in sections 7.5, 
and the first line of \PsiDev\ refers to $\zk=0$, the second to $\zk=1$. 

It is then easy to compute the various sets of matrices discussed 
in section 7.  One finds that 
\eqn\tVDev{\eqalign{ \tV_{ij} &=
\pmatrix{n_i\,n_j&0\cr 0 &n_i\,n_j\cr} 
\hbox{ if $j$ is odd} \cr 
&=\pmatrix{0& n_i\,n_j\cr n_i\,n_j &0 \cr} 
\hbox{  if $j$ is even}\ , \cr }}
\eqn\tNDev{\eqalign
{\tN_x&=\pmatrix{\hN_a&0\cr 0&\hN_{a^c}\cr} \qquad {\rm if}\ \zk=0\cr
&= \pmatrix{0& \hN_a\cr \hN_{a^c}&0 \cr} \qquad {\rm if}\ \zk=1\cr
}}
and 
\eqn\tnDev
{ \tn_x=\cases{ 
\hN_{a}   & if $\zk=0$\cr
\hN_a C   & if $\zk=1$\cr
}}
where the index $c$ in \tNDev\ 
denotes the $\Bbb{Z}_2$ involution of vertices of the 
(ordinary) $D_{2\ell}$ diagram which exchanges the two vertices
of the fork and leaves the other invariant, and 
$C_{ab}=\delta_{ab^c}$. Using these data, one checks \equdim\ and \eqdi.
Finally the matrices $\tM$ restricted to the ``physical" subset,  
i.e. those that do not involve $j=\bar j=2\ell-1$ with labels $\alpha\ne
\beta$, 
are all non negative.  For $j_1,\bj_1$ etc $\ne {h\over 2}= 2\ell-1$:
\eqnn\tMDev
$$\eqalignno{
&\tM_{(j_1,\bj_1)(j_2,\bj_2)}^{\qquad\quad(j_3,\bj_3)}
=\cases{M_{j_1j_2}{}^{j_3} & {\ninepoint if there is 0 or 2 pairs
of $(j,h-j)$ among 
$(j_1,\bj_1),(j_2,\bj_2),(j_3,\bj_3)$} \cr
0 & {\ninepoint otherwise} \cr} \cr 
&\tM_{(j_1,\bj_1)(j_2,\bj_2)}{}^{(2\ell-1,2\ell-1;\,\alpha,\alpha)}=
{1\over\sqrt{2}} M_{j_1j_2}{}^{(2\ell-1,\alpha)} \cr
&\tM_{(2\ell-1,2\ell-1,\za,\za),(2\ell-1,2\ell-1,\zb,\zb)}{}^{(\J)}
=M_{(2\ell-1,\za)(2\ell-1,\zb)}{}^j \cr 
&\tM_{(2\ell-1,2\ell-1,\za,\za),(2\ell-1,2\ell-1,\zb,\zb)}
{}^{(2\ell-1,2\ell-1,\zg,\zg)}=\sqrt{2}
M_{(2\ell-1,\za)(2\ell-1,\zb)}{}^{(2\ell-1,\zg)}\ , \cr}$$
in terms of the ``ordinary'' Pasquier algebra structure constants
$M_{j_1j_2}{}^{j_3}$ for which explicit expressions can be found in 
the Appendix A of \PZa\foot{with unfortunately a misprint which we 
correct here: in the last line of (A.2), the ${1\over 2}$ should read 
${1\over\sqrt{2}}$. }. These expressions of $\tM$ are
in agreement with their connection  with the relative structure 
constants \OPA.

\bigskip


We now turn to the three exceptional cases.

\bigskip
{\bf The case of $E_6$}\par
In that case, it suffices to take $x=(a,b)$, $a=1,\cdots,6$, $b=1,2$ 
and the  $\tV_1{}^x$ equal to the matrices $P_{ab}:=P^{(1)}_{ab}$. 
According to what was stated above in eq \leftN, 
the two sets $\{P_{a1}\}$  and  $\{P_{a2}\}$ 
are separately closed upon the left action of
$N_2$. Moreover, because of symmetries
$P_{13}=P_{62}\,,\ P_{14}=P_{52}\ ,$
$P_{16}=P_{61}\,,\ P_{15}=P_{51}\,,\ P_{32}=P_{23}\,,\ P_{42}=P_{24}$
the two sets may also be regarded as $\{P_{1a}\}$ and 
$\{P_{2a}\}$, $a=1,\cdots 6$ and are separately closed upon right 
action of $N_2$. See figure \ocngraph\ on which each vertex $x$ of
the graph $\widetilde{E_6}$ is assigned its matrix $\tV_{}^x$. 

Using \eigenvem, it is easy to compute the various sets of matrices discussed 
in section 7. One finds, in accordance with \tilint, that 
\eqn\tNDevEs{\eqalign
{\tN_x&=\pmatrix{\hN_a&0\cr 0&\hN_{a}\cr} \qquad {\rm if}\ x=(a,1)\cr
&= \pmatrix{0& \hN_a\cr \hN_{a}& \hN_a\hN_6\cr} \qquad {\rm if}\ x=(a,2)\cr
}}
and $\tn_x$ is given by the last equation \tilint. 
One also computes
$m_j=6,10,14,18,20,20, 20\,,\qquad$ $ 18,14,10,6\,$ for $j=1,\cdots, 11$;  
$\tm_x=6,10,14,10,6,8$ and $10,20,28,20,10,14$
for $x=(a,b=1)$ and $(a,2)$ respectively,  $a=1,\cdots,6$.
Hence one checks \eqdi: $\sum_j m_j=156=\sum_x \tm_x$, and \equdim:
$\sum_j m_j^2=\sum_x \tm_x^2=2512$. 
Finally the matrices $\tM$ factorise into a product of ordinary 
Pasquier matrices
\eqn\factorEs{\tM_{(\I)\,(\J)}{}^{(\K)}=
M_{ij}{}^k\,M_{\bi\bj}{}^{\bk}\ .}

\bigskip
{\bf The case of $E_8$}\par
In that case there are $8\times 8/2=32$ matrices that 
may be taken either as the four sets 
$\{P_{a1}\}$, $\{P_{a2}\}$, $\{P_{a3}\}$ and $\{P_{a8}\}$,
$a=1,2,\cdots, 8$,
or using once again their symmetries, as
$\{P_{1a}\}$, $\{P_{2a}\}$, $\{P_{3a}\}$ and $\{P_{8a}\}$,
$a=1,2,\cdots, 8$. See Fig \ocngraph.

One then computes
\eqn\tNDevEe{\eqalign
{\tN_x&=\pmatrix{\hN_a&0&0&0\cr 0&\hN_{a}&0&0\cr
0&0&\hN_a&0\cr 0&0&0&\hN_{a}\cr} \qquad {\rm if}\ x=(a,1)\cr
&= \pmatrix{0& \hN_a&0&0\cr \hN_{a}&0&\hN_a&0 \cr
0& \hN_a&0&\hN_7\hN_a\cr 0&0&\hN_7\hN_{a}&0 \cr} \qquad {\rm if}\ x=(a,2)\cr
&= \pmatrix{0&0& \hN_a&0\cr 0&\hN_{a}&0& \hN_7\hN_a\cr
\hN_a& 0&(\hN_1+\hN_7)\hN_a&0\cr 
0& \hN_7\hN_a&0& \hN_7\hN_a \cr
} \qquad {\rm if}\ x=(a,3)\cr
&= \pmatrix{0&0&0 &\hN_a\cr0&0&  \hN_7\hN_a&0\cr
0&\hN_7\hN_a&0 &\hN_7\hN_a \cr \hN_a&0&  \hN_7\hN_a&0\cr
} \qquad {\rm if}\ x=(a,8)\cr
}}
and $\tn_x$ as in \tilint. 
Also
$m_j=m_{30-j}=8,14,20,26,32,38,44,48,52,56,60,62,64,64,64\,$ for
$j=1,\cdots, 15$;   
$\tm_x= (8,14,20,26,32,22,12,16) $, $ (14,28,40,52,64,44,22,32)\,, \qquad $ 
$ (20,40,60,$ $78,96,64,32,48)$ and $(16,32,48,64,78,52,26,40)$
for $x=(a,b=1)$,  $(a,2)$,  $(a,3)$ and  $(a,8)$,  
respectively,  $a=1,\cdots,8$.
Hence one checks $\sum_j m_j=1240=\sum_x \tm_x$, 
$\sum_j m_j^2=\sum_x \tm_x^2=63136$. 
Finally the matrices $\tM$ factorise again into a product of ordinary 
Pasquier matrices, like in \factorEs.

\bigskip

\bigskip
{\bf The case of $E_7$}

This case is known to be related to the $D_{10}$ case. 
The $P^{(1)}_{ab}$ matrices of $D_{10}$ were defined in \Vtilde\  by
$$ (P^{(1)}_{ab})_{ij}=\sum_{c\in T= \{1,3,5,7,9,10\}}
n_{ia}{}^c n_{jb}{}^c\ ,$$
with $n$ the solutions of \nim\ pertaining to $D_{10}$.
Using the same matrices, let us now define
the $\tP$ matrices (twisted version of the $P$'s) by
\eqn\Ptilde{ (\tP_{ab})_{ij}=\sum_{c\in \{1,3,5,7,9,10\}}
n_{ia}{}^c n_{jb}{}^{\zeta(c)}\ , } 
with $\zeta$ the usual involution acting on the 
vertices of $T$ 
$$ \{1,3,5,7,9,10\} \mapsto \{1,9,5,7,3,10\} .$$ 

As in section 7 (equation \leftN),  we have the property that upon 
left (resp right) multiplication by $N_2$, 
$ N_2.\tP_{ab}=\sum_{a'} n_{2a}{}^{a'}(\tP_{a'b})\, ,$
(resp $ \tP_{ab}.N_2 =\sum_{b'} (\tP_{ab'}) n_{2b'}{}^{b}\ .$)
Recall that here $n_2$ is the adjacency matrix of $D_{10}$. 
This relation explains the $D_{10}$ pattern of the two chiral
parts of the  Ocneanu graph $\widetilde{E_7}$ on Fig. \ocngraph: 
the red full (resp blue broken) thick line
represents left (resp right) fusion by $N_2$ and connects the matrices
$\tP_{a1}$, (resp $\tP_{1a}$), $a=1,\cdots 10$. 

These matrices have to be supplemented by others to produce the 
full set of matrices $\M^x$ and the second part (the ``coset'' \AO)
of the graph $\widetilde{E_7}$. Using the symmetries
$(\tP_{ab})^T=\tP_{ba}$ etc of the matrices $\tP$, we find that 
starting with matrix $\tP_{12}$, 
left multiplication by $N_2$ produces the  chain
of matrices forming the coset
$$\eqalign{ 
   &\!\! \buildrel{\displaystyle \tP_{16}}\over{\uparrow} \cr
 \tP_{12} \rightarrow \tP_{22}
\rightarrow \tP_{32}=\tP_{18} \rightarrow \tP_{42}=&\tP_{24}
\rightarrow \tP_{14} \rightarrow X:=\tP_{26}-\tP_{24}\ ,\cr}$$
where the splitting of $\tP_{52}$
into the sum  $\tP_{14}+\tP_{16}$ has formed the triple point
of the $E_7$ diagram. 
The matrix $X$ itself may be expressed as a bilinear form in the 
matrices $n$ (relative to $D_{10}$)
\eqn\Xmat{X=\sum_{a,b=2,4,6,8}(\hN_7-\hN_9)_a{}^b\, n_{i1}{}^a n_{j1}{}^b } 
in such a way that the pairs $(a,b)$ that are summed over are
$$ (a,b)\in \{(4,4),(6,6),(8,8),(2,6),(6,2),(4,8),(8,4),(6,8),(8,6)\}\ .$$
One finds,  following downward first 
the $D_{10}$ subgraph, and then the $E_7$ coset
\eqn\tNesev{\eqalign{
&\tN_x=\pmatrix{\hN_x&0\cr 0& n^{(E_7)}_x\cr}\,,\ x=1,8;\qquad 
\tN_9=\pmatrix{\hN_9& 0\cr 0&n^{(E_7)}_9-n^{(E_7)}_3\cr}\,;\cr 
&\tN_{10}=\pmatrix{\hN_{10}& 0\cr 0&n^{(E_7)}_3\cr}
\,,\qquad \tN_x=\pmatrix{0&\cn_{x-10}\cr \cn_{x-10}^T &0\cr}\,, 
\ x=11,\cdots, 17\ .
}}
where $n^{(E_7)}$ denote the $n$-matrices of $E_7$, and $\hN$ 
are relative to $D_{10}$; $\cn_b$, $b=1,\cdots,7$, are seven 
$10\times 7$ rectangular matrices intertwining the $D_{10}$ and
$E_7$ adjacency matrices (see \BPPZ, sec. 3.3, for a formula),   

$$\eqalign{
 &\tn_x=n^{(E_7)}_i\,,\ x=1,8;\ \tn_9=n^{(E_7)}_3\,; \tn_{10}=n^{(E_7)}_9-
n_3^{(E_7)};\, \tn_{11}
=n^{(E_7)}_2\,; \tn_{12}=n^{(E_7)}_1+n^{(E_7)}_3\,;\cr
&\tn_{13}=n^{(E_7)}_{8}\,; \tn_{14}=n^{(E_7)}_{3}+n^{(E_7)}_{5}\,;
\tn_{15}=n^{(E_7)}_{8}\,; \tn_{16}
=n^{(E_7)}_{3}+n^{(E_7)}_{5}\,; 
\tn_{17}=n^{(E_7)}_{7}-n^{(E_7)}_{3}\,.}
$$	
One also computes $m_i=m_{18-i}=7,12,17,22,27,30,33,34,35$ for $1\le i\le 9$,
and $\tm_x=7,12,17,22,27,30,33,34,17,18$; $ 12,24,34,
44,30,16,22$,
so that $\sum_i m_i=\sum \tm_x=399$, $\sum_i m_i^2=\sum_x \tm_x^2=10905$. 
Finally, the $\tM$ matrices may also be computed, and yield non 
negative numbers ($0, {1\over 4}, {1\over 2}, {1\over\sqrt{2}},{3\over 4}, 
1,\sqrt{2}, 2)$, 
which match what was computed on the relative structure constants $d^2$. 
\bigskip
\listrefs

\bye